\documentclass[11pt]{article}
\usepackage[margin=1in]{geometry}
\usepackage{amsmath,amssymb,amsfonts}
\usepackage{graphicx}
\usepackage{tikz}
\usepackage{caption}
\usepackage[colorlinks=true,linkcolor=blue,citecolor=blue,urlcolor=blue]{hyperref}
\usetikzlibrary{arrows.meta,positioning,calc,shapes.geometric}
\usepackage{bookmark}

\newcommand{\ct}[1]{{n_{#1}}}
\newcommand{\ca}[1]{{n^{a}_{#1}}}
\newcommand{\cb}[1]{{n^{b}_{#1}}}
\newcommand{\cm}[1]{{n^{m}_{#1}}}
\newcommand{\Amap}{\cm{A}}
\newcommand{\Bmap}{\cm{B}}
\newcommand{\Cmap}{\cm{C}}
\newcommand{\Dmap}{\cm{D}}
\newcommand{\ncard}[1]{\left\|#1\right\|}
\newcommand{\sgdetector}[4]{%
 \begin{scope}[xshift=#1cm]
  \filldraw[fill=black!10,draw=black] (0,1.7) -- (0.5,2.1) -- (2.5,2.1) -- (2,1.7) -- cycle;   
  \filldraw[fill=black!18,draw=black] (2,0) -- (2.5,0.4) -- (2.5,2.1) -- (2,1.7) -- cycle;     
  \filldraw[fill=white,draw=black] (0,0) rectangle (2,1.7);                                     
  \ifnum#4=0 \else\draw[dashed,black!60] (1,0.85) ++(0,-0.7) -- ++(0,2.5);\fi                
  \draw[-{Stealth[length=2.4mm]},thick,#2] (1,0.85) ++(#4+270:0.7) -- ++(#4+90:2.35);           
  \fill[#2!75!black] (1,0.85) circle (0.34);                                                     
  \node[#2!70!black,below] at (1,-0.05) {#3};
 \end{scope}}
\newcommand{\sgangle}[2]{%
  \draw[black!75,-{Stealth[length=1.6mm]}] (#1,0.85) ++(90:1.5) arc (90:65:1.5);
  \node at ($(#1,0.85)+(77:1.85)$) {#2};}

\title{An event-centric approach to modeling quantum systems without postulating Hilbert space, the Schr\"odinger equation or the Born rule}
\author{Sam Powers and Dejan Stojkovic\\
\small Department of Physics, SUNY at Buffalo, Buffalo, NY 14260-1500, USA}
\date{}

\begin{document}
\maketitle

\begin{abstract}
We define an event-centric framework for modeling quantum systems whose only modeling elements are
classical bits. An event is a correlation between two binary sequences, one representing the observer and
one the observed system, and two events are related by a map, itself a sequence, through the XOR
operation. We postulate that observers can access only the counts of symbols in events and maps, not
their arrangement. Probabilities then follow from counting the compatible microstates, with
indistinguishable microstates interfering through a $\mathbb{Z}_2$ sign, so that epistemic indeterminism
and quantum spin statistics emerge without postulating Hilbert space or the Born rule. Applied to three
scenarios involving spin, the model reproduces the Wigner $d$-matrix and Clebsch-Gordan probabilities in
the sectors studied, with no continuous free or fitted parameters and with discrepancies that vanish as
$O(1/n)$ in the sequence length $n$. In a Bell test it violates the CHSH inequality and approaches the
Tsirelson bound as $n\to\infty$. Because the ontic state is attached to a pair of events, including the
measurement context, the framework instantiates neither the single context-independent state space
assumed by the Kochen-Specker argument nor the ontological-models framework assumed by the
Pusey-Barrett-Rudolph theorem.
\end{abstract}

\section{Introduction}
\label{sec:intro}

Is nature discrete? Certainly, our empirical description of nature is discrete. Over the full span of
humanity's existence, we have thus far collected only a finite amount of information from our
environment. Even if our universe continues on for infinite time, our collective data set will at best
achieve a cardinality of $\aleph_0$, or the countable infinity \cite{Dauben}. This fact stands in
contrast to the uncountable sets employed in our best theoretical descriptions of nature. For example,
the probabilities generated by quantum mechanics (QM) are elements of the real number line. To observe a
physical state associated with these probabilities, one would need an uncountable set of identically
prepared quantum states, not to mention a measurement apparatus capable of performing continuous
measurements. In the modern view of the physical world, both of these conditions are untenable, not merely because
they are unrealistic, but because they are in direct conflict with the discrete nature of matter, as
established by the likes of Boltzmann, Planck, and Einstein.

Each bit of information we collect from our environment is the consequence of one or more discrete
interactions between material objects. These discrete interactions are generically called events. Rather
than focusing on particles or fields, some have found it prudent to instead focus on the event as a
potential building block for models of physical systems \cite{Rovelli,Haag,Froehlich,Drossel,Morgan,Raedt,Giovannetti,Cortes,Markopoulou,Loll,Surya}.
In the event-centric picture, one typically replaces the continuous evolution of particles or fields
with the discrete evolution of causal networks of events, one possible illustration of which is offered
in Figure \ref{fig:network} (left). Perhaps the most intriguing aspect of the event-centric approach is
its apparent relevance in both quantum foundations and quantum gravity research programs.

In quantum foundations, there are still open questions about the reality, or ontological status of the
quantum state \cite{Leifer,Bacciagaluppi}. Do the abstract mathematical structures that arise in quantum
theories actually represent real physical objects? Or, are they a consequence of an observer's limited
information about some more fundamental reality? These questions have spawned many interesting
proposals, some of which lead to radical conclusions about nature, such as the existence of infinite
worlds \cite{DeWitt}. For many, the idea that quantum theory may just be a consequence of hidden
information is the most reasonable. Two claims are easily run together here, and it is worth separating
them at the outset. The first is that a quantum theory may be underwritten by hidden variables. The
second, and stronger, is that the quantum state is itself a state of incomplete information, so that
distinct pure states overlap on the underlying space of real states. It is the second claim which is
excluded by the theorem of Pusey, Barrett and Rudolph \cite{PBR}, within the ontological-models framework
in which that theorem is formulated. The framework developed in this paper makes the first claim and not
the second: it is a hidden information theory in which the hidden information is the microstate underlying
a macrostate rather than the quantum state itself. As formulated here, however, it does not instantiate
the ontological-models framework, because its ontic states are attached to pairs of events together with the relation
between them, rather than to a system on its own. We return to this point, and to what it costs, in
section \ref{sec:psiontic}.
Unfortunately, attempts to construct quantum theories involving hidden information often lead to
important conflicts with experiments. In particular, models which
attempt to ascribe a definite reality to the state of a particle or local field inevitably fail to
correctly predict the statistical correlations observed in Bell tests \cite{Bell,CHSH}. This is precisely
why the event-centric picture is so enticing. Among other advantages, it enables one to construct a
hidden information theory which reproduces the correlations of a Bell test. As we make explicit in
section \ref{sec:whereitenters}, it does so because the measurement settings are part of the ontic state,
so that the measurement-independence assumption of Bell's derivation is not satisfied; in this respect
the framework belongs with the ``all-at-once'' and retrocausal approaches reviewed in
\cite{WhartonArgaman,Adlam}, and its standing with respect to the no-go theorems \cite{KS,PBR} is
discussed in section \ref{sec:contextuality}.

In the more mature approaches to quantum gravity, such as String Theory \cite{Mukhi} and Loop Quantum
Gravity \cite{Ashtekar}, significant progress has been made towards unifying quantum field theories with
general relativity. Though hope remains that the challenges these approaches currently face can be
overcome, it is important for the broader community to continue to develop fresh ideas. Of course,
suggesting that the event could serve as a building block of spacetime is not a radical proposal, nor is
it particularly new. After all, the event is already an essential concept in general and special
relativity, typically thought of as a discrete unit of volume in spacetime. The only real novelty in an
event-centric approach is the assertion that the familiar mathematical structures in general relativity
can be constructed from, or approximated by these discrete elements. Many important advancements along
these lines have already taken place \cite{Markopoulou,Loll,Surya}.

That the event-centric picture offers potential solutions to two of the most important problems facing
modern physics is noteworthy. This, coupled with its compatibility with the discrete nature of empirical
data, suggests that the event may serve as a building block for a next generation theory of physics. The
present work may be viewed as a small step in this direction, where we suggest a general framework in
which simple physical models can be constructed from first principles. In particular, we focus on a
single primitive network element of a larger causal network of events. These primitive network elements
consist of two events, which may share a direct, or indirect causal connection. Direct causal connections
between two events may be interpreted as particles, whereas an indirect causal connection implies that
they are related through a third event. It is the latter scenario which corresponds to experiments
involving entangled particles, such as a Bell test. An example of each of these primitive network
elements is depicted in Figure \ref{fig:network} (right).

We should be clear that the present work by no means represents a complete theory of discrete causal
networks of events. This would require many non-trivial advancements beyond the results obtained thus
far. Rather, the focus of this work will be on establishing a simple yet novel framework in which a new
approach to this idea may be explored. To illustrate the manner in which physics models can be
constructed within this framework, we use the proposed formalism to generate predictions for three
experimental scenarios involving spin. The first involves two rotated Stern-Gerlach detectors that
interact with a single particle with total spin $j$, which was previously addressed in \cite{PowersSG}.
The second scenario involves two aligned Stern-Gerlach detectors that each interact with a separate
component of a single composite spin system, as studied in \cite{PowersSpin}. In the third and final
scenario, two rotated Stern-Gerlach detectors interact with separate components of a single composite
spin system, which enables one to calculate the CHSH test statistic \cite{CHSH}.

The present paper belongs to a broader program in which the same counting construction, built on
binary sequences observed only through their symbol counts, has been applied to quantum spin
\cite{PowersSpin,PowersSG}, to the quantum harmonic oscillator \cite{QHO}, and to the emergence of space,
time and the kinematic structure of special relativity \cite{SR}. Because the framework departs from
quantum mechanics at finite $n$, it is also experimentally testable: its predictions for photon Fock
states have recently been confronted with measurements on cloud photonic quantum processors, which bound
the sequence length $n$, interpreted as the information capacity of the register behind a prepared state,
from below \cite{Fock}.

A question which is unavoidable in the first and third of these scenarios is how the abstract counts
which label a map are to be identified with the physical angle of rotation between two spatial frames.
Section \ref{sec:calibration} fixes this identification, $\tan(\theta/2)=\Bmap/\Amap$, with no free
parameter, and states precisely which parts of the argument are imposed and, consequently, which of the
comparisons with quantum mechanics are genuine tests.

The one ingredient of the formalism that is not fixed by these considerations is the rule by which
indistinguishable ontic states interfere, which enters as an alternating sign attached to a non-local
quantum number (section \ref{sec:interference}). The combinatorial structure of the formalism reduces
the freedom in this rule to the choice of a $\mathbb{Z}_2$ character on a six-dimensional lattice, that is,
to six bits; the calculations reported here constrain two of them, and four remain undetermined (appendix
\ref{app:symmetry}). A companion paper \cite{QHO} examines the same
construction in its simplest setting, where the two events are binary sequences rather than base-4
sequences and the only non-local quantum number is the Hamming distance between them. There it is shown
that, once the transition measure is taken to be the product of a forward and a backward ensemble, the
requirement that the measure account for exactly the $2^n$ configurations of the register, together with
the symmetry under relabeling of the two symbols, leaves exactly one admissible weighting of the hidden
variable, namely the alternating sign, and that the resulting measure is the Krawtchouk finite oscillator
which contracts to the quantum harmonic oscillator as $n\to\infty$. We draw on that result at several
points below: it suggests an information-theoretic origin for the interference rule used here, although
the corresponding statement has not been proved for the larger alphabet of this paper, and it
sharpens the question of which parts of the present construction remain to be derived (appendix
\ref{app:symmetry}). The remaining discussion in this section is dedicated to
the development of concepts that are useful when interpreting the simple but abstract mathematical
elements employed in this framework.

Conceptually, it is beneficial to introduce the observer at this stage. Generically, an observer is an
entity that is capable of assembling information about events. They may then use that information, along
with a physics model, to infer properties of future events. Of course, invoking observers when modeling
physics can be a controversial step. Some may interpret this as a claim that the observer is of some
fundamental importance in nature. This is not the case. We do, however, argue that the observer is of
fundamental importance in our models of nature. While observer-dependent phenomena are well known in
physics, we will push this idea slightly farther. For instance, to obtain information about a given
event, an observer must directly participate in that event. Within the proposed framework, this implies
that events must have a substructure, which is one part system and one part observer.

We are now prepared to make a concrete modeling choice regarding events. To each event we associate an
ordered pair of base-2 sequences of length $n$, where one sequence is associated with the system, and one
with the observer. As will be discussed in section \ref{sec:ontic}, these base-2 sequences are comprised
of the symbols $0$ and $1$. We call them base-2 (or binary) sequences. Here and throughout,
``base-$k$'' refers to an alphabet of $k$ symbols, not to positional numeral notation: the only map from
sequences to numbers used in this work is the counting of symbols. Formally, an event is the ordered pair
$e=(r,s)$ of a reference sequence $r$ and a system sequence $s$, and the primitive network element studied
in this paper is the triple $(e_a,e_b,M)$ of two events together with the map $M=e_a\oplus e_b$ relating
them. To
each ordered pair of base-2 sequences of length $n$, which may be interpreted as base-4 sequences of the
same length, we assign four non-negative integers called counts. Each of these counts represents the
frequency with which a given symbol appears in the assigned sequence. These numbers form the basis of all
physical observables associated with events, one such observable being total spin $j$. A complete
introduction to counts is presented in section \ref{sec:order}.

Also discussed in section \ref{sec:order} is the concept of a map. Just like events, maps are modeled by
ordered pairs of base-2 sequences, which may again be thought of as base-4 sequences. The only difference
being that the base-4 sequences associated with maps also come with the addition modulo two operation
($\oplus$), which elevates them to operators within this developing formalism. Also like events, a map
can be assigned four counts, which are again used as the basis of physical observables. Given a single
event, a map can be used to generate a second event, such that important aspects of the causal
relationship between those events are encoded in the map. However, when taken independently, neither the
initial event nor the map contains sufficient information to completely determine the second event.

To better understand the information stored in the causal relationship between two events, it is useful
to think of ordered pairs of events as base-16 sequences. In other words, an ordered pair of base-4
sequences of length $n$, which we use to model events, can be thought of as a single base-16 sequence of
the same length. As was done with base-4 sequences, we assign 16 counts to these sequences. Of these 16
counts, only 10 can be associated with the two events and the map which relate them. The remaining 6 are
emergent properties of this causal relationship. Conceptually, we distinguish these 6 counts from those
associated with events and maps by invoking the notion of locality. That is, counts associated with
events and maps are taken to be local, while the remaining 6 are taken to be non-local. These non-local
degrees of freedom are discussed at length in sections \ref{sec:indist} and \ref{sec:interference}.

To this point, we have yet to address what was perhaps the most shocking discovery in the history of
physics. Nature appears to be non-deterministic. In the present framework this indeterminism is
epistemic: every ontic state is definite, and the stochasticity of outcomes arises from an epistemic
constraint which we impose on observers. This count-only accessibility condition is a postulate of the
model; it is not derived from the binary ontology. Within this framework, the information observers are permitted to have
about the causal relationship between two events is limited to local counts. This implies that the
information stored in the ordering of symbols within events and maps is hidden from observers. Under
this constraint, one may interpret a particular choice of local counts as specifying the macroscopic
state of a causal network, which is analogous to a quantum state within this framework. For a given
macroscopic state, there will typically be many associated microscopic states, each of which is modeled
by a unique sequence. These microscopic states then represent the possible ontic states, or real physical
states of the underlying causal network.

By limiting the information observers may have about causal networks to macroscopic data, we are forced
to treat them statistically. In other words, the information an observer has about a given causal network
will always take the form of an ensemble of microstates, which we call an ontic state space. The physical
scenarios of interest in this paper all involve two events, each of which is associated with one of the
primitive causal networks depicted in Figure \ref{fig:network} (right). Because these events are
necessarily separated in spacetime, two independent ontic state spaces must be constructed, one for each
event. We may think of these state spaces as being associated with two different observers, or with a
single observer at two different times. Regardless, each one encodes the macroscopic information that is
available to an observer located at that event. Contact is then made with quantum theory by counting the
paths between these state spaces. That is, within this framework, quantum theory concerns the discrete
evolution of statistical ensembles of causal networks.

The discrete nature of the proposed formalism might be of concern for some. After all, our best models of
nature thus far assume continuity, or the existence of uncountable state spaces. Any proposal for a
discrete model should be able to explain why these continuous models work so well. This leads us to one
of the most intriguing features of the proposed framework and the underlying mathematical formalism. In
the limit that the sequence length $n$ goes to $\aleph_0$ (countable infinity), the number of possible
microstates goes to $2^{\aleph_0}$, the cardinality of the continuum (uncountable infinity). Of course, one can always choose a finite $n$ that
is large enough to support a physics model which is sufficiently continuous to agree with empirical data.
Consequently, the question of whether a physics model can be supported by a discrete formalism has already
been answered in the affirmative. It is only a matter of finding the correct approach. Further discussion
on the transition between formalisms employing countable and uncountable sets can be found in
\cite{Raptis,Levi,Hietarinta}.

Many important details have been omitted from this sketch of the proposed framework. In particular, we
have not discussed how it fits within the broader field of quantum foundations. Due to the technical
nature of this discussion, it is difficult to address these points without first providing a more
complete introduction to the formalism. This will be the subject of section \ref{sec:formalism}, which
culminates in a brief discussion about the ontology of probabilities within this framework. The symmetry
structure generated by the transposition operators, which fixes several features of the formalism that
might otherwise appear to be choices, is analyzed separately in appendix \ref{app:symmetry}.
In section \ref{sec:models}, the proposed formalism is used to model the three physical scenarios
mentioned earlier, each of which is associated with one of the primitive network elements depicted in
Figure \ref{fig:network} (right). Section \ref{sec:calibration} is devoted to the calibration of the
rotation angle, which is where the half-angle law is established. In section \ref{sec:results}, these
models are then used to make predictions for physical systems involving several different values of total
spin $j$, which are then compared to non-relativistic QM. Section \ref{sec:contextuality} then examines
the sense in which the framework is contextual, together with its standing with respect to the no-go
theorems of Kochen and Specker \cite{KS} and of Pusey, Barrett and Rudolph \cite{PBR}. We then close
with a brief discussion on future directions of this work in section \ref{sec:discussion}.

\begin{figure}[t]
\centering
\begin{tikzpicture}[scale=0.85,
  ev/.style={circle,draw=black!70,fill=black!35,minimum size=5.5mm,inner sep=0pt},
  evd/.style={circle,draw=black!80,fill=black!55,minimum size=7mm,inner sep=0pt},
  evl/.style={circle,draw=black!30,fill=black!15,minimum size=5mm,inner sep=0pt},
  lk/.style={-{Stealth[length=2mm]},black!25,thick},
  ck/.style={-{Stealth[length=2.2mm]},black!80,thick,dashed}]
 \node[evl] (a1) at (0,0.2) {};   \node[evl] (a2) at (0.9,1.6) {};
 \node[evd] (a3) at (1.9,2.5) {}; \node[evl] (a4) at (1.0,-1.0) {};
 \node[evl] (a5) at (2.2,0.9) {}; \node[evd] (a6) at (2.6,-0.4) {};
 \node[evl] (a7) at (3.4,1.9) {}; \node[evd] (a8) at (4.2,2.6) {};
 \node[evl] (a9) at (4.4,0.6) {}; \node[evl] (a10) at (3.2,-1.2) {};
 \node[evl] (a11) at (5.2,1.5) {};\node[evl] (a12) at (1.6,3.3) {};
 \foreach \x/\y in {a1/a2,a2/a3,a4/a5,a5/a3,a1/a4,a5/a7,a7/a8,a9/a11,a10/a9,a4/a6,a6/a9,a2/a12,a12/a8,a11/a8,a10/a6,a7/a11}
   \draw[lk] (\x) -- (\y);
 \draw[ck] (a6) -- (a3); \draw[ck] (a6) -- (a8); \draw[ck] (a4) -- (a3);
 \begin{scope}[xshift=7.4cm]
   \node[evd] (i1) at (0,1.6) {}; \node[evd] (i2) at (1.7,1.6) {};
   \node[evd] (i3) at (0.85,0) {};
   \draw[ck] (i3) -- (i1); \draw[ck] (i3) -- (i2);
   \draw[rounded corners=8pt,black!70] (-0.45,1.15) rectangle (2.15,2.05);
   \node[evd] (d1) at (3.5,1.6) {}; \node[evd] (d2) at (5.2,1.6) {};
   \node[evd] (d3) at (4.35,0) {};
   \draw[ck] (d3) -- (d1); \draw[ck] (d3) -- (d2);
   \draw[rounded corners=10pt,black!70,rotate around={-42:(4.78,0.8)}]
        ($(4.78,0.8)+(-1.15,-0.45)$) rectangle ($(4.78,0.8)+(1.15,0.45)$);
 \end{scope}
\end{tikzpicture}
\caption{(Left) One possible illustration of a causal network, which takes the form of a directed graph.
The vertices of this graph denote events and the edges denote direct causal connections between events.
An event is a base-4 sequence defined as a correlation between two base-2 sequences, one being the reference sequence representing an observer and the other being an observed entity.
Events are connected by maps (also base-4 sequences), which are just XOR differences between the events.  Among the other things, events can represent  particles and/or spacetime points, while maps can represent any transformation between the initial and final state like rotations, translations, interactions etc.
(Right) The two primitive network elements that will be of interest in this paper. The one on the left
depicts an indirect causal connection between two events, while the one on the right depicts a direct
causal connection.}
\label{fig:network}
\end{figure}

\section{Formalism}
\label{sec:formalism}

In this section, the necessary mathematical elements and associated notation are introduced, which
involve only the most basic ideas from finite groups \cite{Zee}, set theory \cite{Robinson}, and
combinatorics \cite{Faticoni}. For convenience, the notation introduced below is collected in Table
\ref{tab:notation}. Included in this section will be many important technical details that
were not addressed in the conceptual development found in section \ref{sec:intro}. Among these details will be the
construction of an observer's ontic state space, the issue of interference, and the procedure for
calculating probabilities.

\begin{table}[htbp]
\centering
\small
\begin{tabular}{ll}
\hline
symbol & meaning\\
\hline
$n$ & sequence length, also the continuity parameter\\
$X,Y,Z$ & base-4 symbols, ranging over the alphabet $\{A,B,C,D\}$\\
$\ct{XY}$ & count of the base-16 symbol $XY$ (16 of them)\\
$\ca{X}$, $\cb{Y}$ & base-4 counts of Alice's and Bob's event: row and column sums of $\ct{XY}$\\
$\cm{Z}$ & base-4 counts of the map: sums of $\ct{XY}$ over $X\oplus Y=Z$\\
$\alpha,\beta,\gamma$ & map quantum numbers, equations (\ref{eq:alpha}-\ref{eq:gamma})\\
$j_a,m_a,l_a$ & quantum numbers of Alice's event; $j_b,m_b,l_b$ for Bob's\\
$j_{\bar a},m_{\bar a},l_{\bar a}$ & the same with the reversed projection convention, $m_{\bar a}=-m_a$\\
$j_c,m_c,l_c$ & quantum numbers of the composite system\\
$q=(\ell,\lambda)$ & full label; $\ell$ the 10 local, $\lambda$ the 6 non-local quantum numbers\\
$\varepsilon^a(q)$, $L^a(\ell)$, $E^a(u)$ & elementary, local and complete ontic state spaces of Alice\\
$\varepsilon^a_\ast(q)$ & elementary space with unobserved symbol configurations released\\
$\ncard{\ \cdot\ }$ & post-interference cardinality\\
$T^{XY}$ & transposition of the symbols $X$ and $Y$ in Bob's base-4 sequence\\
$R_g$ & the pair operation $Y\mapsto Y\oplus g$, equation (\ref{eq:commuting})\\
\hline
\end{tabular}
\caption{Summary of notation.}
\label{tab:notation}
\end{table}

\subsection{Ontic states}
\label{sec:ontic}

The building block of the proposed formalism is the finite group $\mathbb{Z}_2$. This group consists of
the symbols $0$ and $1$, which are equipped with the XOR (i.e.\ mod 2 addition) operation ($\oplus$). Under this
operation, these symbols obey the following addition table:
\begin{equation}
\begin{split}
0\oplus 0 = 1\oplus 1 = 0\\
1\oplus 0 = 0\oplus 1 = 1
\end{split}
\end{equation}

Ordered lists of these symbols shall be called base-2 sequences, where the set $S(n)$ contains all such
sequences of length $n$. The ontic state space for events is modeled by the Cartesian product of two
copies of this set:\footnote{Throughout, $\otimes$ placed between sets denotes their Cartesian product.}
\begin{equation}
S^2(n)\equiv S(n)\otimes S(n)
\end{equation}

The set $S^2(n)$ contains all base-4 sequences of length $n$, where each base-4 sequence is an ordered
pair of base-2 sequences. To track the ordering of the base-2 sequences within a given pair, we assign
alphanumeric subscripts to each. Typically, one of these base-2 sequences will be associated with the
observer of the event being modeled. We call this base-2 sequence a reference sequence. In this paper,
the reference sequence will be denoted by the subscript $a$ when modeling Alice's event or $b$ when
modeling Bob's, where Alice and Bob are named observers.

The symbols comprising base-4 sequences are members of the Klein four-group
$\mathbb{Z}_2\times\mathbb{Z}_2$. Writing each such symbol as an ordered pair whose first entry is the
reference bit and whose second entry is the system bit, the alphabet is
\begin{equation}
A\equiv(0,0),\qquad B\equiv(1,1),\qquad C\equiv(1,0),\qquad D\equiv(0,1)
\label{eq:alphabet}
\end{equation}
A single subscript records the register a symbol belongs to: $A_a$ is the symbol $A$ assembled from
Alice's reference sequence and her system sequence, and $A_b$ is the same symbol assembled from Bob's.
The subscript is dropped whenever the register is clear from context. This replaces the doubled
subscripts $a1$ and $b2$ used in \cite{PowersSG,PowersSpin}, which carried no information beyond the
identity of the observer.

An example of a base-4 sequence and its base-2 substructure is offered here, where the observer of the
modeled event is Alice:
\begin{equation}
\begin{pmatrix}0\\1\\1\\0\\1\\0\end{pmatrix}_a\otimes
\begin{pmatrix}0\\1\\0\\1\\1\\0\end{pmatrix}=
\begin{pmatrix}A\\B\\C\\D\\B\\A\end{pmatrix}_{a}
\end{equation}

These base-4 symbols, as members of the finite group $\mathbb{Z}_2\times\mathbb{Z}_2$, obey the following
addition table:
\begin{equation}
\begin{split}
A\oplus A=B\oplus B=C\oplus C=D\oplus D=A\\
A\oplus B=B\oplus A=C\oplus D=D\oplus C=B\\
A\oplus C=C\oplus A=B\oplus D=D\oplus B=C\\
A\oplus D=D\oplus A=B\oplus C=C\oplus B=D
\end{split}
\label{eq:z2z2}
\end{equation}

The ontic state space for two causally connected events is modeled by the Cartesian product of two copies
of $S^2(n)$:
\begin{equation}
S^4(n)\equiv S^2(n)\otimes S^2(n)
\end{equation}

The set $S^4(n)$ contains all base-16 sequences of length $n$, where each base-16 sequence is an ordered
pair of base-4 sequences. The 16 symbols comprising these sequences are as follows, where the base-4
symbol on the left is associated with Alice's event and the base-4 symbol on the right is associated with
Bob's event:
\begin{equation}
\begin{array}{llll}
AA\equiv A_{a}\times A_{b}, & AB\equiv A_{a}\times B_{b}, & AC\equiv A_{a}\times C_{b}, & AD\equiv A_{a}\times D_{b}\\
BA\equiv B_{a}\times A_{b}, & BB\equiv B_{a}\times B_{b}, & BC\equiv B_{a}\times C_{b}, & BD\equiv B_{a}\times D_{b}\\
CA\equiv C_{a}\times A_{b}, & CB\equiv C_{a}\times B_{b}, & CC\equiv C_{a}\times C_{b}, & CD\equiv C_{a}\times D_{b}\\
DA\equiv D_{a}\times A_{b}, & DB\equiv D_{a}\times B_{b}, & DC\equiv D_{a}\times C_{b}, & DD\equiv D_{a}\times D_{b}
\end{array}
\end{equation}

\subsection{Ordering ontic states}
\label{sec:order}

Subsets of $S^4(n)$ are used to model the information observers have about causal networks consisting of
two events. To define these subsets, we must first establish an ordering for $S^4(n)$. Because we are
modeling a system whose observable behavior is non-deterministic, this ordering must be partial rather than total. This implies that
the labels chosen to identify a particular sequence will not be unique in general.

The ordering parameters used within this formalism are the frequencies with which symbols appear in sequences. These frequencies are called counts. In the case of a base-4 sequence, the counts are: $\ct{A}$, $\ct{B}$, $\ct{C}$, and $\ct{D}$.
Taken on their own, these counts may be viewed as discrete variables with the following properties:
\begin{equation}
\ct{A},\ct{B},\ct{C},\ct{D}\ge 0
\end{equation}
\begin{equation}
\ct{A}+\ct{B}+\ct{C}+\ct{D}=n
\end{equation}

Because base-16 sequences are ordered pairs of base-4 sequences, the base-4 counts are recovered from
the base-16 counts by summing over one index. We write $n_{XY}$ for the count of the base-16 symbol
$XY$, and $\ca{X}$, $\cb{X}$ for the base-4 counts of Alice's and Bob's events. Then
\begin{equation}
\ca{X}=\sum_{Y}n_{XY},\qquad \cb{Y}=\sum_{X}n_{XY}
\label{eq:marginals}
\end{equation}
where here and throughout $X$ and $Y$ range over the alphabet $\{A,B,C,D\}$. Equivalently: arrange the
sixteen counts $n_{XY}$ in a $4\times4$ array whose rows are indexed by Alice's symbol and whose columns
are indexed by Bob's. The base-4 counts of Alice's event are then the row sums of that array, and those
of Bob's event are the column sums.

The base-4 counts in equation (\ref{eq:marginals}) are associated with the events at Alice's and Bob's detector.
There are four more base-4 counts of interest, which are related to the map connecting Alice's and Bob's
event. Maps between base-4 sequences are also base-4 sequences, equipped with the element-wise mod2
addition operation ($\oplus$). An example of one such map is offered here:
\begin{equation}
\begin{pmatrix}A\\B\\A\\C\\B\\A\end{pmatrix}_{a}\oplus
\begin{pmatrix}A\\C\\B\\B\\D\\A\end{pmatrix}_{map}=
\begin{pmatrix}A\\D\\B\\D\\C\\A\end{pmatrix}_{b}
\end{equation}

The counts of the map are read off the same array, but summed along the level sets of $\oplus$ rather
than along rows or columns. Writing $\cm{Z}$ for the base-4 counts of the map, and using the
$\mathbb{Z}_2\times\mathbb{Z}_2$ addition table given in equation (\ref{eq:z2z2}),
\begin{equation}
\cm{Z}=\sum_{X\oplus Y=Z}n_{XY}
\label{eq:mapcounts}
\end{equation}
Written out, equation (\ref{eq:mapcounts}) gives
\begin{equation}
\begin{array}{ll}
\Amap=\ct{AA}+\ct{BB}+\ct{CC}+\ct{DD}, & \Cmap=\ct{AC}+\ct{BD}+\ct{CA}+\ct{DB}\\[2pt]
\Bmap=\ct{AB}+\ct{BA}+\ct{CD}+\ct{DC}, & \Dmap=\ct{AD}+\ct{BC}+\ct{CB}+\ct{DA}
\end{array}
\label{eq:mapexplicit}
\end{equation}
Equations (\ref{eq:marginals}) and (\ref{eq:mapcounts}) are the only bookkeeping rules required to pass
between the base-4 and base-16 descriptions.

When modeling a physical system, it is necessary to express all relevant quantities in terms of counts.
For the models presented in this paper, physical quantities will typically be linear functions of counts,
which we call quantum numbers. While there is no general methodology for identifying the proper map
between counts and quantum numbers, some compositions of counts have properties of particular
significance when modeling physics. An example of one such quantity is the Hamming distance between two
base-2 sequences, which is equivalent to the sums $\ca{C}+\ca{D}$ and
$\cb{C}+\cb{D}$. As one might infer from the name, the Hamming distance qualifies as a metric,
which means that it satisfies a triangle inequality. This quantity is used to model the quantum number
for total spin within this paper, though it may also be used to model spatial degrees of freedom. This
leads to an interesting geometric picture that accompanies all of the models presented here. It is worth
emphasizing at this stage that the Hamming distance is \emph{not} the quantity which will be used to
model the angle of rotation between two spatial frames. As we show in section \ref{sec:calibration}, that
role is played by a ratio of map counts rather than by a count itself.

Another important property of quantum numbers is related to the concept of locality, as discussed in
section \ref{sec:intro}. Within the proposed models, a quantum number is taken to be local if it is a property of an
event or map. That is, if it can be expressed in terms of the base-4 counts associated with an individual
event, or a map which connects two events. For base-16 sequences, there will always be ten such quantum
numbers, one of which is the sequence length $n$. This leaves six quantum numbers that may be thought of
as non-local, which play an important role in driving interference. This split is not a choice. Alice's,
Bob's and the map's base-4 counts are, respectively, the row sums, the column sums, and the sums along the
level sets of $\oplus$ of the $4\times4$ array of counts $n_{XY}$; together these three margins define a
linear map of rank ten, whose kernel is a rank-six integer lattice of moves invisible to the observer. The
hidden degrees of freedom live on this lattice, and it carries a rich structure: the operations generating
it, introduced in section \ref{sec:indist}, form the alternating group $A_4$, and the twenty non-local
quantum numbers defined there are the line sums of the affine plane $AG(2,4)$ (appendix \ref{app:symmetry}).

\subsection{Ontic state spaces}
\label{sec:spaces}

The experiments to be modeled in this paper will always involve two events. Conceptually, it is
beneficial to think of these events as measurements performed by the observers Alice and Bob, each of
which is assigned a detector. Depending on the time ordering of these events, which may or may not be
definite, we can think of Alice and Bob as two separate observers, or the same observer at two separate
times. Either way, the experiments being modeled in this paper require two sources of information to
fully characterize the outcome.

The ontic state spaces $E^a$ and $E^b$, which are subsets of $S^4(n)$, contain the information each
observer has about the experiment being performed, provided some specific physical context. One
identifies this context by placing each of the 16 degrees of freedom, or quantum numbers, into one of
four categories: random variable, conditioning variable, local nuisance variable, or non-local nuisance
variable. We store these variables in the sets $R$, $U$, $W$, and $\Lambda$, respectively, where
$R\cup U\cup W\cup\Lambda=Q$, or the set of all possible combinations of quantum numbers. Individual
elements of these sets, which we denote as $r\in R$, $u\in U$, $w\in W$, and $\lambda\in\Lambda$, are each
$N$-tuples, where $N$ is equal to the number of variables stored in that set.

One example of a conditioning variable is $n$, or the total length of the sequences used to model the
system of interest. This quantity may be viewed as a continuity parameter in the proposed models,
allowing probabilities to become arbitrarily smooth. Other conditioning variables may include the total
spin of a particle, or the relative rotation between two spatial frames. Broadly speaking, they are
quantities which may be selected or controlled in a typical experimental setting. Random variables, on
the other hand, are the physical observables which are not selected or controlled. Instead, they may take
on any value permitted by the conditioning variables. Finally, nuisance variables are the physical
quantities which are not observed in the experiment being performed. These quantities are summed over
when calculating probabilities.

With these definitions in hand, we may inspect the composition of the ontic state spaces $E^a$ and $E^b$.
The simplest way to approach this task is to begin with elementary subsets of $S^4(n)$. Given a complete
set of 16 quantum numbers, which enables one to determine all 16 underlying counts, we introduce the
elementary state spaces $\varepsilon^a(q)\subset E^a$ and $\varepsilon^b(q)\subset E^b$. Here and
below we abbreviate the full assignment of quantum numbers by a single label,
\begin{equation}
q\equiv(\ell,\lambda),\qquad \ell\equiv(r,u,w)
\label{eq:labels}
\end{equation}
so that $\ell$ collects the ten local quantum numbers and $\lambda$ the six non-local ones. These elementary state spaces each contain all base-16
sequences compatible with the given quantum numbers, as well as a fixed base-4 sequence associated with
Alice's ($\varepsilon^a$) or Bob's ($\varepsilon^b$) event. That is, in $\varepsilon^a$ the configuration
of the base-4 symbols in Alice's event is held fixed. Alternatively, in $\varepsilon^b$ it is the
configuration of the base-4 symbols in Bob's event that is held fixed. Thus, the base-16 sequences that
comprise an observer's elementary state space are those which are compatible with the given quantum
numbers, as well as the base-4 sequence associated with their event. An example of two such elementary
state spaces are as follows, where all base-16 sequences share the same set of 16 quantum numbers
($\ct{AA}=2$, $\ct{BA}=1$, $\ct{BB}=1$, $\ct{CD}=1$, $\ct{DC}=1$, all other counts zero):
\begin{equation}
\varepsilon^a=\begin{pmatrix}A\\B\\D\\C\\B\\A\end{pmatrix}_{a}\otimes
\left\{\begin{pmatrix}A\\B\\C\\D\\A\\A\end{pmatrix}_{b},\begin{pmatrix}A\\A\\C\\D\\B\\A\end{pmatrix}_{b}\right\},\quad
\varepsilon^b=\left\{\begin{pmatrix}A\\B\\D\\C\\B\\A\end{pmatrix}_{a},\begin{pmatrix}A\\B\\D\\C\\A\\B\end{pmatrix}_{a},\begin{pmatrix}B\\B\\D\\C\\A\\A\end{pmatrix}_{a}\right\}\otimes
\begin{pmatrix}A\\B\\C\\D\\A\\A\end{pmatrix}_{b}
\label{eq:elementary}
\end{equation}

It is worth emphasizing that the elementary ensembles depicted in equation (\ref{eq:elementary}) are
complete. That is, they include all base-16 sequences compatible with a fixed set of quantum numbers and
the event associated with Alice or Bob. Their cardinalities, $2$ and $3$, agree with equation
(\ref{eq:epscard}) below. When constructing these elementary state spaces, it may be useful
to think of Alice's and Bob's fixed event as a template, which can be used to guide the construction of
the other observer's possible events. It is also important to note that holding a base-4 sequence fixed
is not equivalent to actually knowing the configuration of the symbols. This implies that $\varepsilon^a$
and $\varepsilon^b$ each have many unique compositions, one for each permutation of the fixed base-4
sequence. Though the composition of these elementary state spaces will vary, their cardinalities remain
invariant under these permutations. These cardinalities have a simple closed form. Given the complete set of 16 quantum numbers, one first
determines the 16 underlying counts $n_{XY}$ by expressing the quantum numbers as a $16\times16$
coefficient matrix and inverting it. Because the base-4 sequence of one event is held fixed while the
other is free to vary, what remains is simply to distribute the symbols of the free event among the
positions occupied by each symbol of the fixed one. This yields a product of multinomial coefficients:
\begin{equation}
|\varepsilon^a(q)|=\prod_X\frac{\ca{X}!}{\prod_Y n_{XY}!},\qquad
|\varepsilon^b(q)|=\prod_Y\frac{\cb{Y}!}{\prod_X n_{XY}!}
\label{eq:epscard}
\end{equation}
In terms of the $4\times4$ array of counts, $|\varepsilon^a|$ is a product of multinomials taken over
rows and $|\varepsilon^b|$ a product taken over columns. Equation (\ref{eq:epscard}) is equivalent to
the three step construction used in \cite{PowersSG,PowersSpin}, in which one forms the total
$\phi(v)=n!/\prod_{XY}n_{XY}!$ and then corrects for the fixed sequence by a factor
$\prod_X\ca{X}!/n!$; the factors of $n!$ cancel, leaving equation (\ref{eq:epscard}).

Holding the base-4 symbols comprising events fixed is directly related to their ontological status.
Formally, all observed properties of events are taken to be definite states of reality. Thus, the base-4
symbols within an event whose counts are observed during the course of an experiment may not vary in
configuration within the associated observer's state space. Given that all four base-4 symbols are held
fixed in Alice's ($\varepsilon^a$) and Bob's ($\varepsilon^b$) elementary state spaces, the assumption
being made is that all four quantum numbers associated with each of their events are either random
variables or conditioning variables. In the case that one or more of these quantum numbers is treated as
a local nuisance variable, a union over these elementary state spaces must be defined. In particular, we
must define a union over all configurations of the base-4 symbols associated with unobserved quantum
numbers. Let $\chi_a(w)\subseteq\{A,B,C,D\}$ denote the set of symbols in Alice's event whose counts
are unobserved, and $\chi_b(w)$ the same for Bob's. For example, if $\ca{A}$ and $\ca{B}$ are treated as
nuisance variables then $\chi_a(w)=\{A,B\}$. We write the resulting unions as
\begin{equation}
\varepsilon^a_\ast(q)\equiv\bigcup_{\chi_a(w)}\varepsilon^a(q),\qquad
\varepsilon^b_\ast(q)\equiv\bigcup_{\chi_b(w)}\varepsilon^b(q)
\label{eq:starred}
\end{equation}
where the star is a reminder that the configuration of the unobserved symbols has been released.

We are now prepared to define Alice's ($E^a$) and Bob's ($E^b$) complete state spaces, given some choice
of the conditioning variables $u$. With the abbreviations (\ref{eq:labels}) and (\ref{eq:starred}),
these are simply the unions of the starred elementary spaces over every label compatible with $u$:
\begin{equation}
E^{a}(u)\equiv\bigcup_{q\in Q(u)}\varepsilon^{a}_\ast(q),\qquad
E^{b}(u)\equiv\bigcup_{q\in Q(u)}\varepsilon^{b}_\ast(q)
\label{eq:Espaces}
\end{equation}
where $Q(u)\subset Q$ is the set of labels consistent with the chosen conditioning variables.

As we stated at the beginning of this section, the experiments being modeled in this paper will always
require two sources of information to fully characterize the outcome. For this reason, it is actually the
product of Alice's and Bob's state spaces that is of interest when calculating probabilities. In
particular, we are interested in a modified Cartesian product of $E^a(u)$ and $E^b(u)$, which we denote
as $E^a(u)\bar{\otimes}E^b(u)$, such that all ordered pairs share the same set of local quantum numbers.
Put more plainly, we are interested in the joint state space in which Alice and Bob agree on all local
quantum numbers for each possible experimental outcome. The only remaining issues to address are that of
indistinguishable states and interference, which are closely related to one another.

\subsection{Indistinguishable ontic states}
\label{sec:indist}

Within the models proposed here, the information an observer is permitted to have regarding a particular
causal network is limited to local quantum numbers. This implies that any base-16 sequences within $E^a$
or $E^b$ that share the same set of local quantum numbers must be associated with indistinguishable ontic
states. For a given choice of local quantum numbers, the associated set of indistinguishable states can
be defined as a union over all possible combinations of the non-local quantum numbers like so:
\begin{equation}
L^{a}(\ell)\equiv\bigcup_{\lambda\in Q(\ell)}\varepsilon^{a}(\ell,\lambda),\qquad
L^{b}(\ell)\equiv\bigcup_{\lambda\in Q(\ell)}\varepsilon^{b}(\ell,\lambda)
\label{eq:La}
\end{equation}

Within the local state spaces $L^a$ and $L^b$, there are two distinct types of indistinguishability, both
of which can be understood in terms of permutations. The first type arises due to the permutation
symmetry of the base-16 sequences within the elementary state spaces $\varepsilon^a$ and
$\varepsilon^b$. In the case of $\varepsilon^a$, the base-16 sequences comprising this set are related to
one another through permutations of Bob's base-4 sequence, which can be seen in equation
(\ref{eq:elementary}). Importantly, these permutations leave all base-16 counts and Alice's fixed base-4
sequence invariant. Likewise, the base-16 sequences within the elementary state space $\varepsilon^b$ are
related to one another through permutations of Alice's base-4 sequence. Again, such that the base-16
counts and Bob's fixed base-4 sequence are left invariant.

The second type of indistinguishability arises due to variations in the non-local quantum numbers
$\lambda$. Once again, this indistinguishability is related to permutations. However, unlike the ones
discussed in the previous paragraph and depicted in equation (\ref{eq:elementary}), these permutations do
not leave all base-16 counts invariant. Instead, they lead to variations in the non-local quantum numbers
$\lambda$, while leaving all local quantum numbers unchanged. In other words, these permutations are
symmetries of local quantum numbers, but not of non-local quantum numbers.

Both of these situations can be elucidated further by considering individual transpositions of base-4
symbols. By transposition, we mean the act of swapping two different base-4 symbols within a base-4
sequence. For clarity, we will assume that these operations are being performed on a randomly chosen
element within $L^a$, as defined in equation (\ref{eq:La}). In total, there are six unique transpositions
possible: $(A_{b},B_{b})$, $(A_{b},C_{b})$, $(A_{b},D_{b})$, $(B_{b},C_{b})$, $(B_{b},D_{b})$,
and $(C_{b},D_{b})$. In each of these cases, single instances of the symbols indicated are exchanged for
one another within Bob's base-4 sequence. To each of these transpositions we associate an operator, which
we denote as $T^{AB}$, $T^{AC}$, $T^{AD}$, $T^{BC}$, $T^{BD}$, and
$T^{CD}$, respectively.

In some cases, applying these operators will not lead to any variations in the base-16 counts, which
corresponds to the local symmetries of the first type. This occurs when the base-16 symbols, or operands,
acted on by a given transposition operator share the same symbol in Alice's base-4 sequence. In all,
there are 6 operators that may each act on one of 4 operands, leading to 24 unique operations of this
type. All base-16 sequences within the elementary state space $\varepsilon^a$ are related to one another
through some composition of these 24 distinct transposition operations. An example of one such operation
is as follows:
\begin{equation}
\begin{pmatrix}AA\\BD\\CA\\CC\\BB\\AA\end{pmatrix}\xrightarrow{T^{AC}}
\begin{pmatrix}AA\\BD\\CC\\CA\\BB\\AA\end{pmatrix}
\end{equation}

Local symmetries of the second type are slightly more complicated. This is because any single
transposition operator cannot simultaneously change the base-16 counts while also leaving the local
quantum numbers invariant. An example of why this is the case can be seen in the following, where we pay
special attention to the map relating Alice's and Bob's base-4 sequence:
\begin{equation}
\begin{pmatrix}A\\B\\C\\D\\B\\A\end{pmatrix}_{a}\oplus\begin{pmatrix}A\\A\\B\\B\\A\\B\end{pmatrix}_{map}
=\begin{pmatrix}A\\B\\D\\C\\B\\B\end{pmatrix}_{b}
\xrightarrow{T^{CD}}
\begin{pmatrix}A\\B\\C\\D\\B\\A\end{pmatrix}_{a}\oplus\begin{pmatrix}A\\A\\A\\A\\A\\B\end{pmatrix}_{map}
=\begin{pmatrix}A\\B\\C\\D\\B\\B\end{pmatrix}_{b}
\end{equation}

Clearly, the base-4 counts associated with the map relating Alice's and Bob's base-4 sequence changed as a
result of this operation. To avoid this, transposition operators associated with the second type of local
symmetry must always act in pairs. For example, when the operator $R_B\equiv T^{AB}T^{CD}$, defined in equation (\ref{eq:commuting}) below, acts on the
base-16 symbols $AA$, $BB$, $CD$, and $DC$, we have the following:
\begin{equation}
\begin{pmatrix}A\\B\\C\\D\\B\\A\end{pmatrix}_{a}\oplus\begin{pmatrix}A\\A\\B\\B\\A\\B\end{pmatrix}_{map}
=\begin{pmatrix}A\\B\\D\\C\\B\\B\end{pmatrix}_{b}
\xrightarrow{R_B}
\begin{pmatrix}A\\B\\C\\D\\B\\A\end{pmatrix}_{a}\oplus\begin{pmatrix}B\\B\\A\\A\\A\\B\end{pmatrix}_{map}
=\begin{pmatrix}B\\A\\C\\D\\B\\B\end{pmatrix}_{b}
\label{eq:pairop}
\end{equation}

As we claimed, the operation depicted in equation (\ref{eq:pairop}) indeed leads to the conservation of
all local quantum numbers, while generating a change in the base-16 counts. In general, local symmetries
of the second type are generated by the operators $R_B$, $R_C$ and $R_D$ of equation
(\ref{eq:commuting}) together with the non-commuting ordered pairs of transpositions. In total, there are 36 unique pairs that can be created using the 6 basic operators.
However, 6 of these are just repetitions of the same operator, which do not contribute to local
symmetries of the second type:
\begin{equation}
T^{AB}T^{AB},\ T^{AC}T^{AC},\ T^{AD}T^{AD},\
T^{BC}T^{BC},\ T^{BD}T^{BD},\ T^{CD}T^{CD}
\end{equation}

The remaining 30 can be divided into two classes, which are distinguished from one another through their
behavior under commutation. Six of them commute, and because of that commutativity they represent only 3
distinct operators. These three have a compact description: each is the operation which adds a fixed
group element to Bob's symbol at a pair of positions. Accordingly we name them by that element,
\begin{equation}
R_B\equiv T^{AB}T^{CD},\qquad R_C\equiv T^{AC}T^{BD},\qquad R_D\equiv T^{AD}T^{BC}
\label{eq:commuting}
\end{equation}
so that $R_g$ implements $Y\mapsto Y\oplus g$ at two positions of Bob's base-4 sequence, for
$g\in\{B,C,D\}$. This notation is used throughout the remainder of the paper in place of the ordered
pairs of transpositions. Adopting it brings the previous total of 30 down to 27. The remaining 24 operators are those which do not commute, each of which
involves the transposition of only 3 of the 4 base-4 symbols:
\begin{equation}
\begin{split}
T^{BC}T^{BD},\ T^{BD}T^{CD},\ T^{CD}T^{BC},\ T^{BD}T^{BC},\ T^{CD}T^{BD},\ T^{BC}T^{CD}\\
T^{AC}T^{AD},\ T^{AD}T^{CD},\ T^{CD}T^{AC},\ T^{AD}T^{AC},\ T^{CD}T^{AD},\ T^{AC}T^{CD}\\
T^{AB}T^{AD},\ T^{AD}T^{BD},\ T^{BD}T^{AB},\ T^{AD}T^{AB},\ T^{BD}T^{AD},\ T^{AB}T^{BD}\\
T^{AB}T^{AC},\ T^{AC}T^{BC},\ T^{BC}T^{AB},\ T^{AC}T^{AB},\ T^{BC}T^{AC},\ T^{AB}T^{BC}
\end{split}
\label{eq:noncommuting}
\end{equation}

The base-16 symbols being operated on, or the operands, are just as important as the operators
themselves. To generate the desired local symmetry transformation, an operator must act on the
appropriate collection of base-16 symbols. Our initial focus will be on the collections that obey a
symmetry under the action of the commutative operators defined in equation (\ref{eq:commuting}). Each of
these collections consists of 4 base-16 symbols. For each one, we define a non-local quantum number
consisting of a sum of the counts associated with each base-16 symbol. When the appropriate operator acts
on the symbols associated with a given non-local quantum number, its numerical value will decrease by 1.
Simultaneously, a different non-local quantum number will increase by 1. In general, non-local quantum
numbers always come in pairs, such that the action of the appropriate operator leaves the sum of this
pair invariant. These non-local quantum numbers are defined here:
\begin{equation}
\begin{split}
\nu^0\equiv\tfrac{1}{4}(\ct{AA}+\ct{BB}+\ct{CD}+\ct{DC}),\quad &\mu^0\equiv\tfrac{1}{4}(\ct{AB}+\ct{BA}+\ct{CC}+\ct{DD})\\
\nu^1\equiv\tfrac{1}{4}(\ct{AD}+\ct{BC}+\ct{CA}+\ct{DB}),\quad &\mu^1\equiv\tfrac{1}{4}(\ct{AC}+\ct{BD}+\ct{CB}+\ct{DA})\\
\nu^2\equiv\tfrac{1}{4}(\ct{AA}+\ct{CC}+\ct{DB}+\ct{BD}),\quad &\mu^2\equiv\tfrac{1}{4}(\ct{AC}+\ct{CA}+\ct{DD}+\ct{BB})\\
\nu^3\equiv\tfrac{1}{4}(\ct{BA}+\ct{DC}+\ct{AD}+\ct{CB}),\quad &\mu^3\equiv\tfrac{1}{4}(\ct{BC}+\ct{DA}+\ct{AB}+\ct{CD})\\
\nu^4\equiv\tfrac{1}{4}(\ct{AA}+\ct{DD}+\ct{BC}+\ct{CB}),\quad &\mu^4\equiv\tfrac{1}{4}(\ct{AD}+\ct{DA}+\ct{BB}+\ct{CC})\\
\nu^5\equiv\tfrac{1}{4}(\ct{BA}+\ct{CD}+\ct{AC}+\ct{DB}),\quad &\mu^5\equiv\tfrac{1}{4}(\ct{BD}+\ct{CA}+\ct{AB}+\ct{DC})
\end{split}
\label{eq:nu}
\end{equation}
\begin{equation}
\begin{split}
\kappa^0\equiv\tfrac{1}{4}(\ct{AA}+\ct{DC}+\ct{CB}+\ct{BD}),\quad &\kappa^1\equiv\tfrac{1}{4}(\ct{AC}+\ct{DA}+\ct{CD}+\ct{BB})\\
\kappa^2\equiv\tfrac{1}{4}(\ct{BA}+\ct{CC}+\ct{AD}+\ct{DB}),\quad &\kappa^3\equiv\tfrac{1}{4}(\ct{BC}+\ct{CA}+\ct{AB}+\ct{DD})
\end{split}
\label{eq:kappa}
\end{equation}
\begin{equation}
\begin{split}
\omega^0\equiv\tfrac{1}{4}(\ct{AA}+\ct{CD}+\ct{DB}+\ct{BC}),\quad &\omega^1\equiv\tfrac{1}{4}(\ct{AD}+\ct{CA}+\ct{BB}+\ct{DC})\\
\omega^2\equiv\tfrac{1}{4}(\ct{BA}+\ct{DD}+\ct{AC}+\ct{CB}),\quad &\omega^3\equiv\tfrac{1}{4}(\ct{BD}+\ct{DA}+\ct{AB}+\ct{CC})
\end{split}
\label{eq:omega}
\end{equation}

These 20 non-local quantum numbers can be divided into two classes. The first, which are those defined in
equation (\ref{eq:nu}), are unique in that they are each associated with exactly 1 of the 3 commutative
operators. These associations are as follows, where quantum numbers appear in parentheses along with
their partner under the indicated operator:
\begin{equation}
\begin{split}
R_B&:\ (\nu^0,\mu^0),\ (\nu^1,\mu^1)\\
R_C&:\ (\nu^2,\mu^2),\ (\nu^3,\mu^3)\\
R_D&:\ (\nu^4,\mu^4),\ (\nu^5,\mu^5)
\end{split}
\label{eq:nupartners}
\end{equation}

When one of these operations occurs, the only $\nu$'s or $\mu$'s impacted are those involved in the
operation. Thus, these operations generate $\Delta\nu^i=\pm1$ and $\Delta\mu^i=\mp1$, while leaving all
other $\nu$'s and $\mu$'s invariant. The second class of quantum numbers, which are those defined in
equations (\ref{eq:kappa}) and (\ref{eq:omega}), are each associated with all 3 of the commutative
operators. This implies that each of these non-local quantum numbers has three partners, one for each
operator:
\begin{equation}
\begin{split}
R_B&:\ (\kappa^0,\kappa^3),\ (\kappa^1,\kappa^2),\ (\omega^0,\omega^3),\ (\omega^1,\omega^2)\\
R_C&:\ (\kappa^0,\kappa^1),\ (\kappa^2,\kappa^3),\ (\omega^0,\omega^2),\ (\omega^1,\omega^3)\\
R_D&:\ (\kappa^0,\kappa^2),\ (\kappa^1,\kappa^3),\ (\omega^0,\omega^1),\ (\omega^2,\omega^3)
\end{split}
\end{equation}

Like the operations in equation (\ref{eq:nupartners}), the only $\kappa$'s or $\omega$'s impacted by these
operations are those actually involved, leaving all others invariant. This independence does not however
span the two classes of quantum numbers we have outlined here. Any operation involving a $\nu$ and $\mu$
will induce a change of $\pm\frac{1}{2}$ in each of the $\kappa$'s and $\omega$'s. Similarly, operations
involving $\kappa$'s or $\omega$'s will induce a change of $\pm\frac{1}{2}$ in 8 of the 12 $\nu$'s and
$\mu$'s.

The final issue we must address in this section is that of the non-commutative operations. These
operations each involve 3 base-16 symbols, rather than 4. The full set of non-local quantum numbers
involved in these operations, of which there are 32, can be found in appendix \ref{app:noncomm}. As with
those previously discussed, each of these non-local quantum numbers has a partner under the action of a
given non-commutative operator. In total, there are 16 partnerships of interest, each of which is
associated with exactly six of the non-commutative operators in equation (\ref{eq:noncommuting}). As was
the case with $\nu$'s/$\mu$'s and $\kappa$'s/$\omega$'s, the non-local quantum numbers defined in appendix
\ref{app:noncomm} lack independence from operations in which they are not directly involved. Likewise,
the $\nu$'s/$\mu$'s and $\kappa$'s/$\omega$'s will vary under the action of the non-commutative
operations. This complicated interdependence of the non-local quantum numbers introduced in this section
will be relevant in the following section, where these quantum numbers will be used to drive
interference. As we will see, care must be taken to ensure that the six non-local quantum numbers chosen
to populate $\Lambda$ lead to a stable calculation.

Needless to say, the local state spaces $L^a$ and $L^b$ have a rich set of symmetries. Though much of what
has been discussed in this section remains physically obscure, several interesting features are of note.
To begin with, we point out that a local state space is always associated with a specific choice of the
local quantum numbers. In other words, one must assume a particular observable outcome for the experiment
being modeled. Within this context, the symmetries of a local state space represent a means of organizing
all of the unobservable states that contribute to that particular observable outcome. The fact that these
symmetries come in two distinct forms is directly related to the nature of the hidden information.
Symmetries of the first type arise because the information stored in the ordering of the symbols
comprising a given sequence is hidden from observers. While symmetries of the second type arise because
the numerical values of the non-local quantum numbers are hidden from observers. Put more simply, if no
information were hidden from observers, no two ontic states would be indistinguishable within the
proposed formalism. Thus, symmetries are directly connected to the concept of hidden information within
this framework.

\subsection{Interference}
\label{sec:interference}

In section \ref{sec:indist}, we identified quantum numbers that qualify as non-local within this
formalism. We found them by studying the symmetries of local state spaces, which are associated with the
transposition operators introduced in equations (\ref{eq:commuting}-\ref{eq:noncommuting}). We may now
use these results to elucidate the phenomenon of interference, or situations in which base-16 sequences
within a local state space annihilate one another.

Given two indistinguishable states from $L^a$ or $L^b$, which we label with the subscripts 0 and 1, the
total difference between their non-local quantum numbers is as follows, where $\lambda$ is a 6-tuple:
\begin{equation}
f(\lambda_0,\lambda_1)\equiv\sum_{k=0}^{5}\left(\lambda^k_0-\lambda^k_1\right)
\label{eq:parity}
\end{equation}
The signed sum is essential: individual components $\lambda^k$ can change by $\pm\frac12$ under a single
operation (section \ref{sec:indist}), so the parity of $\sum_k|\lambda^k_0-\lambda^k_1|$ would in general
differ from that of $f$. For example, the operation $R_B$ of equation (\ref{eq:pairop}) changes
$(\nu^0,\kappa^1,\omega^2)$ by $(-1,-\frac12,+\frac12)$, so that $f$ is odd while the sum of absolute
values is even.

Within the local state spaces defined in equation (\ref{eq:La}), two states will
annihilate one another if $f(\lambda_0,\lambda_1)$ is odd. These annihilations can be accounted for when
calculating the post-interference cardinalities of the local state spaces $L^a$ and $L^b$ in the following
way, where we use the notation $\|\cdot\|$ to distinguish this operation from the pre-interference
cardinality:
\begin{equation}
\ncard{L^a(\ell)}=\left|\sum_{\lambda_i\in Q(\ell)}(-1)^{\sum_k(\lambda^k_0-\lambda^k_i)}|\varepsilon^a(\ell,\lambda_i)|\right|
\label{eq:Lainter}
\end{equation}
\begin{equation}
\ncard{L^b(\ell)}=\left|\sum_{\lambda_i\in Q(\ell)}(-1)^{\sum_k(\lambda^k_0-\lambda^k_i)}|\varepsilon^b(\ell,\lambda_i)|\right|
\label{eq:Lbinter}
\end{equation}

The absolute value operation on the right hand side of equations (\ref{eq:Lainter}) and
(\ref{eq:Lbinter}) is included to ensure that the resulting cardinalities are positive. This operation is
necessary due to the arbitrary sign convention used in the interference term, which arises from the choice
of $\lambda_0$. It should also be noted that the method of accounting for interference presented here
differs from the one used in \cite{PowersSG,PowersSpin}, though both methods are technically equivalent.

The alternating sign in equations (\ref{eq:Lainter}) and (\ref{eq:Lbinter}) is introduced here as a
rule, and it is worth recording what is known about its origin. In the companion paper \cite{QHO} the
same construction is carried out for the smallest alphabet, in which Alice's and Bob's events are binary
sequences and their pairing is a base-4 sequence. The elementary cardinalities there are exactly those of
equation (\ref{eq:epscard}), the transition measure is the product of Alice's and Bob's signed sums, and
the only quantity that is not fixed by the local counts is the Hamming distance $d_{ab}$ between the two
sequences (equivalently their overlap $t$, the number of positions at which both carry a $1$), which is
the base-2 counterpart of the non-local quantum numbers of section \ref{sec:indist}. For that model it is proved that, given the bilinear
forward--backward form of the measure, two conditions determine the weighting $w(t)$ of the hidden
variable completely: that the measure summed over outcomes account for exactly the $2^n$ configurations of
the register, and that it be invariant under relabeling of the two symbols. The unique solution is
$w(t)=\pm(-1)^t$; every other choice of sign, at any value of $t$, exceeds the capacity $2^n$ by a
computable positive amount, and the unsigned sum exceeds it by the factor $\binom{n}{n^a_1}$, where $n^a_1$ is the number of $1$'s in Alice's sequence. The
alternating sign is thus, in base 2, the bookkeeping that prevents observationally indistinguishable
configurations from being counted more than once, and the resulting measure is the square of a Krawtchouk
polynomial, which contracts to the quantum harmonic oscillator as $n\to\infty$. We take this as
evidence that the sign in equations (\ref{eq:Lainter}) and (\ref{eq:Lbinter}) may ultimately be a
consequence of the restriction of the observer to counts rather than an independent postulate. In the
present paper, however, it is an input: we have not proved the corresponding statement for the base-16
alphabet used here, where the hidden variable is a
six-component object rather than a single integer; what is known about the space of admissible rules,
and what an extension of the base-2 argument would require, is set out in appendix \ref{app:symmetry}.

The parity (odd/even) of the quantity defined in equation (\ref{eq:parity}) contains information about the
transposition operations through which two indistinguishable states are related. More precisely, it tells
us the parity of the total number of these operations, which are those associated with the operators
introduced in equations (\ref{eq:commuting}) and (\ref{eq:noncommuting}). Exactly which transposition
operations the parity of equation (\ref{eq:parity}) is sensitive to depends solely on the 6 non-local
quantum numbers chosen to populate $\Lambda$. Though a complete overview of these choices is beyond the
scope of this work, we offer here a brief discussion of the three most relevant considerations when making
this choice.

First, we must ensure that the variations in $\sum_k\lambda^k$ within a local state space are always
integers. Otherwise, the cardinalities of local state spaces, as they are defined in equations
(\ref{eq:Lainter}) and (\ref{eq:Lbinter}), may become complex numbers. Non-local quantum numbers were
defined in a way that ensures they always vary in integer increments under operations in which they are
directly involved. However, the interdependence we discussed in section \ref{sec:indist} leads to the
possibility of non-integer variations. These non-integer variations can be corrected for within the full
sum by selecting combinations of non-local quantum numbers which collectively have integer variations.
This is an important selection criterion when choosing non-local quantum numbers to populate $\Lambda$.

Second, the number and type of operations the parity of equation (\ref{eq:parity}) is sensitive to must be
considered. Again, this depends solely on the choice of the 6 non-local quantum numbers. Depending on this
choice, operations involving the operators introduced in equations (\ref{eq:commuting}) and
(\ref{eq:noncommuting}) may or may not induce a variation in the sum $\sum_k\lambda^k$. Of the operations
that do induce a variation in this quantity, some will lead to odd variations and some will lead to even
variations. With respect to interference, it is the operations which yield odd variations in the sum
$\sum_k\lambda^k$ which are of interest. Again, exactly which operations these will be depends on the
choice of non-local quantum numbers, though no choice will ever encompass all possible operations.

Third, the choice of 6 non-local quantum numbers must form a complete set of base-16 quantum numbers when
paired with the 10 local quantum numbers. This is a basic requirement of this formalism as it enables one
to map quantum numbers to base-16 counts. The simplest way to verify that a given choice of 16 quantum
numbers is complete is to express them in matrix form, and attempt to invert the matrix. If it is
invertible, then this choice meets the condition of completeness.

Even with these considerations in mind, there still remain many valid choices for the 6 non-local quantum
numbers. For the models presented here, we make the following choice, which we store in the set $\Lambda$:
\begin{equation}
\Lambda=(\nu^0,\nu^1,\nu^4,\nu^5,\kappa^1,\omega^2)
\label{eq:lambdachoice}
\end{equation}

What has been presented in this section is by no means the complete picture associated with the phenomenon
of interference. However, it is sufficient to justify our choice of the 6 non-local quantum numbers given
in equation (\ref{eq:lambdachoice}). First, this choice guarantees that variations in the sum
$\sum_k\lambda^k$ will always be integers. Second, this choice maximizes the sensitivity of equation
(\ref{eq:parity}) to the underlying transposition operations. This group of operations can be found in
appendix \ref{app:chosen}. Lastly, this choice will always form a complete set when paired with the 10
local quantum numbers. With this choice made, the only remaining task is to calculate the probability of
observing a particular experimental outcome. The structure underlying these three criteria, and the
precise extent to which equation (\ref{eq:lambdachoice}) remains a choice at all, are analyzed in
appendix \ref{app:symmetry}.

\subsection{Calculating probabilities}
\label{sec:probs}

We may now address the main point of this formalism, which is to calculate the probability $P(r|u)$. As
discussed in section \ref{sec:spaces}, the mathematical structure of interest is the joint state space
$E^a(u)\bar{\otimes}E^b(u)$. Recall that this joint state space includes all possible ordered pairs of
states in $E^a(u)$ and $E^b(u)$ which share the same set of local quantum numbers. This implies that
$E^a(u)\bar{\otimes}E^b(u)$ may be expressed as a union over the Cartesian product of the local state
spaces $L^a$ and $L^b$ like so:
\begin{equation}
E^a(u)\bar{\otimes}E^b(u)=\bigcup_{(r,w)\in Q(u)}\ \bigcup_{\chi_{a}(w),\chi_{b}(w)}L^a(\ell)\otimes L^b(\ell)
\end{equation}

To calculate the probability of observing a particular choice of $r$, given $u$, we must first determine
the post-interference cardinality of the Cartesian product space $L^a(\ell)\otimes L^b(\ell)$. By the
combinatorial rule of product, this quantity is the scalar product of equations (\ref{eq:Lainter}) and
(\ref{eq:Lbinter}):
\begin{equation}
\ncard{L^a(\ell)\otimes L^b(\ell)}=\ncard{L^a(\ell)}\ncard{L^b(\ell)}
\label{eq:prodcard}
\end{equation}

To account for the union over local state space configurations, the quantity in equation
(\ref{eq:prodcard}) must be scaled by the number of ways the unobserved symbols may be arranged. These
are again multinomial coefficients, taken over the symbol sets $\chi_a(w)$ and $\chi_b(w)$ introduced in
equation (\ref{eq:starred}):
\begin{equation}
G_a(\ell)\equiv\frac{\big(\sum_{X\in\chi_a}\ca{X}\big)!}{\prod_{X\in\chi_a}\ca{X}!},\qquad
G_b(\ell)\equiv\frac{\big(\sum_{Y\in\chi_b}\cb{Y}\big)!}{\prod_{Y\in\chi_b}\cb{Y}!}
\label{eq:Gab}
\end{equation}

Thus, the total number of states associated with a particular choice of $r$, within the joint state space
$E^a(u)\bar{\otimes}E^b(u)$, can be found by summing over the product of equations
(\ref{eq:prodcard}) and (\ref{eq:Gab}) for all possible combinations of the local nuisance variables $w$:
\begin{equation}
\Upsilon(r,u)=\sum_{w\in Q(r,u)}G_{a}(\ell)G_{b}(\ell)\ncard{L^a(\ell)}\ncard{L^b(\ell)}
\label{eq:upsilon}
\end{equation}

This result is then normalized by the total post-interference cardinality of Alice's and Bob's joint state
space, which can be obtained by summing over equation (\ref{eq:upsilon}) for all possible combinations of
the random variables $r$:
\begin{equation}
\ncard{E^a(u)\bar{\otimes}E^b(u)}=\sum_{r\in Q(u)}\Upsilon(r,u)
\end{equation}

We are left with the following expression for the probability of observing a particular choice of $r$,
given $u$:
\begin{equation}
P(r|u)=\frac{\Upsilon(r,u)}{\sum_{r\in Q(u)}\Upsilon(r,u)}
\label{eq:prob}
\end{equation}

We are now in a position to say something precise about the general nature of probabilities within the
proposed formalism. First, they arise due to the presence of hidden information, implying that they are
epistemic in origin. It is worth being explicit about what the hidden information is: it is the ordering
of the symbols within a sequence together with the values of the non-local quantum numbers, that is, the
microstate underlying a given macrostate, and not the quantum state itself. Second, the state spaces
associated with distinct choices of observables are always disjoint. That is, no single base-16 sequence
will ever appear in state spaces associated with two different sets of observables. This property
distinguishes this proposal from the well known Spekkens toy model \cite{Spekkens}, and it is the source
of the contextuality examined in section \ref{sec:contextuality}. It should not be read as a statement
about the PBR theorem \cite{PBR}: disjointness across choices of observables is a statement about
measurement contexts, whereas the PBR theorem concerns preparations. The relation of the framework to that
theorem is set out in section \ref{sec:psiontic}.

Third, the act of measurement, or the performance of an experiment, may be interpreted as the revelation
of the preexisting state of the system under study. Some modification to one's definition of a
``system'' may, however, be necessary. By ``system'', we mean two causally connected events. Of course, this
interpretation has many wide ranging implications, most of which are beyond the narrow scope of this
paper. However, it is important to note that this particular version of realism does not compromise this
framework's status with respect to quantum contextuality theorems like the Kochen-Specker theorem
\cite{KS}. Rather, one may view the event-centric picture as a natural implication of quantum
contextuality and the closely related issue of Bell's inequality \cite{Bell,CHSH}. The latter point is
made explicit in section \ref{sec:bellresults}, where we report a clear violation of the equivalent CHSH
inequality.

The final point we make is related to the frequentist interpretation of probabilities. The size of the
conditioning variable $n$ has a significant impact on the size of state spaces, and thus on the continuity
of probabilities. In the case that $n$ is small, it is conceivable that Alice and Bob could perform enough
experiments as to have observed all possible ontic states, where we assume that no ontic state ever occurs
more than once. In this case, the predicted probability and their measurement results should match
exactly, which is in alignment with the frequentist interpretation. A secondary consequence of a small $n$
is a loss of statistical independence, where we again assume that no state occurs more than once. That is,
Alice's and Bob's past measurement outcomes would influence what they know about future experiments,
ultimately leading to the possibility of completely deterministic experiments. In practice, however, we
will assume that $n$ is extremely large, and possibly infinite. In the case that $n$ becomes infinite, the
cardinality of ontic state spaces will generally become uncountable, with only a few extreme cases in
which they remain countable. Because any record of measurements is discrete, and thus countable,
statistical independence is then ensured, even if every state occurs only once.

\section{Concrete examples}
\label{sec:models}

The purpose of this section is to illustrate the manner in which the proposed framework and associated
formalism can be used to model non-trivial physical systems, which are otherwise modeled using QM. In
total, we will introduce three models involving spin, the last of which can be applied to a Bell test.
Importantly, this section should not be viewed as a complete and general analysis of the modeled systems,
nor the complete physical picture as it relates to the underlying formalism. Again, it is simply an
illustration of how the proposed framework can be applied to specific physical systems. Further
investigation and development are necessary before these models can be fully understood and validated.

\subsection{Local quantum numbers}
\label{sec:localqn}

We begin the process of introducing physics models with a general discussion on local quantum numbers.
These quantum numbers must be functions of the base-4 counts defined in equations (\ref{eq:marginals}) and (\ref{eq:mapcounts}). Though there
are many possible choices, certain compositions of base-4 counts are of special interest. Throughout this
section, we will motivate and define the various local quantum numbers of interest in the models proposed
here.

The first local quantum number is the sequence length $n$. As discussed in section \ref{sec:spaces}, this
quantity may be interpreted as a continuity parameter, allowing probabilities to become arbitrarily
smooth. This is because as $n$ grows larger, so too will the underlying state spaces. For example, the
number of states in the set $S^4(n)$ is $16^n$. The sequence length $n$ may also be used to normalize
other quantum numbers, allowing those to become arbitrarily smooth as well. Because all base-4 sequences
within a given ontic state must be the same length, $n$ may be defined in terms of the base-4 counts
associated with events, or maps:
\begin{equation}
\begin{split}
n&\equiv\ca{A}+\ca{B}+\ca{C}+\ca{D}\\
n&\equiv\cb{A}+\cb{B}+\cb{C}+\cb{D}\\
n&\equiv\Amap+\Bmap+\Cmap+\Dmap
\end{split}
\label{eq:nmap}
\end{equation}

The next set of quantum numbers of interest, which we denote as $j_{a}$ and $j_{b}$, are proportional to
the Hamming distance between the base-2 sequences comprising Alice's and Bob's event. Because the Hamming
distance obeys a triangle inequality, we interpret these quantum numbers as the total spin observed at
each event. These quantities are defined as follows, where the factor of $\frac{1}{2}$ is included to
align with the traditional units of angular momentum:
\begin{equation}
j_{a}\equiv\frac{\ca{C}+\ca{D}}{2},\qquad j_{b}\equiv\frac{\cb{C}+\cb{D}}{2}
\end{equation}

Closely related to $j_{a}$ and $j_{b}$ are the associated spin projection quantum numbers $m_{a}$ and
$m_{b}$. Of course, these quantities must obey the ranges $-j_{a}\le m_{a}\le j_{a}$ and
$-j_{b}\le m_{b}\le j_{b}$, respectively. This leaves us with two possible definitions, which are
proportional to the differences $\ct{C}-\ct{D}$ or $\ct{D}-\ct{C}$. We choose the former, leading to the
following definitions:
\begin{equation}
m_{a}\equiv\frac{\ca{C}-\ca{D}}{2},\qquad m_{b}\equiv\frac{\cb{C}-\cb{D}}{2}
\label{eq:mdef}
\end{equation}
As shown in \cite{PowersSpin}, this physical interpretation of quantum numbers is consistent with angular momentum composition in quantum mechanics and also with Clebsch-Gordan coefficients. 

The final quantum number associated with events does not yet have a clear physical interpretation. Because
of this, the chosen definition should be viewed as operational and subject to change. We denote these
provisional quantities as $l_{a}$ and $l_{b}$ and choose a definition which is analogous to $m_{a}$ and
$m_{b}$:
\begin{equation}
l_{a}\equiv\frac{\ca{A}-\ca{B}}{2},\qquad l_{b}\equiv\frac{\cb{A}-\cb{B}}{2}
\end{equation}

In all of the models considered here, the quantum numbers $n$, $j$, $m$, and $l$ will be those used to
order the set $S^2(n)$, or the ontic state space for events. What remains is to define the quantum numbers
associated with maps. For the time being, we denote these three quantum numbers using the generic symbols
$\alpha$, $\beta$, and $\gamma$, which we define as follows:
\begin{align}
\alpha&\equiv\Bmap+\Cmap\label{eq:alpha}\\
\beta&\equiv\Bmap+\Dmap\label{eq:beta}\\
\gamma&\equiv\Cmap+\Dmap\label{eq:gamma}
\end{align}

When introducing each model, a physical interpretation will be offered for these abstract quantities when
appropriate. There will also be several additional quantum numbers defined within those sections, which we
refrain from introducing here for the sake of clarity. We stress that $\alpha$, $\beta$ and
$\gamma$ are counts, and that the physical angle of rotation is \emph{not} taken to be proportional
to any of them. The relationship between map counts and the rotation angle is the subject of section
\ref{sec:calibration}.

\subsection{Spin systems in rotated frames}
\label{sec:rotmodel}

The physical system of interest in this section consists of two Stern-Gerlach detectors which may be
rotated with respect to one another, as depicted in Figure \ref{fig:sg1}. In this case, an experiment
consists of two events with a direct causal connection and a definite time ordering. That is, Alice's
event precedes Bob's, or Bob's event precedes Alice's. For the purposes of this section, we will assume
that Alice's event precedes Bob's. Within this context, one may imagine a single particle of spin $j$
first interacting with Alice's detector and then with Bob's.

\begin{figure}[t]
\centering
\begin{tikzpicture}[scale=0.95]
 \sgdetector{0}{red}{Alice ($a$)}{0}
 \sgdetector{5}{blue}{Bob ($b$)}{-25}
  \sgangle{6}{$\theta$}
 \draw[fill=white] (2.18,0.88) rectangle (2.32,1.32);
 \draw[fill=white] (4.93,0.88) rectangle (5.07,1.32);
 \draw[dashed,black!60,-{Stealth[length=2mm]}] (2.32,1.1) -- (4.93,1.1);
\end{tikzpicture}
\caption{Two Stern-Gerlach detectors with a relative angle of rotation $\theta$ are assigned to separate
observers named Alice and Bob. Within each detector occurs a single event, which share a direct causal
connection to one another.}
\label{fig:sg1}
\end{figure}

An obvious requirement of this model is that the spin quantum number associated with Alice's event be the
same as the one associated with Bob's, or that $j_{a}=j_{b}$. This can be accomplished by choosing
$\gamma=0$. In terms of the underlying base-16 counts, this constraint has the following impact,
which can be inferred from equations (\ref{eq:mapexplicit}) and (\ref{eq:gamma}):
\begin{equation}
\ct{AC}+\ct{BD}+\ct{CA}+\ct{DB}+\ct{AD}+\ct{BC}+\ct{CB}+\ct{DA}=0
\end{equation}

Thus, the proposed constraint reduces the total number of degrees of freedom within this model to 8,
rather than 16. After applying this constraint, the quantum numbers associated with events can be
expressed in terms of base-16 counts like so, where we note that $j_{a}$ and $j_{b}$ are now identical:
\begin{equation}
n=\ct{AA}+\ct{BB}+\ct{CC}+\ct{DD}+\ct{AB}+\ct{BA}+\ct{CD}+\ct{DC}
\end{equation}
\begin{equation}
j_{a}=\frac{\ct{CC}+\ct{CD}+\ct{DC}+\ct{DD}}{2},\qquad j_{b}=\frac{\ct{CC}+\ct{DC}+\ct{CD}+\ct{DD}}{2}
\end{equation}
\begin{equation}
m_{a}=\frac{\ct{CC}+\ct{CD}-\ct{DC}-\ct{DD}}{2},\qquad m_{b}=\frac{\ct{CC}+\ct{DC}-\ct{CD}-\ct{DD}}{2}
\end{equation}
\begin{equation}
l_{a}=\frac{\ct{AA}+\ct{AB}-\ct{BA}-\ct{BB}}{2},\qquad l_{b}=\frac{\ct{AA}+\ct{BA}-\ct{AB}-\ct{BB}}{2}
\end{equation}

A further consequence of setting $\gamma=0$ is that $\alpha=\beta=\Bmap$, and from
equation (\ref{eq:nmap}) that
\begin{equation}
\Amap+\Bmap=n
\label{eq:ABn}
\end{equation}
This implies that the maps of interest within this model will only contain the symbols $A$ and $B$. An
example of one such map is offered here:
\begin{equation}
\begin{pmatrix}A\\B\\C\\C\\B\\A\end{pmatrix}_{a}\oplus\begin{pmatrix}A\\A\\B\\B\\B\\A\end{pmatrix}_{map}
=\begin{pmatrix}A\\B\\D\\D\\A\\A\end{pmatrix}_{b}
\end{equation}

The relative angle of rotation between Alice's and Bob's detector is a function of the two surviving map
counts $\Amap$ and $\Bmap$ alone. In section \ref{sec:calibration} we show that this function is fixed,
with no free parameters, by the requirement that the angles of successive rotations add, together with
the spin-$\frac12$ sector, and that it takes the form
\begin{equation}
\tan\left(\frac{\theta}{2}\right)\equiv\frac{\Bmap}{\Amap}=\frac{\Bmap}{n-\Bmap}
\label{eq:calib}
\end{equation}
As $\Bmap$ runs over the integers $0,1,\dots,n$, equation (\ref{eq:calib}) sweeps $\theta$ over the full
range $[0,\pi]$, and this range can be made arbitrarily smooth by increasing $n$.

The 7 quantum numbers we have defined thus far constitute the full set of local quantum numbers necessary
for the proposed model. The only remaining degree of freedom is the non-local quantum number $\nu^0$, or
its partner $\mu^0$, as they are defined in equation (\ref{eq:nu}). In fact, $\nu^0$ and $\mu^0$ are the
only quantum numbers among those defined in equations (\ref{eq:nu}-\ref{eq:omega}) that remain non-local
under the constraint $\gamma=0$. In this case, we will use $\nu^0$ as the final quantum number. This
quantity, which is closely related to $\mu_{a1,b2}$ in \cite{PowersSG}, is restated here:
\begin{equation}
\nu^0=\frac{1}{4}(\ct{AA}+\ct{BB}+\ct{CD}+\ct{DC})
\end{equation}

What remains is to decide which of the local quantum numbers to treat as random variables, conditioning
variables, and local nuisance variables. Given the physical context, the random variable of interest is
$m_{b}$, or the spin projection quantum number associated with Bob's event. The conditioning variables
are $n$, $j$, $m_{a}$, and $\theta$, where $j=j_{a}=j_{b}$. Finally, the local nuisance variables will
be $l_{a}$ and $l_{b}$. With these choices, we are left with the following definitions for the sets $R$,
$U$, $W$, and $\Lambda$:
\begin{equation}
R=(m_{b}),\quad U=(n,j,m_{a},\theta),\quad W=(l_{a},l_{b}),\quad \Lambda=(\nu^0)
\end{equation}

\subsection{Entangled spin systems}
\label{sec:entmodel}

The physical system of interest in this section again consists of two Stern-Gerlach detectors, though we
now require that these detectors be aligned ($\theta=0$), as depicted in Figure \ref{fig:sg2}. The
experiments under consideration involve two events with an indirect causal connection and an indefinite
time ordering. That is, Alice's event may precede Bob's, Bob's event may precede Alice's, or they may
occur simultaneously. Of course, this implies that one can no longer imagine Alice's and Bob's events as
being connected by a single particle, as in section \ref{sec:rotmodel}. Instead, they must be associated
with particles that are related through a third event, which we may associate with an observer named
Charlie.

\begin{figure}[t]
\centering
\begin{tikzpicture}[scale=0.95]
 \sgdetector{0}{red}{Alice ($a$)}{0}
 \sgdetector{6.4}{blue}{Bob ($b$)}{0}
 \draw[fill=white] (2.18,0.88) rectangle (2.32,1.32);
 \draw[fill=white] (6.33,0.88) rectangle (6.47,1.32);
 \fill[black!65] (4.33,1.1) circle (0.28);
 \draw[{Stealth[length=2mm]}-{Stealth[length=2mm]},black!55] (2.42,1.1) -- (3.92,1.1);
 \draw[{Stealth[length=2mm]}-{Stealth[length=2mm]},black!55] (4.74,1.1) -- (6.23,1.1);
\end{tikzpicture}
\caption{Two aligned Stern-Gerlach detectors are assigned to separate observers named Alice and Bob.
Within each detector occurs a single event, which share an indirect causal connection to one another.}
\label{fig:sg2}
\end{figure}

As was the case in section \ref{sec:rotmodel}, the model introduced here involves setting one of the
quantities defined in equations (\ref{eq:alpha}-\ref{eq:gamma}) equal to zero. For reasons that will
become clear shortly, we set $\alpha=0$. In terms of the underlying base-16 counts, this constraint
has the following impact, which can be inferred from equations (\ref{eq:mapexplicit}) and
(\ref{eq:alpha}):
\begin{equation}
\ct{AB}+\ct{BA}+\ct{CD}+\ct{DC}+\ct{AC}+\ct{CA}+\ct{BD}+\ct{DB}=0
\end{equation}

The proposed constraint reduces the total number of degrees of freedom within this model to 8, rather than
16. After applying this constraint, the quantum numbers associated with Alice's and Bob's events can be
expressed in terms of base-16 counts like so:
\begin{equation}
n=\ct{AA}+\ct{BB}+\ct{CC}+\ct{DD}+\ct{AD}+\ct{BC}+\ct{CB}+\ct{DA}
\end{equation}
\begin{equation}
j_{a}=\frac{\ct{CC}+\ct{CB}+\ct{DA}+\ct{DD}}{2},\qquad j_{b}=\frac{\ct{CC}+\ct{BC}+\ct{AD}+\ct{DD}}{2}
\label{eq:jent}
\end{equation}
\begin{equation}
m_{a}=\frac{\ct{CC}+\ct{CB}-\ct{DA}-\ct{DD}}{2},\qquad m_{b}=\frac{\ct{CC}+\ct{BC}-\ct{AD}-\ct{DD}}{2}
\end{equation}
\begin{equation}
l_{a}=\frac{\ct{AA}+\ct{AD}-\ct{BC}-\ct{BB}}{2},\qquad l_{b}=\frac{\ct{AA}+\ct{DA}-\ct{CB}-\ct{BB}}{2}
\label{eq:lent}
\end{equation}

The constraint $\alpha=0$ also implies that $\beta=\gamma=\Dmap$. Thus, the maps of
interest within this model will only contain the symbols $A$ and $D$. An example of one such map is
offered here:
\begin{equation}
\begin{pmatrix}A\\B\\C\\C\\B\\A\end{pmatrix}_{a}\oplus\begin{pmatrix}A\\A\\D\\A\\A\\D\end{pmatrix}_{map}
=\begin{pmatrix}A\\B\\B\\C\\B\\D\end{pmatrix}_{b}
\end{equation}

An important consequence of restricting maps to the symbols $A$ and $D$ is that the base-2 sequences
carrying the subscripts $a$ and $b$ will always be identical. In other words, Alice's and Bob's events
will always share a common reference sequence within this model. This leaves just three unique base-2
sequences in each ontic state within this model: Alice's system sequence (labeled $1$ below), Bob's
system sequence (labeled $2$), and their common reference sequence (labeled $(a,b)$). We may interpret
these as the vertices of a triangle:
\begin{equation}
\begin{tikzpicture}[baseline=(current bounding box.center),scale=1.0]
 \coordinate (v1) at (0,1.5); \coordinate (v2) at (2.2,1.5); \coordinate (v3) at (1.1,0);
 \draw[black!45] (v1)--(v2)--(v3)--cycle;
 \fill (v1) circle (1.4pt); \fill (v2) circle (1.4pt); \fill (v3) circle (1.4pt);
 \node[above left] at (v1) {$1$}; \node[above right] at (v2) {$2$};
 \node[below] at (v3) {$(a,b)$};
 \node at (1.1,1.75) {$j_{c}$};
 \node at (0.15,0.75) {$j_{a}$}; \node at (2.05,0.75) {$j_{b}$};
\end{tikzpicture}
\label{eq:triangle}
\end{equation}

Equations (\ref{eq:jent}-\ref{eq:lent}) represent the full set of local quantum numbers associated with two
legs of this triangle. Those associated with the third leg are given here:
\begin{align}
j_{c}&\equiv\frac{\ct{BC}+\ct{DA}+\ct{AD}+\ct{CB}}{2}\\
m_{c}&\equiv\frac{\ct{BC}+\ct{DA}-\ct{AD}-\ct{CB}}{2}\\
l_{c}&\equiv\frac{\ct{AA}+\ct{CC}-\ct{BB}-\ct{DD}}{2}
\end{align}

The quantum numbers $j_{a}$, $j_{b}$, and $j_{c}$ are each proportional to the Hamming distance between
the base-2 sequences indicated by their respective subscripts. This implies that we may think of them as
the edge lengths for the triangle depicted in equation (\ref{eq:triangle}), which means that they obey the
following selection rule of angular momentum addition:
\begin{equation}
|j_{a}-j_{b}|\le j_{c}\le j_{a}+j_{b}
\label{eq:triangleineq}
\end{equation}

Of the three quantities defined above, $j_{c}$ is the only one that contains information not already
included in equations (\ref{eq:jent}-\ref{eq:lent}). In fact, it is proportional to the number of $D$'s
that appear in the map relating Alice's and Bob's events, which will be important in the following
section. The quantum numbers $m_{c}$ and $l_{c}$, on the other hand, can each be expressed in terms of
$m_{a}$, $m_{b}$, $l_{a}$, and $l_{b}$ in the following way:
\begin{align}
m_{c}&=-m_{a}+m_{b}=l_{b}-l_{a}\label{eq:m12}\\
l_{c}&=l_{a}+m_{b}=l_{b}+m_{a}
\end{align}

Equation (\ref{eq:m12}) contains the expression $m_{c}=-m_{a}+m_{b}$, which is reminiscent of
conservation of angular momentum. In this form, the event at Bob's detector is associated with a composite
system, while the event at Alice's detector is associated with a component of that system. To consider
alternatives to this situation, we need a way to modify the signs of the spin projection quantum numbers.
In general, this can be accomplished by commuting the base-2 sequences within Alice's and/or Bob's events.
Under these operations, the numerical values associated with the counts $\ct{C}$ and $\ct{D}$ are
exchanged, while those associated with $\ct{A}$ and $\ct{B}$ are left invariant. By the definition of spin
projection quantum numbers offered in equation (\ref{eq:mdef}), this exchange results in a sign change for
$m$'s, while leaving all other quantum numbers associated with events unchanged.

Unfortunately, commuting base-2 sequences within events complicates the quantum numbers associated with
maps, as well as the non-local quantum numbers discussed in section \ref{sec:indist}. While this may
indeed be the proper course in the long run, a simpler solution will suffice for the models of interest in
this paper. In section \ref{sec:localqn}, recall that we had two choices for the definition of the spin
projection quantum numbers. Rather than choosing $\ct{C}-\ct{D}$, we could have instead chosen
$\ct{D}-\ct{C}$. By allowing this choice to vary between the definition of Alice's and Bob's spin
projection quantum numbers, we can capture the effect of the commutation operation, while leaving the
quantum numbers we have already defined alone. For the purposes of this section and the next, we will
choose this alternative definition for Alice's spin projection quantum number. We distinguish between
these two definitions by placing a bar over Alice's subscript:
\begin{equation}
j_{\bar a}=j_{a}=\frac{\ca{D}+\ca{C}}{2},\qquad
m_{\bar a}=-m_{a}=\frac{\ca{D}-\ca{C}}{2},\qquad
l_{\bar a}=l_{a}=\frac{\ca{A}-\ca{B}}{2}
\label{eq:swap}
\end{equation}

With this modification, the relations for $m_{c}$ and $l_{c}$ become the following:
\begin{align}
m_{c}&=m_{\bar a}+m_{b}=l_{b}-l_{\bar a}\label{eq:m12b}\\
l_{c}&=l_{\bar a}+m_{b}=l_{b}-m_{\bar a}\label{eq:l12b}
\end{align}

Thus, the spin addition selection rule $m_{c}=m_{\bar a}+m_{b}$ now joins equation (\ref{eq:triangleineq})
as an intrinsic property of the spin quantum numbers $j$ and $m$, under the proposed constraints. This
result is central to the issue of entanglement, which demands a perfect correlation between Alice's and
Bob's measurement outcomes when $\theta=0$. Importantly, equations (\ref{eq:m12b}) and (\ref{eq:l12b}) are
both a direct consequence of Alice's and Bob's events sharing a common reference sequence. Note that this
result can be traced all the way back to the choice to set $\alpha=0$.

Before moving on, we again note that $j_{c}$ is actually proportional to the number of $D$'s that appear
in the maps connecting Alice's and Bob's events. In this way, it plays a role analogous to the map count
$\Bmap$ introduced in the previous section. This association between the total spin of a composite system
and the number of $D$'s that appear in maps will persist in the following section. In fact, it will
supersede its association with the Hamming distance between Alice's and Bob's system sequences.

Due to the identities in equations (\ref{eq:m12b}) and (\ref{eq:l12b}), only seven of the ten quantum
numbers introduced thus far are necessary to form a complete set. Our choices will be $n$, $j_{\bar a}$,
$m_{\bar a}$, $j_{b}$, $m_{b}$, $j_{c}$, and $l_{c}$. The only remaining degree of freedom is the
non-local quantum number $\nu^4$, or its partner $\mu^4$, as defined in equation (\ref{eq:nu}). We will
use $\nu^4$ in this case, which is closely related to $k$ in \cite{PowersSpin}. This non-local quantum
number is restated here:
\begin{equation}
\nu^4=\frac{1}{4}(\ct{AA}+\ct{DD}+\ct{BC}+\ct{CB})
\end{equation}

In this model, the random variables are $m_{\bar a}$ and $m_{b}$, the conditioning variables are $n$,
$j_{\bar a}$, $j_{b}$, and $j_{c}$, and the local nuisance variable is $l_{c}$. Notably, the quantum number
$m_{c}$ does not explicitly enter this probability calculation. However, it does constrain the allowed
combinations of $m_{\bar a}$ and $m_{b}$ through equation (\ref{eq:m12b}). We may interpret this as a
consequence of $m_{c}$ being associated with a third event, which is not explicitly modeled. We now have
the following definitions of the sets $R$, $U$, $W$, and $\Lambda$:
\begin{equation}
R=(m_{\bar a},m_{b}),\quad U=(n,j_{\bar a},j_{b},j_{c}),\quad W=(l_{c}),\quad \Lambda=(\nu^4)
\end{equation}

\subsection{Bell Test}
\label{sec:bellmodel}

In this section, we will combine the previous two models to produce one which will enable us to model a
Bell test (Figure \ref{fig:sg3}). This can be accomplished by combining the maps used in the previous two
sections under the addition mod2 operation. One of these maps is comprised of $A$'s and $D$'s and encodes
the total spin of the composite system. The other is comprised of $A$'s and $B$'s and encodes the relative
rotation between the two detectors. These two maps sum to form a single unique base-4 sequence, which
encodes both of these degrees of freedom. One example of this construction is given here:
\begin{equation}
\begin{pmatrix}A\\B\\B\\A\\B\\A\end{pmatrix}\oplus\begin{pmatrix}A\\A\\D\\D\\A\\A\end{pmatrix}
=\begin{pmatrix}A\\B\\C\\D\\B\\A\end{pmatrix}_{map}
\end{equation}

\begin{figure}[t]
\centering
\begin{tikzpicture}[scale=0.95]
 \sgdetector{0}{red}{Alice ($a$)}{0}
 \sgdetector{6.4}{blue}{Bob ($b$)}{-25}
 \sgangle{7.4}{$\theta_{ab}$}
 \draw[fill=white] (2.18,0.88) rectangle (2.32,1.32);
 \draw[fill=white] (6.33,0.88) rectangle (6.47,1.32);
 \fill[black!65] (4.33,1.1) circle (0.28);
 \draw[{Stealth[length=2mm]}-{Stealth[length=2mm]},black!55] (2.42,1.1) -- (3.92,1.1);
 \draw[{Stealth[length=2mm]}-{Stealth[length=2mm]},black!55] (4.74,1.1) -- (6.23,1.1);
\end{tikzpicture}
\caption{Two Stern-Gerlach detectors with a relative angle of rotation $\theta_{ab}$ are assigned to
separate observers named Alice and Bob. Within each detector occurs a single event, which share an
indirect causal connection to one another.}
\label{fig:sg3}
\end{figure}

The number of $B$'s that appear in the map encoding rotation is equivalent to the sum
$\ct{B}+\ct{C}=\alpha$ in the combined map. Similarly, the number of $D$'s in the map encoding the
total spin of the composite system is equivalent to the sum $\ct{D}+\ct{C}=\gamma$ in the combined
map. Likewise, the number of $A$'s in the map encoding rotation is $n-\alpha$. We are thus left with
the following definitions of the total spin of the composite spin system and of the angle between Alice's
and Bob's detector within this model:
\begin{equation}
j_c=\frac{1}{2}\gamma
\end{equation}
\begin{equation}
\tan\left(\frac{\theta_{ab}}{2}\right)=\frac{\alpha}{n-\alpha}
\label{eq:calibbell}
\end{equation}

Equation (\ref{eq:calibbell}) is the direct generalisation of equation (\ref{eq:calib}): in the pure
rotation model of section \ref{sec:rotmodel} one has $\gamma=0$, so that $\alpha=\Bmap$ and
$n-\alpha=\Amap$, and the two expressions coincide.

We retain the symbol $j_c$ for the total spin of the composite system, but its definition has changed:
in section \ref{sec:entmodel} $j_c$ was half the Hamming distance between Alice's and Bob's system
sequences, whereas here it is half the number of $D$'s in the spin-encoding map, $\gamma/2$. Once rotation
is allowed the two no longer coincide, and the total spin of the composite system can no longer be
interpreted as a Hamming distance. However, the triangle inequality
defined in equation (\ref{eq:triangleineq}) still holds. This is because the rotation maps, which involve
$B$'s, always leave $j_{\bar a}$ and $j_{b}$ invariant.

The change from $\theta$ to $\theta_{ab}$, on the other hand, indicates that the new quantity is a function
of the Hamming distance $\alpha$ between Alice's and Bob's reference sequences, as indicated by the subscripts.
Actually, this is also the case for $\theta$, but in section \ref{sec:rotmodel}, $\theta$ was a function of
the Hamming distance between Alice's and Bob's complete base-4 sequences, rather than just the reference
sequences.

These new quantum numbers can be paired with $n$, $j_{\bar a}$, $m_{\bar a}$, $l_{\bar a}$, $j_{b}$, $m_{b}$, and
$l_{b}$, which together constitute 9 of the 10 local quantum numbers associated with the experiments of
interest in this section. The 10th local quantum number will be $\beta$, as defined in equation
(\ref{eq:beta}). Like the quantum numbers $l_{\bar a}$ and $l_{b}$, the physical interpretation of
$\beta$ is unclear at this time. So, it will be treated as a nuisance variable within probability
calculations. Its allowed range follows directly from the definitions
(\ref{eq:alpha}-\ref{eq:gamma}):
\begin{equation}
|\alpha-\gamma|\le\beta\le\alpha+\gamma,\qquad
\beta\equiv\alpha+\gamma\ (\mathrm{mod}\ 2)
\label{eq:betarange}
\end{equation}
Equation (\ref{eq:betarange}) is stated purely in terms of counts, and therefore requires no reference
to the angle.

Unlike the models introduced in sections \ref{sec:rotmodel} and \ref{sec:entmodel}, this model requires
all 6 non-local quantum numbers to form a complete set, which will be those chosen in equation
(\ref{eq:lambdachoice}):
\begin{equation}
\begin{split}
\nu^0\equiv\tfrac{1}{4}(\ct{AA}+\ct{BB}+\ct{CD}+\ct{DC}),\quad
&\nu^1\equiv\tfrac{1}{4}(\ct{AD}+\ct{BC}+\ct{CA}+\ct{DB})\\
\nu^4\equiv\tfrac{1}{4}(\ct{AA}+\ct{DD}+\ct{BC}+\ct{CB}),\quad
&\nu^5\equiv\tfrac{1}{4}(\ct{BA}+\ct{CD}+\ct{AC}+\ct{DB})\\
\kappa^1=\tfrac{1}{4}(\ct{AC}+\ct{DA}+\ct{CD}+\ct{BB}),\quad
&\omega^2=\tfrac{1}{4}(\ct{BA}+\ct{DD}+\ct{AC}+\ct{CB})
\end{split}
\end{equation}

The final issue we must address before modeling a Bell test concerns the spin projection quantum number
for the composite system. Typically, Bell tests are performed on composite systems with total spin 0. As a
result, the spatial frame used to define this quantity is irrelevant in the resulting experiment. That is,
only the relative angle between Alice's and Bob's detector has any physical significance. However, if the
composite system has non-zero total spin, the spatial frame associated with this event does become
relevant. In this case, we must keep track of Alice's and Bob's angle of rotation with respect to this
third spatial frame, which we may associate with an observer named Charlie. Because Charlie's event is not
explicitly included in this model, we are limited to scenarios in which this spatial orientation does not
matter. With this in mind, we may now generalize the definition of $m_c$, the spin projection quantum
number of the composite system, to the present model. At $\theta_{ab}=0$ it must reduce to the definition
of section \ref{sec:entmodel}:
\begin{equation}
(\theta_{ab}=0):\quad m_c=\frac{\ct{BC}+\ct{DA}-\ct{AD}-\ct{CB}}{2}
\end{equation}

In the case that $\theta_{ab}\ne0$, four additional base-16 counts will begin to contribute to $m_c$.
These are $\ct{AC}$, $\ct{CA}$, $\ct{BD}$, and $\ct{DB}$. To ensure that
$-j_c\le m_c\le j_c$, two of these additional base-16 counts must have a coefficient of
$-\frac{1}{2}$ and two must have a coefficient of $+\frac{1}{2}$. This implies that we have six unique
choices when defining $m_c$, one for each combination of coefficients. Of these six choices, only two
will yield the appropriate behavior for the full range of $\theta_{ab}$. These are defined as follows,
where the distinguishing superscripts will be explained shortly:
\begin{align}
m^a_c&\equiv\tfrac{1}{2}(\ct{BC}+\ct{DA}+\ct{BD}+\ct{DB}-\ct{AD}-\ct{CB}-\ct{AC}-\ct{CA})\\
m^b_c&\equiv\tfrac{1}{2}(\ct{BC}+\ct{DA}+\ct{AC}+\ct{CA}-\ct{AD}-\ct{CB}-\ct{BD}-\ct{DB})
\end{align}

The difference between these two quantities can be best understood by considering cases in which they are
non-zero. As an example, consider the case in which $j_c=1$, $m^a_c=m^b_c=+1$,
$j_{\bar a}=j_{b}=+\frac{1}{2}$. When $\theta_{ab}=0$, the only non-zero base-16 counts involved in the
definition of $m^a_c$ and $m^b_c$ are $\ct{BC}$ and $\ct{DA}$, which both equal 1. When
$\theta_{ab}=\pi$, these counts go to zero. To ensure that $m^a_c$ and $m^b_c$ are preserved
under this limit, we must have $\ct{BD}=\ct{DB}=1$ ($m^a_c=+1$) or $\ct{AC}=\ct{CA}=1$
($m^b_c=+1$). Therefore, when $\theta_{ab}=\pi$, $m^a_c=+1$ implies $m_{\bar a}=+\frac{1}{2}$ and
$m_{b}=-\frac{1}{2}$ with certainty, while $m^b_c=+1$ implies $m_{\bar a}=-\frac{1}{2}$ and
$m_{b}=+\frac{1}{2}$ with certainty.

In this example the distinction between $m^a_c$ and $m^b_c$ outlined above holds for arbitrary
rotations. If $m^a_c$ is used, the spin projection quantum number observed at Alice's detector is
always $+\frac{1}{2}$, while Bob's may vary. Alternatively, using $m^b_c$ will ensure
$m_{b}=+\frac{1}{2}$, while $m_{\bar a}$ may vary. Physically, the distinction between $m^a_c$ and
$m^b_c$ can be understood in terms of spatial frames. That is, using $m^a_c$ implies the spin
projection quantum number for the composite system has been defined with respect to Alice's spatial frame,
while $m^b_c$ implies it has been defined with respect to Bob's. Of course, when
$m^a_c=m^b_c=0$, this distinction becomes unimportant.

As in section \ref{sec:entmodel}, the spin projection quantum number for the composite system will not
explicitly enter the probability calculation. Instead, it will act as a constraint on the quantum numbers
that do enter this calculation. In terms of these quantum numbers, $m^a_c$ and $m^b_c$ can be
expressed as follows:
\begin{align}
m^a_c&=\frac{n}{4}+\frac{m_{\bar a}}{2}-\frac{l_{\bar a}}{2}-2(2\omega^2+\nu^0+\nu^1-\nu^4-\nu^5)
\label{eq:magamma}\\
m^b_c&=-\frac{n}{4}+\frac{m_{b}}{2}+\frac{l_{b}}{2}+\frac{1}{2}(\alpha-\gamma)
+2(2\kappa^1-\nu^0+\nu^1+\nu^4-\nu^5)
\end{align}
where $\alpha$ is related to $\theta_{ab}$ through the calibration (\ref{eq:calibbell}) and $\gamma=2j_c$,
so that these relations are stated purely in terms of counts.

In the following section, we will use $m^a_c$ to constrain the permitted ranges of the relevant
quantum numbers. Thus, equation (\ref{eq:magamma}) will supersede equation (\ref{eq:m12b}) as the
selection rule governing this more general calculation. The only remaining task of this section is the
assignment of quantum numbers to their respective categories, which are as follows:
\begin{equation}
R=(m_{\bar a},m_{b}),\ U=(n,j_{\bar a},j_{b},\theta_{ab},j_c),\ W=(l_{\bar a},l_{b},\beta),\
\Lambda=(\nu^0,\nu^1,\nu^4,\nu^5,\kappa^1,\omega^2)
\end{equation}

Before proceeding we record a structural observation which will be useful in section
\ref{sec:bellresults}. A Bell test is performed on a composite system with $j_c=0$, which by
$j_c=\frac{1}{2}\gamma$ forces $\gamma=0$ and hence $\Cmap=\Dmap=0$. Consequently the
eight counts $\ct{AC}$, $\ct{BD}$, $\ct{CA}$, $\ct{DB}$, $\ct{AD}$, $\ct{BC}$, $\ct{CB}$, $\ct{DA}$ all
vanish, $\beta=\alpha=\Bmap$ by equation (\ref{eq:betarange}), and $m^a_c$ vanishes
identically by equation (\ref{eq:magamma}). The singlet Bell model therefore inherits exactly the count
structure of section \ref{sec:rotmodel}, with the sign convention (\ref{eq:swap}) applied at Alice's
event and with both $m_{\bar a}$ and $m_{b}$ promoted to random variables. This is a useful consistency
check on the formalism, and it is the reason that the results of sections \ref{sec:rotresults} and
\ref{sec:bellresults} are so tightly linked.

\section{Calibration of the rotation angle}
\label{sec:calibration}

Sections \ref{sec:rotmodel} and \ref{sec:bellmodel} both require a rule which converts the counts
labeling a rotation map into a physical angle. Nothing introduced so far fixes this rule. The counts
$(\Amap,\Bmap)$ are the only data a rotation map carries, and the function taking those data to $\theta$
is a genuine physical input which must be determined rather than assumed. In this section we fix it. The
argument has two parts, which we present separately and with an explicit statement of what each does and
does not establish.

\subsection{Composition of rotations}

The first part is structural. Within this formalism the rotation maps of section \ref{sec:rotmodel} form a
one parameter family labeled by the pair of counts $(\Amap,\Bmap)$ subject to $\Amap+\Bmap=n$. A
physically acceptable angle assignment $\theta(\Amap,\Bmap)$ must be compatible with the composition of
successive rotations: if a map taking Alice's frame to Bob's is followed by a map taking Bob's frame to a
third frame, the assigned angles must add.

Associate to each rotation map the complex number
\begin{equation}
z\equiv\Amap+i\Bmap
\label{eq:complexmap}
\end{equation}
and to the composition of two maps the product $z_1z_2$, rescaled so that the composite again has counts
summing to $n$. Under this rule the arguments add exactly,
\begin{equation}
\arg(z_1z_2)=\arg(z_1)+\arg(z_2)
\end{equation}
so the unique angle-like quantity carried by a map, and the only one which is additive under composition,
is $\arg(z)=\arctan(\Bmap/\Amap)$.

It remains to fix the proportionality between $\arg(z)$ and the physical angle $\theta$. This is settled by
the fact that the objects being rotated are spinors: a rotation by $\theta$ acts on a spin-$\frac{1}{2}$
system through the half angle $\theta/2$. Identifying the additive map angle with the spinor half angle
gives $\theta/2=\arctan(\Bmap/\Amap)$, or
\begin{equation}
\boxed{\ \tan\left(\frac{\theta}{2}\right)=\frac{\Bmap}{\Amap}\ }
\label{eq:calibbox}
\end{equation}

We are careful not to overstate this argument. The composition rule (\ref{eq:complexmap}) is an additional
structure imposed on the map counts; it is not derived from the $\oplus$ operation, since the counts of a
composition of two maps under $\oplus$ depend on the overlap of the two sequences and not on the counts
alone. What the argument establishes is that \emph{if} map angles are to be additive under composition,
then the assignment must be of the form (\ref{eq:calibbox}), and that no free parameter survives. We
regard this as motivation, and turn now to the independent verification which fixes the same law.

\subsection{The \texorpdfstring{spin-$\frac{1}{2}$}{spin-1/2} sector}

The second part is a direct calculation. Consider the model of section \ref{sec:rotmodel} with
$j=\frac{1}{2}$ and $m_{a}=+\frac{1}{2}$. In this sector the block of counts carrying the symbols $C$ and
$D$ contains a single unit, so each local state space consists of a single elementary state space (a
single value of $\nu^0$) and interference is absent. The sum over the nuisance variables $l_a,l_b$ in equation (\ref{eq:upsilon}) can then be carried
out in closed form. Writing $M=n-1$ and $B'$ for the number of $B$ symbols in the map that fall in the
$\{A,B\}$ block of Alice's event, one has $G_a|\varepsilon^a|=G_b|\varepsilon^b|=M!/(\ct{AA}!\ct{AB}!\ct{BA}!\ct{BB}!)$,
and the Vandermonde identity $\sum_{a+b=N}\binom{N}{a}^2=\binom{2N}{N}$ gives
\begin{equation}
\Upsilon(B')=\binom{M}{B'}^{2}\binom{2B'}{B'}\binom{2M-2B'}{M-B'},\qquad
\Upsilon\!\left(m_b=+\tfrac12\right)=\Upsilon(\Bmap),\quad
\Upsilon\!\left(m_b=-\tfrac12\right)=\Upsilon(\Bmap-1).
\label{eq:halfclosed}
\end{equation}
The ratio of the two probabilities is therefore exactly
\begin{equation}
\frac{P(m_b=-\frac12)}{P(m_b=+\frac12)}=\frac{\Bmap^{3}\,(2\Amap-1)}{\Amap^{3}\,(2\Bmap-1)}
=\frac{\Bmap^{2}}{\Amap^{2}}\left[1+\frac{\Amap-\Bmap}{2\Amap\Bmap-\Amap}\right],
\label{eq:halfratio}
\end{equation}
and taking $n\to\infty$ at fixed $\Bmap/n$ yields
\begin{equation}
\lim_{n\to\infty}P\left(m_{b}=-\tfrac{1}{2}\ \middle|\ n,\tfrac{1}{2},+\tfrac{1}{2},\Bmap\right)
=\frac{\Bmap^2}{\Amap^2+\Bmap^2},\qquad
\lim_{n\to\infty}P\left(m_{b}=+\tfrac{1}{2}\right)=\frac{\Amap^2}{\Amap^2+\Bmap^2}
\label{eq:halflimit}
\end{equation}
The corresponding QM predictions are $\sin^2(\theta/2)$ and $\cos^2(\theta/2)$. Equating them requires
\begin{equation}
\sin^2\left(\frac{\theta}{2}\right)=\frac{\Bmap^2}{\Amap^2+\Bmap^2},\qquad
\cos^2\left(\frac{\theta}{2}\right)=\frac{\Amap^2}{\Amap^2+\Bmap^2}
\end{equation}
which is equivalent to equation (\ref{eq:calibbox}). The two arguments agree, and the calibration is fixed
with no adjustable constants.

Equation (\ref{eq:halfratio}) also gives the finite $n$ correction in closed form. Under the calibration,
$(\Amap-\Bmap)/(2\Amap\Bmap)=\cot\theta/n$, so that
\begin{equation}
P\left(m_b=-\tfrac12\right)=\sin^2\frac{\theta}{2}+\frac{\sin 2\theta}{8n}+O(n^{-2}),\qquad
\max_\theta|\Delta|=\frac{1}{8n}+O(n^{-2}),
\label{eq:halfcorrection}
\end{equation}
with the maximum attained at $\theta=\pi/4$ and $3\pi/4$. For $\theta<\pi/2$ the model therefore
slightly overestimates the probability of a spin flip, and for $\theta>\pi/2$ it slightly underestimates it. This is the residual plotted in the lower panel of
Figure \ref{fig:f5} and the first column of Table \ref{tab:conv}.

Two comments are in order. First, the calibration is fixed by the spin-$\frac{1}{2}$ sector alone. It is
therefore a genuine test, not a fit, that the same rule reproduces the Wigner $d$-functions for $j=1$, as
demonstrated in section \ref{sec:rotresults}; the $m_a=0$ case in particular involves interference between
two ontic states and is not implied by the spin-$\frac12$ result. The entangled-spin results of
section \ref{sec:entresults} involve no rotation and do not depend on the calibration at all. The CHSH
statistic of section
\ref{sec:bellresults} is in a different position: as shown at the end of section \ref{sec:bellmodel}, the
singlet Bell model has exactly the count structure of the spin-$\frac12$ rotation model, so once the
calibration reproduces $\sin^2(\theta/2)$ the singlet correlation $-\cos\theta$ and the Tsirelson value
$2\sqrt2$ follow algebraically. We present the CHSH result as a consistency check of the construction
rather than as an independent prediction. Second, equation (\ref{eq:halflimit}) holds only
in the limit; at finite $n$ the model departs from QM by an amount which is $O(1/n)$ throughout, exactly
so in the spin-$\frac12$ sector by equation (\ref{eq:halfcorrection}) and numerically for $j=1$. The
discreteness of the underlying formalism is therefore not hidden by the
calibration, it is simply relegated to a correction which vanishes in the large-$n$ limit. This is exactly
the behavior one would want from a discrete model of a continuous theory.

It is also worth recording the equivalent statement in terms of the standard half angle identities. Writing
$\Amap=n-\Bmap$, equation (\ref{eq:calibbox}) gives
\begin{equation}
\cos\theta=\frac{\Amap^2-\Bmap^2}{\Amap^2+\Bmap^2},\qquad
\sin\theta=\frac{2\Amap\Bmap}{\Amap^2+\Bmap^2}
\label{eq:pythag}
\end{equation}
so that $(\Amap,\Bmap)$ generate the Pythagorean parametrisation of the unit circle. The discrete set of
angles accessible at a given $n$ is therefore the set of angles subtended by Pythagorean-like pairs
$(n-\Bmap,\Bmap)$, which becomes dense in $[0,\pi]$ as $n\to\infty$.

\section{Results}
\label{sec:results}

In this section, we compare the output of equation (\ref{eq:prob}) to the predictions of QM for each of the
models introduced in section \ref{sec:models}. All numbers quoted below were obtained by exact integer
evaluation of equations (\ref{eq:Lainter}-\ref{eq:prob}); no floating point arithmetic enters except in the
final division and in the evaluation of the QM comparison formulas.

The overall picture is as follows. For the model of section \ref{sec:rotmodel}, we consider systems with
total spin $\frac{1}{2}$ and 1. Under the calibration (\ref{eq:calibbox}) the deviation from QM
vanishes in the limit of large $n$, for every value of $j$, $m_{a}$ and $\theta$ examined; it is
$O(1/n)$ analytically for $j=\frac12$ and numerically consistent with $O(1/n)$ for $j=1$. No
tuning parameter is introduced or required. For the model introduced in section \ref{sec:entmodel}, which
involves no rotation and is therefore unaffected by the calibration, we consider constituent spin systems
with total spin $\frac{1}{2}$ and 1, and composite spin systems with total spin 0, 1, and 2. In all but one
case, exact agreement is found between QM and the proposed model. In the remaining case, the difference
between the two models tends to zero as $n$ becomes large. Finally, the model introduced in section
\ref{sec:bellmodel} is used to perform a CHSH calculation, which converges to the Tsirelson bound
$2\sqrt{2}$. The logical status of each of these comparisons is set out at the end of section
\ref{sec:calibration}.

\subsection{Results: spin systems in rotated frames}
\label{sec:rotresults}

The physical system of interest in this section is a particle with total spin $j$, which first interacts
with Alice's Stern-Gerlach detector, followed by Bob's. The probability we are interested in calculating is
for the spin projection quantum number observed by Bob, which we denote as $m_{b}$. In QM, this
probability is given by the square of Wigner's d-matrix formula, which is a function of $j$, $m_{a}$,
$m_{b}$, and $\theta$:
\begin{equation}
\begin{split}
d^j_{m_{b},m_{a}}(\theta)=\sum_q(-1)^{m_{b}-m_{a}+q}
\frac{(j+m_{a})!(j-m_{a})!(j+m_{b})!(j-m_{b})!}{(j+m_{a}-q)!q!(m_{b}-m_{a}+q)!(j-m_{b}-q)!}\\
\times\left(\cos\left(\frac{\theta}{2}\right)\right)^{2j+m_{a}-m_{b}-2q}
\left(\sin\left(\frac{\theta}{2}\right)\right)^{m_{b}-m_{a}+2q}
\end{split}
\label{eq:wigner}
\end{equation}

For the model proposed in section \ref{sec:rotmodel}, the expression which generates the probability of
observing a particular value of $m_{b}$ is given by the following, where $l_{a}$, $l_{b}$, and $\nu^0$
are summed over as outlined in section \ref{sec:probs}:
\begin{equation}
P(m_{b}|n,j,m_{a},\theta)=\frac{\Upsilon(n,j,m_{a},m_{b},\theta)}
{\sum_{m_{b}\in Q(n,j,m_{a},\theta)}\Upsilon(n,j,m_{a},m_{b},\theta)}
\label{eq:probrot}
\end{equation}

\begin{figure}[t]
\centering
\includegraphics[width=0.6\textwidth]{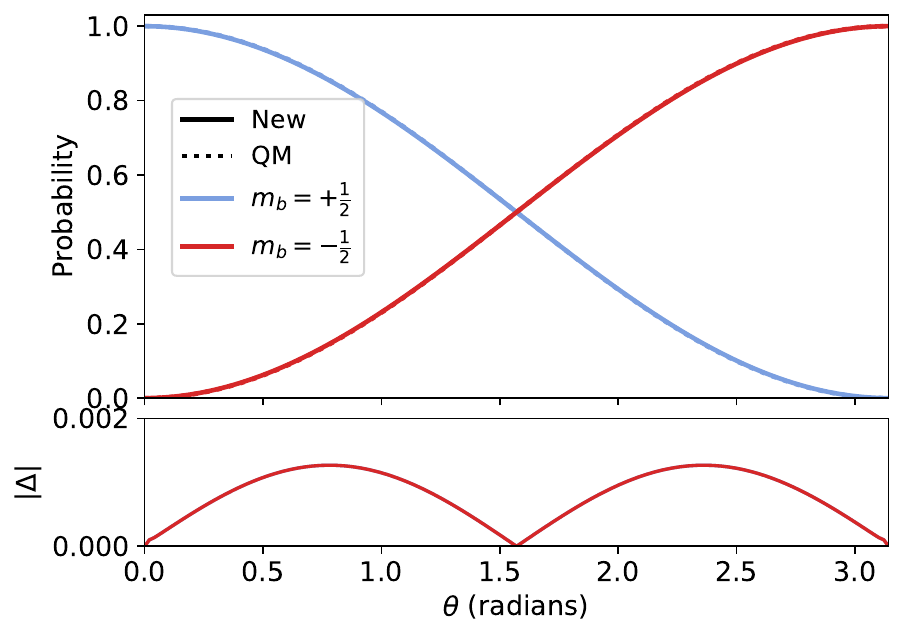}
\caption{A comparison of the proposed model with QM for $n=100$, $j=\frac{1}{2}$,
$m_{a}=+\frac{1}{2}$, where $|\Delta|$ is the magnitude of the difference between equations
(\ref{eq:wigner}) and (\ref{eq:probrot}).}
\label{fig:f5}
\end{figure}

In Figure \ref{fig:f5}, a comparison of equations (\ref{eq:wigner}) and (\ref{eq:probrot}) is offered for
$j=\frac{1}{2}$, $n=100$, and a full range of $\theta$. The two curves are indistinguishable at the
resolution of the plot. The residual, shown in the lower panel, vanishes at $\theta=0$, $\frac{\pi}{2}$
and $\pi$, and peaks near $\frac{\pi}{4}$ and $\frac{3\pi}{4}$ at $1.27\times10^{-3}$, in agreement with the
closed form (\ref{eq:halfcorrection}), whose leading term is $1/(8n)=1.25\times10^{-3}$.

\begin{figure}[t]
\centering
\includegraphics[width=0.6\textwidth]{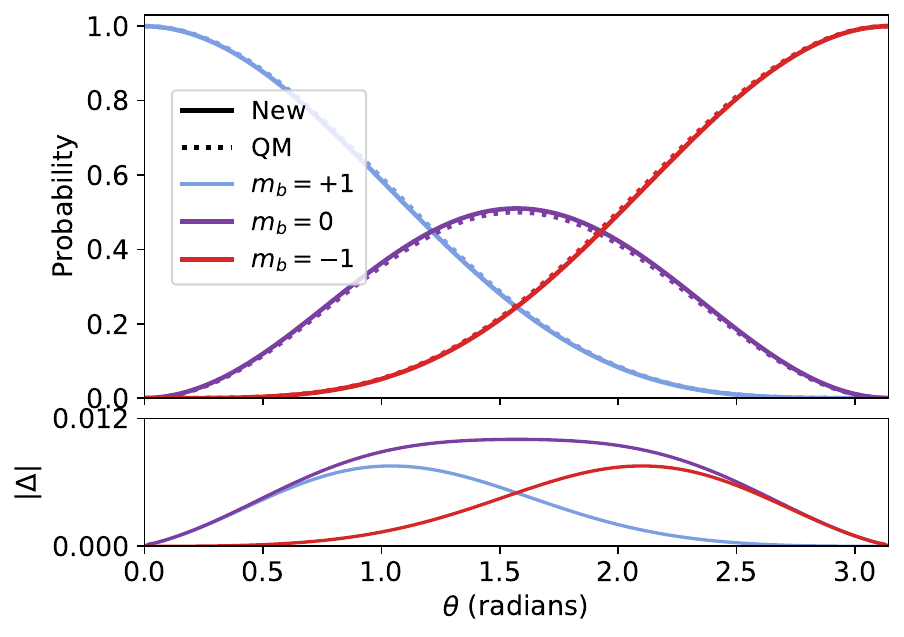}
\caption{As in Figure \ref{fig:f5}, but for $n=100$, $j=1$, $m_{a}=+1$.}
\label{fig:f6}
\end{figure}

\begin{figure}[t]
\centering
\includegraphics[width=0.6\textwidth]{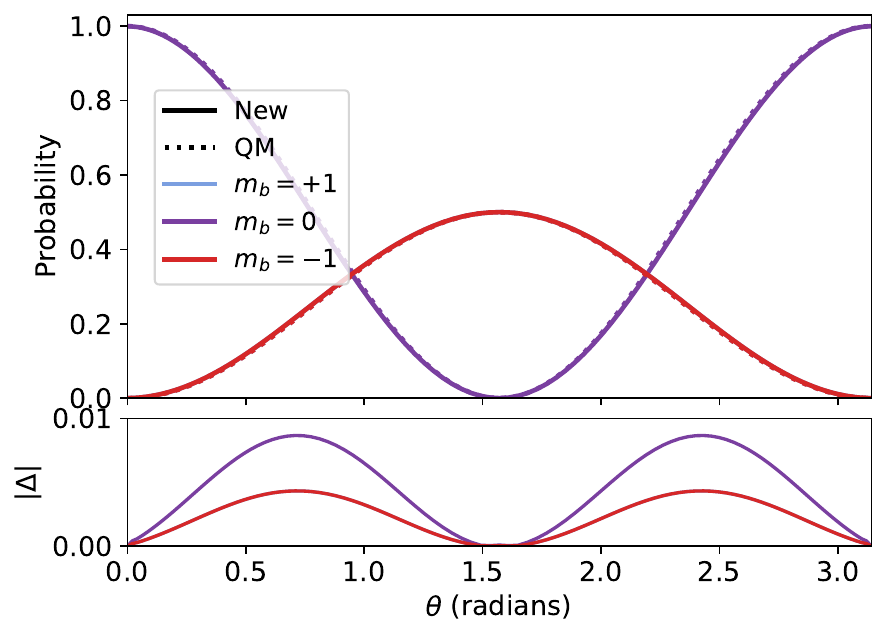}
\caption{As in Figure \ref{fig:f5}, but for $n=100$, $j=1$, $m_{a}=0$. By symmetry the curves for
$m_b=+1$ and $m_b=-1$ coincide, so only the latter is visible.}
\label{fig:f7}
\end{figure}

In Figures \ref{fig:f6} and \ref{fig:f7}, identical comparisons are offered for physical systems with total
spin 1 and spin projections $m_{a}=+1$ and $m_{a}=0$, respectively; no further adjustment of any kind
enters. Generally, for fixed $n$, the
deviation between the proposed model and QM increases as a function of total spin, but at fixed $j$ it
decreases with $n$.

\begin{figure}[t]
\centering
\includegraphics[width=0.6\textwidth]{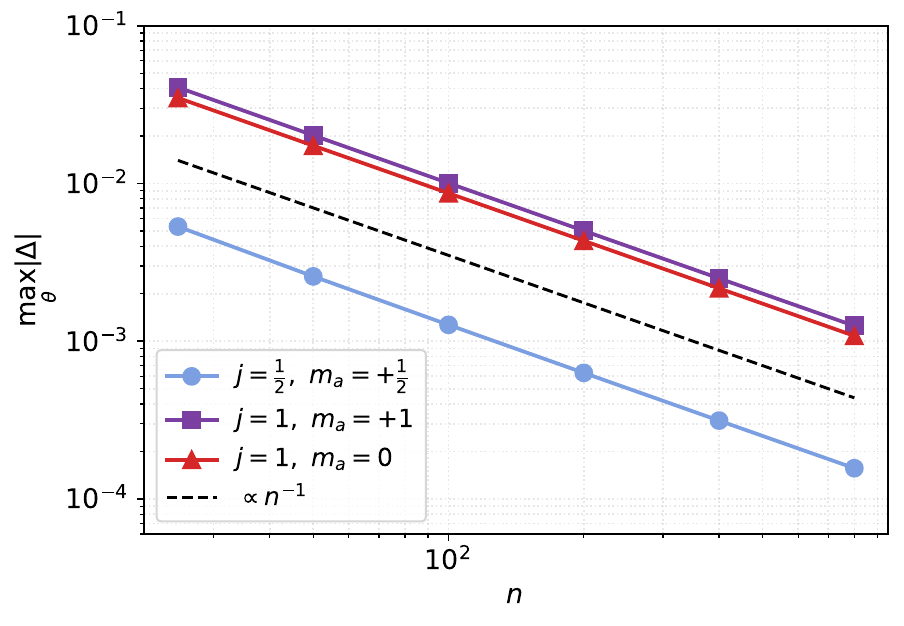}
\caption{Maximum deviation from QM as a function of $n$, for the three systems of Figures
\ref{fig:f5}--\ref{fig:f7}. The dashed line is proportional to $n^{-1}$.}
\label{fig:f8}
\end{figure}

The rate at which QM is recovered is made quantitative in Figure \ref{fig:f8}, where we plot the maximum
over $\theta$ of $|\Delta|$ as a function of $n$. For each of the three systems considered this quantity
halves with every doubling of $n$ over the entire range examined, as expected for an $O(1/n)$ correction;
for $j=\frac12$ this is the analytic result (\ref{eq:halfcorrection}), while for $j=1$ it is a numerical
observation. Representative
values are collected in Table \ref{tab:conv}.

\begin{table}[t]
\centering
\begin{tabular}{r|ccc}
$n$ & $j=\frac{1}{2},\ m_{a}{=}{+}\frac{1}{2}$ & $j=1,\ m_{a}{=}{+}1$ & $j=1,\ m_{a}{=}0$\\
\hline
25  & 0.00533 & 0.04065 & 0.03487\\
50  & 0.00258 & 0.02019 & 0.01736\\
100 & 0.00127 & 0.01005 & 0.00868\\
200 & 0.00063 & 0.00501 & 0.00433\\
400 & 0.00031 & 0.00250 & 0.00216\\
800 & 0.00016 & 0.00125 & 0.00108\\
\end{tabular}
\caption{$\max_\theta|\Delta|$ between the proposed model and QM under the calibration
(\ref{eq:calibbox}), the maximum being taken over every integer value of $\Bmap$ (every integer for
$n\le200$, a grid of 121 values for $n=400$ and $800$). Each column halves with every doubling of $n$;
the first column approaches $1/(8n)$, equation (\ref{eq:halfcorrection}), the $O(n^{-2})$ excess being
visible at the smallest $n$.}
\label{tab:conv}
\end{table}

A sample calculation for this model can be found in appendix \ref{app:c1}.

\subsection{Results: entangled spin systems}
\label{sec:entresults}

The physical system of interest in this section involves two spin systems with total spin $j_{\bar a}$ and
$j_{b}$, which are members of a composite system with total spin $j_{c}$. Each constituent spin system
interacts with either Alice's or Bob's Stern-Gerlach detector, which are spatially aligned. Because no
rotation is involved, the results of this section are independent of the calibration established in section
\ref{sec:calibration} and are unchanged from \cite{PowersSpin}. We include them for completeness. The
probability we are interested in calculating is for the spin projection quantum numbers observed by Alice
and Bob, which we denote as $m_{\bar a}$ and $m_{b}$. In QM, this probability is given by the square of the
Clebsch-Gordan coefficient:
\begin{equation}
\begin{split}
\langle j_{\bar a}j_{b}j_{c}m_{c}|j_{\bar a}j_{b}m_{\bar a}m_{b}\rangle=
\sqrt{\frac{(2j_{c}+1)(j_{\bar a}+j_{b}-j_{c})!(j_{c}+j_{\bar a}-j_{b})!(j_{c}+j_{b}-j_{\bar a})!}{(j_{\bar a}+j_{b}+j_{c}+1)!}}\\
\times\sum_z\frac{(-1)^z\sqrt{(j_{\bar a}+m_{\bar a})!(j_{\bar a}-m_{\bar a})!(j_{b}+m_{b})!(j_{b}-m_{b})!(j_{c}+m_{c})!(j_{c}-m_{c})!}}
{z!(j_{\bar a}+j_{b}-j_{c}-z)!(j_{\bar a}-m_{\bar a}-z)!(j_{b}+m_{b}-z)!(j_{c}-j_{b}+m_{\bar a}+z)!(j_{c}-j_{\bar a}-m_{b}+z)!}
\end{split}
\label{eq:cg}
\end{equation}

For the model proposed in section \ref{sec:entmodel}, the expression which generates the probability of
observing a particular combination of $m_{\bar a}$ and $m_{b}$ is given by the following, where $l_{c}$ and
$\nu^4$ are summed over as outlined in section \ref{sec:probs}, and $m_{c}$ acts as a constraint:
\begin{equation}
P(m_{\bar a},m_{b}|n,j_{\bar a},j_{b},j_{c})=\frac{\Upsilon(n,j_{\bar a},m_{\bar a},j_{b},m_{b},j_{c})}
{\sum_{m_{\bar a},m_{b}\in Q(n,j_{\bar a},j_{b},j_{c})}\Upsilon(n,j_{\bar a},m_{\bar a},j_{b},m_{b},j_{c})}
\label{eq:probent}
\end{equation}

\begin{figure}[t]
\centering
\includegraphics[width=0.55\textwidth]{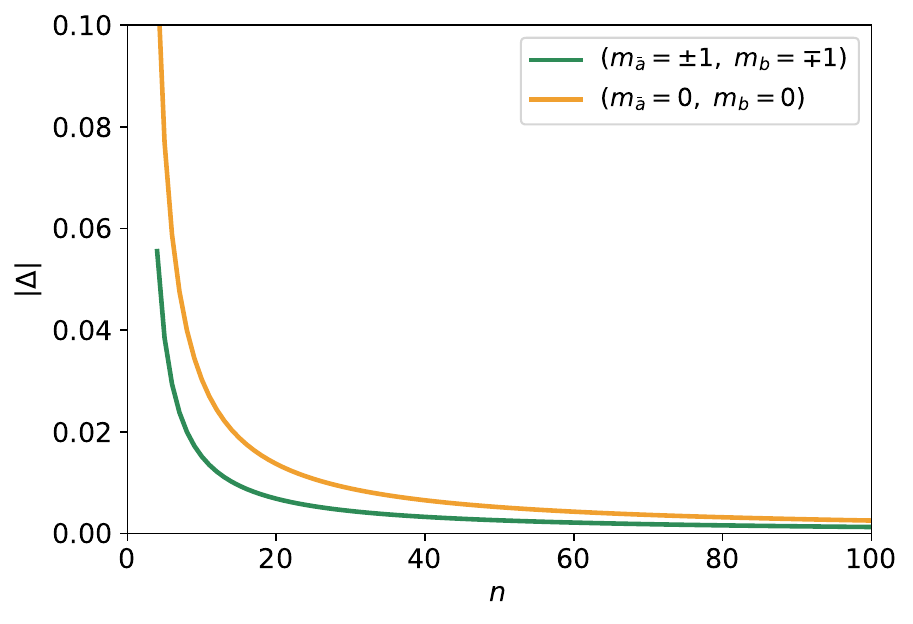}
\caption{A comparison of equations (\ref{eq:cg}) and (\ref{eq:probent}) over the range $4\le n\le100$,
where $j_{\bar a}=1$, $j_{b}=1$, $j_{c}=1$, $m_{c}=0$, and $|\Delta|$ is the magnitude of the difference
between the two models; the curves are $1/(4n-7)$ and $1/[2(4n-7)]$, equation (\ref{eq:entclosed}).}
\label{fig:f9}
\end{figure}

In the case that $j_{\bar a}=j_{b}=\frac{1}{2}$ and $n\ge2$, the proposed model and QM agree exactly for
$j_{c}=1$ and $j_{c}=0$. For $j_{\bar a}=j_{b}=1$ and $n\ge4$, perfect agreement between the two models is
again found for $j_{c}=2$, $j_{c}=1$, and $j_{c}=0$, in all but one case. The only case in which a
difference exists is for $j_{c}=1$ and $m_{c}=0$, where QM assigns zero probability to the outcome
$(m_{\bar a},m_b)=(0,0)$ while the model gives, exactly,
\begin{equation}
P(0,0\,|\,n,1,1,1)=\frac{1}{4n-7},
\label{eq:entclosed}
\end{equation}
which is derived in closed form in appendix \ref{app:entclosed}, has been confirmed by exact
evaluation for $4\le n\le 30$, and is the upper curve of Figure \ref{fig:f9}; the two
outcomes $(\pm1,\mp1)$ each absorb half of this deficit, giving the lower curve. The difference tends to
zero as $n$ becomes large. Thus, with $n=100$, the differences between the proposed model
and QM are much less than 1\% for the systems considered here. A sample calculation for this model can be
found in appendix \ref{app:c2}.

\subsection{Results: Bell test}
\label{sec:bellresults}

The physical system of interest in this section involves two spin systems with total spin
$j_{\bar a}=j_{b}=\frac{1}{2}$, which are members of a composite system with total spin $j_c=0$. Each
constituent spin system interacts with either Alice's or Bob's Stern-Gerlach detector, which may be rotated
with respect to one another. The model introduced in section \ref{sec:bellmodel} can be applied to this
physical system to calculate the expectation values of the product $4m_{\bar a}m_{b}$, as a function of the
angle between Alice's and Bob's detectors. These expectation values can then be used to check if the
proposed model indeed violates the CHSH inequality \cite{CHSH}, and if so, how it compares to the
prediction of QM. We begin this calculation by writing the general expression for the probability that
Alice and Bob observe a particular combination of $m_{\bar a}$ and $m_{b}$, where $l_{\bar a}$, $l_{b}$,
$\beta$, $\nu^0$, $\nu^1$, $\nu^4$, $\nu^5$, $\kappa^1$, and $\omega^2$ are summed over as outlined
in section \ref{sec:probs}, and $m^a_c$ acts as a constraint:
\begin{equation}
P(m_{\bar a},m_{b}|n,j_{\bar a},j_{b},\theta_{ab},j_c)=
\frac{\Upsilon(n,j_{\bar a},m_{\bar a},j_{b},m_{b},\theta_{ab},j_c)}
{\sum_{m_{\bar a},m_{b}\in Q(n,j_{\bar a},j_{b},\theta_{ab},j_c)}\Upsilon(n,j_{\bar a},m_{\bar a},j_{b},m_{b},\theta_{ab},j_c)}
\end{equation}

Given $j_{\bar a}=j_{b}=\frac{1}{2}$, $j_c=m^a_c=0$, and some choice of the angle $\theta_{ab}$, the
expectation value of the product $4m_{\bar a}m_{b}$ is given by the following:
\begin{equation}
\mathcal{E}(n,\theta_{ab})=\sum_{m_{\bar a},m_{b}}4m_{\bar a}m_{b}
P\left(m_{\bar a},m_{b}\middle|n,\tfrac{1}{2},\tfrac{1}{2},\theta_{ab},0\right)
\label{eq:Eexp}
\end{equation}

Using the structural observation at the end of section \ref{sec:bellmodel}, together with the limit
(\ref{eq:halflimit}) and the identity (\ref{eq:pythag}), this expectation value can be evaluated in closed
form in the large $n$ limit:
\begin{equation}
\lim_{n\to\infty}\mathcal{E}(n,\theta_{ab})
=-\frac{\Amap^2-\Bmap^2}{\Amap^2+\Bmap^2}=-\cos\theta_{ab}
\label{eq:Elimit}
\end{equation}
which is precisely the singlet correlation of QM. In fact the structural observation gives more than the
limit. With $j_c=0$ the outcomes $(m_{\bar a},m_b)$ of equal sign correspond to a $D$ symbol in Alice's
event facing a $C$ in Bob's, and hence to $B'=\alpha-1$ in equation (\ref{eq:halfclosed}), while outcomes
of opposite sign correspond to $B'=\alpha$. Writing $r$ for the ratio (\ref{eq:halfratio}) with
$\Bmap=\alpha$, the correlation at finite $n$ is exactly
\begin{equation}
\mathcal{E}(n,\theta_{ab})=\frac{r-1}{r+1},\qquad
r=\frac{\alpha^{3}\,(2n-2\alpha-1)}{(n-\alpha)^{3}\,(2\alpha-1)},
\label{eq:Eexact}
\end{equation}
and by equation (\ref{eq:halfcorrection})
\begin{equation}
\mathcal{E}(n,\theta_{ab})=-\cos\theta_{ab}+\frac{\sin 2\theta_{ab}}{4n}+O(n^{-2}).
\label{eq:Ecorrection}
\end{equation}
The approach to the QM correlation is therefore $O(1/n)$, and at $\theta_{ab}$ corresponding to
$\alpha/n=0.2$ the coefficient $\sin(2\theta_{ab})/4=0.2076$ agrees with the limiting value obtained by
direct numerical evaluation of $n\,|\mathcal{E}(n,\theta_{ab})+\cos\theta_{ab}|$ at large $n$. The sign of the correction is such that the model is always
\emph{less} anticorrelated than QM for $\theta_{ab}<\pi/2$ and less correlated for $\theta_{ab}>\pi/2$.

The expectation value (\ref{eq:Eexp}) can be used to calculate the CHSH test statistic like so, where each
angle is associated with a unique combination of detector orientations:
\begin{equation}
\mathcal{S}(n,\theta_{ab},\theta_{ab'},\theta_{a'b},\theta_{a'b'})=
|\mathcal{E}(n,\theta_{ab})-\mathcal{E}(n,\theta_{ab'})+\mathcal{E}(n,\theta_{a'b})+\mathcal{E}(n,\theta_{a'b'})|
\label{eq:chsh}
\end{equation}

In a Bell test, Alice and Bob each have two possible detector orientations, which are typically defined
with respect to some fixed coordinate system. However, it is only the relative angle between the two
detectors that is physically meaningful when $m^a_c=0$, as discussed at the end of section
\ref{sec:bellmodel}. This relative angle is precisely the quantity modeled by $\theta_{ab}$, as defined in
equation (\ref{eq:calibbell}). The detector orientations we shall use in this analysis, as defined with
respect to some fixed coordinate system, are as follows:
\begin{equation}
\theta_a=0^\circ,\ \theta_{a'}=90^\circ,\ \theta_b=45^\circ,\ \theta_{b'}=135^\circ
\end{equation}
With these detector orientations, the four possible angles between Alice's and Bob's detectors are:
\begin{equation}
\theta_{ab}=45^\circ,\ \theta_{ab'}=135^\circ,\ \theta_{a'b}=45^\circ,\ \theta_{a'b'}=45^\circ
\label{eq:angles}
\end{equation}

The angles required by equation (\ref{eq:angles}) are realized by the integer count values which best
satisfy the calibration (\ref{eq:calibbell}). Since
$\tan(22.5^\circ)=\sqrt{2}-1$ and $\tan(67.5^\circ)=\sqrt{2}+1$, these are
\begin{equation}
\alpha(45^\circ)=\left[n\left(1-\tfrac{1}{\sqrt{2}}\right)\right],\qquad
\alpha(135^\circ)=\left[\tfrac{n}{\sqrt{2}}\right]
\end{equation}
where $[\ ]$ denotes rounding to the nearest integer. For $n=100$ this gives $\alpha=29$ and
$\alpha=71$, and equation (\ref{eq:chsh}) yields
\begin{equation}
\mathcal{S}(100,45^\circ,135^\circ,45^\circ,45^\circ)=2.846
\label{eq:S100}
\end{equation}
Thus, the proposed model violates the CHSH inequality ($\mathcal{S}\le2$) by a wide margin, and does so at
a value close to the QM maximum of $2\sqrt{2}$ \cite{Tsirelson}.

\begin{figure}[t]
\centering
\includegraphics[width=0.62\textwidth]{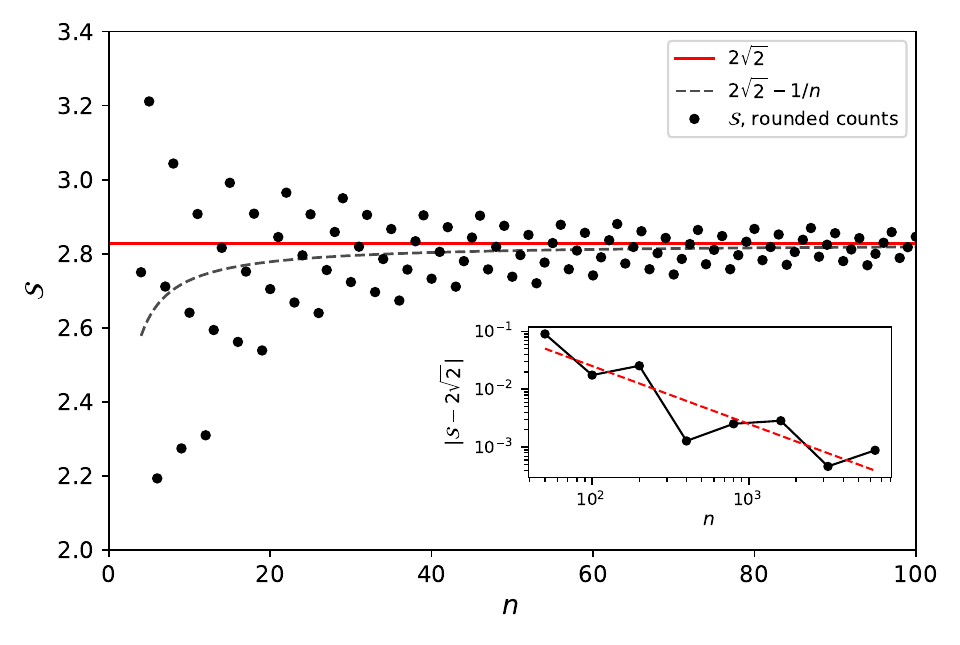}
\caption{A plot of equation (\ref{eq:chsh}) for every value of $n$ in the range $4\le n\le100$, using
the independently rounded counts $\alpha(45^\circ)$ and $\alpha(135^\circ)$ for the angles specified in
equation (\ref{eq:angles}). The red line indicates the Tsirelson bound of
$2\sqrt{2}$, and the gray dashed curve is $2\sqrt2-1/n$, the maximum of $\mathcal{S}$ over geometrically
consistent settings in the continuous extension of the count parameter, equation (\ref{eq:Smax}). The inset shows $|\mathcal{S}-2\sqrt{2}|$ over the range
$50\le n\le6400$, with a dashed reference line proportional to $n^{-1}$. Points above the red line reflect the rounding of the four
relative angles to values that are not jointly realizable, not an excess over the quantum correlations:
at every $n$ the model lies $1/n+O(n^{-2})$ below the quantum value at the same four angles, equation
(\ref{eq:SvsQM}).}
\label{fig:f10}
\end{figure}

The value (\ref{eq:S100}) exceeds $2\sqrt{2}=2.828$, and it is important to understand why this is not a
violation of the Tsirelson bound. The two rounded counts realize the angles
$\theta_{ab}=2\arctan(29/71)=44.44^\circ$ and $\theta_{ab'}=2\arctan(71/29)=135.56^\circ$, and three
copies of the former do not compose to the latter: $3\times44.44^\circ=133.3^\circ$. No four detector
orientations produce the quadruple of relative angles $(44.44^\circ,135.56^\circ,44.44^\circ,44.44^\circ)$,
so the Tsirelson bound, which is a statement about realizable settings, does not apply to it. Indeed the
quantum correlation $-\cos\theta$ evaluated at the same four angles gives $2.856$, and the model's value
lies below it by $1/n$ up to $O(n^{-2})$ corrections. This is general: by equation (\ref{eq:Ecorrection}), at any quadruple of
angles the model gives
\begin{equation}
\mathcal{S}_{\rm model}=\mathcal{S}_{\rm QM}(\text{same angles})-\frac{1}{n}+O(n^{-2}),
\label{eq:SvsQM}
\end{equation}
since the four corrections $\sin 2\theta/4n$ sum to $1/n$ at the CHSH angles. Maximizing the
right-hand side of (\ref{eq:Eexact}) over geometrically consistent orientations, that is over four
angles $\phi_a,\phi_{a'},\phi_b,\phi_{b'}$ with $\theta_{xy}=|\phi_x-\phi_y|$, and treating $\alpha$
in equation (\ref{eq:Eexact}) as a continuous variable (at finite $n$ four exactly consistent settings
generally cannot all be realized with integer counts, see below), we find
\begin{equation}
\max\mathcal{S}_{\rm cont}(n)=2\sqrt{2}-\frac{1}{n}+O(n^{-2}),
\label{eq:Smax}
\end{equation}
which we have confirmed numerically at $n=50$, $100$ and $400$ ($2.8078$, $2.8183$, $2.8259$); the
maximum is attained at the standard orientations $\phi_a=0$, $\phi_{a'}=90^\circ$, $\phi_b=45^\circ$,
$\phi_{b'}=135^\circ$. The subscript is a reminder that equation (\ref{eq:Smax}) is a statement about
the continuous extension of the count parameter, not the maximum over the integer-count model itself. In
that extension the model approaches the Tsirelson bound from below, and a finite-$n$ network is slightly
\emph{less} correlated than quantum mechanics permits, by $1/n$ in the CHSH statistic. At finite $n$,
however, the physically accessible orientations form a discrete set which is not closed under composition
(see below), so the exact four-setting CHSH experiment is represented only approximately, and the CHSH
statistic evaluated at realizable integer counts carries an additional discretization error of order
$1/n$ which can have either sign (Figure \ref{fig:f10}). Whether the $1/n$ deficit offers an experimental
handle on $n$ is a question we leave for future work.

Due to the granularity of the angles accessible at finite $n$, the precise behavior of the CHSH test
statistic as $n$ becomes large can be difficult to identify by only studying a small range of $n$'s. In
Figure \ref{fig:f10}, the CHSH test statistic evaluated at the independently rounded counts is plotted
for the range $4\le n\le100$ (for $n\le3$ the rounded counts are degenerate). The scatter visible at small $n$, in both directions about $2\sqrt2$, is a
direct consequence of this granularity: the integer count values which best realize the angles
(\ref{eq:angles}) approximate them only to $O(1/n)$, and the resulting quadruple is not exactly
realizable. As $n$ grows, $\mathcal{S}$ converges onto $2\sqrt{2}$, as shown in the inset, with
$|\mathcal{S}-2\sqrt{2}|$ falling as $O(1/n)$.

A related structural point should be recorded. The angles available at a given $n$ are the discrete set
$\{2\arctan[\alpha/(n-\alpha)]\}$, and this set is not closed under addition: the composition of two
realizable rotations is in general not realizable at the same $n$. A four-setting Bell experiment, which
requires four orientations on a common circle, is therefore represented exactly only in the limit
$n\to\infty$. This is the same limitation noted for the composition rule (\ref{eq:complexmap}) in
section \ref{sec:calibration}.

A sample calculation for this model can be found in appendix \ref{app:c3}.

\section{Contextuality and the status of the quantum state}
\label{sec:contextuality}

Section \ref{sec:bellresults} reported a violation of the CHSH inequality, which is already a proof that
the framework cannot be reproduced by any non-contextual hidden variable model. That statement is
negative, and it is worth asking what the positive content is: where exactly does contextuality enter the
formalism, what plays the role of a context, and how much contextuality is present. This section addresses
those questions. The numerical statements below were obtained from the same exact integer evaluation used
throughout section \ref{sec:results}.

\subsection{Where contextuality enters}
\label{sec:whereitenters}

The ontic states of this formalism are base-16 sequences, and a base-16 sequence is an ordered pair of
events \emph{together with} the map that relates them. The map is not an inert label. In the model of
section \ref{sec:rotmodel} its counts fix the relative angle between Alice's and Bob's detectors, and in
section \ref{sec:entmodel} they fix the total spin of the composite system. The ontic state is therefore
never a property of a system considered on its own; it is a property of a system together with the
arrangement in which it is interrogated.

This is made precise by the categorization introduced in section \ref{sec:spaces}. A context, in the
present framework, is the assignment of the sixteen quantum numbers to the four categories $R$, $U$, $W$
and $\Lambda$, together with a choice of values for the conditioning variables $u$. The ontic state space
$E^{a}(u)\bar{\otimes}E^{b}(u)$ is constructed relative to that assignment, and as noted in section
\ref{sec:probs} the state spaces belonging to distinct contexts are disjoint: no base-16 sequence ever
appears in two of them.

The consequence for the Kochen-Specker theorem \cite{KS} is direct, and it is worth stating carefully
because it is easy to misread this framework as a naive realist model of the kind the theorem excludes.
The theorem concerns a valuation defined on the observables of a single system, for instance the squared
spin components $S_{\hat n}^2$ of a spin-1 particle along a finite set of directions $\hat n$, and shows
that no such valuation can be non-contextual. In the present framework there is no such valuation to
begin with, because there is no such observable: a value exists only as a count at an event, and every
event involves one apparatus orientation. The quantity ``$S_{\hat n}^2$ of the particle,'' considered
apart from an event with a detector along $\hat n$, is not an element of the formalism. The framework
therefore does not instantiate the single context-independent ontic space assumed by the standard
Kochen-Specker valuation. We emphasize that this context dependence is built into the ontology by
construction rather than derived as a dynamical consequence: the framework does not satisfy the premise
of the theorem, which is different from explaining contextuality from deeper principles. Put differently, every ontic state assigns a definite value to all sixteen quantum
numbers of the two-event context in which it lives, but value definiteness in the Kochen-Specker sense,
which would require definite values for the observables of all contexts at once, is exactly what is given
up. What is retained is the definiteness of the quantities realized in the events that occur. This is the
sense in which the framework is relational, in the spirit of \cite{Rovelli}, and the sense in which
measurements in different contexts are different random variables, as in the contextuality-by-default
framework of Dzhafarov and Kujala \cite{Dzhafarov}. The Kochen-Specker argument requires one set of
hidden states on which the value assignments belonging to different measurement contexts can be compared;
here there is no such set, because changing the context changes which base-16 sequences exist at all.
Contextuality is thus not an emergent constraint that the models are found to satisfy after the fact. It
is built into the definition of the ontic state space.

The same structural feature accounts for the framework's standing with respect to Bell's theorem
\cite{Bell,CHSH}, and it is worth naming the assumption involved. Bell's derivation requires a variable
$\lambda$ shared by the two wings whose distribution is independent of the settings $a$ and $b$, an
assumption variously called measurement independence, free choice or no-conspiracy, together with locality
of the outcome functions. In the present framework the probabilities take the form
\begin{equation}
P(m_{\bar a},m_b\,|\,\theta_a,\theta_b)=\sum_{\lambda}P(m_{\bar a},m_b\,|\,\lambda,\theta_{ab})\,
p(\lambda\,|\,\theta_{ab}),
\label{eq:measdep}
\end{equation}
where $\lambda$ runs over the ontic states (base-16 sequences) of the context, and the relative
orientation is carried by the map counts, which are part of $\lambda$ itself. The distribution
$p(\lambda|\theta_{ab})$ is supported on the state space $E^a(u)\bar{\otimes}E^b(u)$, which is
constructed conditional on $\theta_{ab}$, and distinct settings have disjoint supports. The model
therefore does not satisfy the measurement-independence assumption $p(\lambda|a,b)=p(\lambda)$, and the
Bell inequality does not follow from the assumptions of the standard local hidden-variable derivation. We
do not posit a dynamical retrocausal mechanism for this: the ontic state is defined for the complete
two-event network, including the relation that specifies the measurement context. This places the
framework alongside the ``all-at-once'', Lagrangian-schema and retrocausal approaches reviewed
in \cite{WhartonArgaman,Adlam}, in which the ontic state is a property of the whole experiment including its
settings, rather than alongside models that violate locality. Two consequences should be acknowledged. First,
the correlations are non-signalling (section \ref{sec:nosignalling}) even though the settings enter the
ontic state, and it is a known feature of such models that this combination requires a form of fine-tuning
in the sense of \cite{WoodSpekkens}; the present formalism achieves it through the exact symmetry of the
singlet state space under $m_{\bar a}\leftrightarrow -m_{\bar a}$, and we do not claim to have explained
why nature should be so tuned. Second, the singlet Bell model does not contain separate setting variables
for Alice and Bob at all, only the relative map $\theta_{ab}$, and it has exactly the count structure of the
single-particle rotation model of section \ref{sec:rotmodel}. The CHSH violation is therefore no more
mysterious, and no more of an independent prediction, than the reproduction of $\sin^2(\theta/2)$ in that
model; what Bell's theorem asks, namely how such correlations are produced between separated wings with
independently chosen settings, is answered here only in the sense just described.

The disjointness of state spaces across contexts is what distinguishes this proposal from the Spekkens
toy model \cite{Spekkens}. It bears also on the ontological status of the quantum state, but for a
different reason and with a different meaning, which we take up in section \ref{sec:psiontic}.

\subsection{Contexts are subgroups of the alphabet}
\label{sec:contextsubgroups}

The three models of section \ref{sec:models} are distinguished by which of the map quantum numbers of
equations (\ref{eq:alpha}-\ref{eq:gamma}) is set to zero, and this admits a clean algebraic reading.
Setting $\gamma=0$ restricts the map to the symbols $\{A,B\}$; setting $\alpha=0$ restricts it to
$\{A,D\}$; setting $\beta=0$ would restrict it to $\{A,C\}$. By the addition table (\ref{eq:z2z2}) each of
these pairs is closed under $\oplus$, so
\begin{equation}
\{A,B\},\qquad \{A,C\},\qquad \{A,D\}
\label{eq:threesubgroups}
\end{equation}
are precisely the three subgroups of order two of the Klein four-group, and they are exhaustive. A context
in which the map is confined to a single subgroup is one in which a single kind of relationship between
Alice's and Bob's event is available: rotation in the case of $\{A,B\}$, composite spin in the case of
$\{A,D\}$. These three subgroups are permuted transitively by $\mathrm{Aut}(K)=GL(2,2)\cong S_3$, the
group which appendix \ref{app:symmetry} identifies as the automorphism group of the alphabet and
whose cyclic subgroup appears in the operator algebra $A_4$. No context is algebraically preferred; what
a particular model does is to select one of them.

These three sectors intersect only trivially. Any two of the subgroups in (\ref{eq:threesubgroups}) meet
in $\{A\}$ alone, so a base-16 sequence lying in two sectors at once must have a map consisting entirely
of $A$'s. By the addition table that is the identity map, for which Alice's and Bob's events are the same
sequence, the relative angle vanishes and the composite spin vanishes. The overlap of any two contexts is
therefore the single degenerate arrangement in which there is nothing to measure. This is the disjointness
of section \ref{sec:probs} in explicit algebraic form.

The Bell test of section \ref{sec:bellmodel} is exactly the situation in which two of these mutually
exclusive contexts are demanded at once. A Bell test requires a composite spin system, which needs the
symbol $D$, and rotated detectors, which need the symbol $B$; and $\{A,B,D\}$ is not closed under $\oplus$,
since $B\oplus D=C$. The full four-symbol alphabet is therefore forced. This is visible in the structure of
the models themselves: in sections \ref{sec:rotmodel} and \ref{sec:entmodel} the set $\Lambda$ contains a
single non-local quantum number, whereas in section \ref{sec:bellmodel} all six are required. We may
summarize:

\begin{quote}
\emph{Contextuality in this framework is the statement that the map alphabet cannot be reduced to a single
subgroup of $\mathbb{Z}_2\times\mathbb{Z}_2$ when incompatible measurements are combined. Interference, and
with it the violation of the CHSH inequality, appears precisely when the alphabet is forced to be
complete.}
\end{quote}

\subsection{The empirical model is non-signalling}
\label{sec:nosignalling}

Contextuality is only meaningful for an empirical model whose marginals are well defined, so it must be
checked that the correlations of section \ref{sec:bellresults} cannot be used to signal. Within the model
of section \ref{sec:bellmodel} with $j_c=0$, the quantity to examine is the marginal probability of
Alice's outcome as a function of Bob's setting,
\begin{equation}
P\!\left(m_{\bar a}\,\middle|\,n,\theta_{ab}\right)=\sum_{m_b}
P\!\left(m_{\bar a},m_b\,\middle|\,n,\tfrac{1}{2},\tfrac{1}{2},\theta_{ab},0\right)
\end{equation}
This marginal can be evaluated exactly for every $n$ and $\alpha$. By the structural observation at the
end of section \ref{sec:bellmodel}, the four outcomes fall into two classes. Outcomes of equal sign,
$(m_{\bar a},m_b)=(\pm\frac12,\pm\frac12)$, pair Alice's $D$ with Bob's $C$ or Alice's $C$ with Bob's
$D$; both symbols $DC$ and $CD$ contribute a $B$ to the map, so $B'=\alpha-1$ in equation
(\ref{eq:halfclosed}). Outcomes of opposite sign pair $C$ with $C$ or $D$ with $D$, and $B'=\alpha$.
Since $\Upsilon$ depends on the outcome only through $B'$,
\begin{equation}
\Upsilon(+,+)=\Upsilon(-,-)=\Upsilon(\alpha-1),\qquad \Upsilon(+,-)=\Upsilon(-,+)=\Upsilon(\alpha),
\end{equation}
and therefore
\begin{equation}
P\!\left(m_{\bar a}=\pm\tfrac12\,\middle|\,n,\theta_{ab}\right)
=\frac{\Upsilon(\alpha-1)+\Upsilon(\alpha)}{2\left[\Upsilon(\alpha-1)+\Upsilon(\alpha)\right]}=\frac12
\end{equation}
for every $n$ and every $\alpha$, independently of Bob's setting. The same argument applies with Alice and
Bob interchanged. We have also confirmed this by exact evaluation for $n=12$, $25$ and $50$ over a range
of $\alpha$. The empirical model is therefore non-signalling, and the CHSH violation reported in
section \ref{sec:bellresults} is a genuine statement about contextuality rather than an artefact of a
model in which the marginals depend on the distant setting.

\subsection{How much contextuality: the contextual fraction}
\label{sec:contextualfraction}

Contextuality admits a quantitative measure. Given an empirical model $e$, one may ask for the largest
weight $\lambda$ with which it can be written as a mixture
\begin{equation}
e=\lambda\,e^{NC}+(1-\lambda)\,e'
\end{equation}
of a non-contextual model $e^{NC}$ and an arbitrary non-signalling remainder $e'$. The contextual
fraction is $CF=1-\lambda_{\max}$ \cite{AbramskyBrandenburger,ABM}. This quantity is not merely bounded
by the violation of a Bell inequality. It is \emph{equal} to the maximum, taken over all Bell
inequalities for the scenario in question, of the normalized violation \cite{ABM}, where an inequality
with local bound $b$ and algebraic non-signalling maximum $N$ is said to be violated to normalized degree
$(V-b)/(N-b)$ by a model attaining the value $V$.

For the two-party, two-setting, two-outcome scenario relevant here the local polytope is completely
characterized: apart from the positivity facets, which no probability model can violate, its facets are
exactly the eight CHSH inequalities obtained from equation (\ref{eq:chsh}) by changes of sign
\cite{Fine}. Each of them has $b=2$ and $N=4$, so that
\begin{equation}
CF=\max_{i}\ \frac{\mathcal{S}_i-2}{2}
\label{eq:cf}
\end{equation}
the maximum running over those eight inequalities. Evaluating all eight for the proposed model at the
detector orientations of equation (\ref{eq:angles}), we find that the maximum is attained by precisely
the combination written in equation (\ref{eq:chsh}), for every value of $n$ examined. Equation
(\ref{eq:cf}) therefore reduces to $CF=(\mathcal{S}-2)/2$, with $\mathcal{S}$ the quantity already
computed in section \ref{sec:bellresults}, and the contextual fraction of the model is determined
exactly rather than bounded. Quantum mechanics saturates the Tsirelson bound at
$\mathcal{S}=2\sqrt{2}$ and hence has $CF=\sqrt{2}-1\approx0.4142$. The resulting values are collected
in Table \ref{tab:cf}, together with the quantum values at the same rounded angles.

\begin{table}[htbp]
\centering
\begin{tabular}{r|cccccc}
$n$ & 100 & 400 & 800 & 1600 & 3200 & 6400\\
\hline
$\theta_{ab}$ realized & $44.44^\circ$ & $44.92^\circ$ & $44.92^\circ$ & $45.05^\circ$ & $44.98^\circ$ & $45.01^\circ$\\
$\mathcal{S}$ (model) & 2.8460 & 2.8297 & 2.8310 & 2.8256 & 2.8289 & 2.8275\\
$\mathcal{S}_{\rm QM}$ (same angles) & 2.8562 & 2.8322 & 2.8322 & 2.8262 & 2.8292 & 2.8277\\
$n\,(\mathcal{S}_{\rm QM}-\mathcal{S})$ & 1.016 & 1.004 & 1.002 & 1.001 & 1.0005 & 1.0002\\
$CF$ (model) & 0.4230 & 0.4149 & 0.4155 & 0.4128 & 0.4144 & 0.4138\\
$CF$ (consistent settings, eq. \ref{eq:Smax}) & 0.4091 & 0.4130 & 0.4136 & 0.4139 & 0.4141 & 0.4141\\
\end{tabular}
\caption{The contextual fraction (\ref{eq:cf}) of the proposed model, evaluated at the independently
rounded counts realising the detector orientations of equation (\ref{eq:angles}), compared with the
quantum value $\mathcal{S}_{\rm QM}=|{-3}\cos\theta_{ab}+\cos\theta_{ab'}|$ at the same realized angles.
The model lies below the quantum value by $1/n+O(n^{-2})$ at every $n$, equation (\ref{eq:SvsQM}). The last row is
the contextual fraction obtained from the maximum of $\mathcal{S}$ over geometrically consistent settings
in the continuous extension of the count parameter, equation (\ref{eq:Smax}). The quantum value is $\sqrt{2}-1=0.4142$.}
\label{tab:cf}
\end{table}

Two notions of contextual fraction should be kept apart. Over geometrically consistent settings, in the
continuous extension of the count parameter (last row of Table \ref{tab:cf}),
\begin{equation}
CF_{\rm cont}(n)=\sqrt2-1-\frac{1}{2n}+O(n^{-2}),
\end{equation}
which approaches the quantum value from below. The values evaluated at independently rounded integer
counts (the row labeled ``model'') are instead a diagnostic of discretization. The scatter in that row is
the angular granularity discussed in section \ref{sec:bellresults}: at finite $n$ the independently rounded counts
realize a quadruple of angles which is not jointly realizable by four detector orientations, and the
quantum value at those same angles fluctuates about $2\sqrt2$ in the same way. The entries above
$\sqrt2-1$ at $n=100$ and $n=800$ are of this origin and do not indicate contextuality in excess of
quantum mechanics: at every $n$ the model is less contextual than quantum mechanics at the same angles by
$1/(2n)+O(n^{-2})$. The framework is thus not merely
contextual; in the limit of large $n$ it is contextual to exactly the degree that quantum mechanics is,
and its finite $n$ departures from that degree are of the same order as every other finite $n$ correction
reported in this work.

\subsection{The ontological status of the quantum state}
\label{sec:psiontic}

The introduction raised the question of whether the mathematical objects of quantum theory represent
physical things, or an observer's incomplete information about them. It is natural to ask where the
present framework stands with respect to the theorem of Pusey, Barrett and Rudolph \cite{PBR}, and the
answer requires some care, because the question turns out not to be well posed in the form in which it is
usually asked.

The theorem is formulated within the ontological-models framework of Harrigan and Spekkens
\cite{HarriganSpekkens}. That framework posits a single space $\Lambda$ of ontic states; a preparation of
the quantum state $\psi$ induces a distribution $\mu_\psi(\lambda)$ on $\Lambda$; a measurement $M$
induces response functions $\xi_M(k|\lambda)$; and, crucially, $\lambda$ is the state of the system after
preparation and before measurement, so that $\mu_\psi$ does not depend on which $M$ will later be
performed. Within this framework a model is called $\psi$-ontic if distinct pure states have disjoint
supports and $\psi$-epistemic otherwise. The $\psi$-ontic models further divide into $\psi$-complete
models, in which $\lambda$ is $\psi$ and nothing more, and $\psi$-supplemented (or $\psi$-incomplete)
models, in which $\lambda$ carries additional hidden variables, Bohmian mechanics being the standard
example \cite{Leifer}. Granted an assumption of preparation independence, the PBR theorem shows that
$\psi$-epistemic models contradict quantum mechanics.

The present framework does not fit this template, and the reason is structural rather than incidental. An
ontic state is a base-16 sequence, and a base-16 sequence is an ordered pair of events together with the
map relating them. The map carries the measurement setting: in section \ref{sec:rotmodel} its counts are
the relative angle between the preparing and measuring apparatus, and in section \ref{sec:entmodel} they
are the total spin of the composite system being interrogated. The sequence also determines the outcome
$m_b$. There is therefore no object in the formalism corresponding to $\lambda$ in the ontological-models
sense, namely a state of the system alone, defined before the measurement and independently of its
setting. The state space $E^a(u)\bar{\otimes}E^b(u)$ exists only relative to a context $u$ which includes
that setting.

It follows that the $\psi$-ontic/$\psi$-epistemic dichotomy cannot be applied directly. To see that
appeals to disjointness do not settle the matter, consider a spin-$\frac12$ system prepared along Alice's
axis and the same system prepared along an axis at an angle $\theta$ to it. In quantum mechanics these
preparations are non-orthogonal, with overlap $\cos^2(\theta/2)$. In the present framework the two
situations are distinguished by the map count $\Bmap$, and since $\Bmap$ is a function of the sequence,
no sequence belongs to both. But $\Bmap$ is the relation between Alice's event and Bob's, and within a
two-event model ``prepared at angle $\theta$ to Alice's axis and measured along Bob's'' and ``prepared
along Alice's axis and measured at angle $\theta$'' are the same statement. The disjointness exhibited
is disjointness across measurement contexts, precisely the kind which section \ref{sec:probs} noted
carries no implication for the PBR theorem. More generally, since every quantum number, including the
outcome, is a function of the sequence, any two distinct macrostates have disjoint microstate supports;
this is true of any model whose ontic states are complete records of an experiment, and it is not the
content of the PBR notion.

What can be said is the following. First, the framework as formulated here lies outside the scope of the PBR theorem, and
of $\psi$-ontology theorems generally, for the same reason that retrocausal and all-at-once models do:
those theorems assume that the ontic state is independent of the future measurement setting, and here it
is not \cite{Leifer,LeiferPusey,WhartonArgaman}. The theorem neither constrains the framework nor is
evaded by it in any non-trivial sense. Second, within any fixed context, distinct macrostates are carried
by disjoint sets of microstates, and the probabilities of section \ref{sec:probs} arise from counting the
microstates compatible with a macrostate. In this restricted sense the framework is a hidden-variable
theory of the $\psi$-supplemented type relative to a context: the macrostate is a function of the
microstate, and the hidden information is the ordering of symbols within sequences together with the
values of the non-local quantum numbers. The same reading is reached independently in the base-2 setting
of \cite{QHO}, where the hidden variable over which interference sums, the Hamming distance between the
two sequences, is a quantum number of the transition as a whole, defined jointly by both endpoints and
unknowable from either alone; interference there is the mandatory bookkeeping over this relational
variable, and the present paper's non-local quantum numbers are its base-16 generalisation. Third, a context-independent notion of the quantum state of a
system considered by itself does not exist in this framework. That is the same feature which makes the
framework contextual in the sense of section \ref{sec:whereitenters}, and it is what makes the question
``is $\psi$ ontic?'' unanswerable as posed rather than answered in the affirmative.

We record finally that preparation independence, the auxiliary assumption of \cite{PBR}, which requires
the joint ontic state of two independently prepared systems to factorise, is not obviously formulable in a
framework whose ontic states belong to networks of events rather than to systems. The models developed
here, which treat a single primitive network element, are not equipped to settle the question, and we do
not rely on it.

\section{Discussion}
\label{sec:discussion}

We have presented here a fresh approach to modeling physics, which assumes the existence of a causal
network consisting only of events and their relationships. The focus of this work has been only the
simplest possible elements of this network, which each consist of two causally connected events. To model
these exceedingly simple physical elements, we chose equally simple mathematical elements. Using only
ordered lists of 0's and 1's, a robust formalism was developed and shown to have many non-trivial features
of interest with respect to modeling physics. These include effective (epistemic) indeterminism, a rich and diverse set of
symmetries, interference, conservation laws, and much more. Perhaps equally important, this formalism is
accompanied by a vivid conceptual picture, consisting of causal networks of events, observers, their
statistical ensembles, and the evolution of those ensembles.

With the calibration $\tan(\theta/2)=\Bmap/\Amap$ of section \ref{sec:calibration}, the model reproduces
the quantum probabilities of every sector studied here in the large-$n$ limit, with corrections which
vanish as $O(1/n)$ and with no continuous free parameters anywhere in the construction. The half angle in
equation (\ref{eq:calibbox}) is the same half angle that distinguishes $SU(2)$ from $SO(3)$. We have not,
of course, derived quantum mechanics in general; what has been shown is agreement in the spin sectors
examined.

A second theme is structural. Appendix \ref{app:symmetry} shows that several features which were
previously presented as choices are not choices at all: ten local and six non-local quantum numbers is
forced by the rank of the three-margin map, the two classes of symmetry operation identified in section
\ref{sec:indist} are exactly the translations and the three-cycles of the alternating group $A_4$, and
the twenty non-local quantum numbers are the line sums of an affine plane of order four. What remains a
genuine choice is the interference rule, and that too is now quantified: it is a $\mathbb{Z}_2$ character
on a rank six lattice, so there are sixty four candidates, of which the calculations reported here
constrain two bits. The construction therefore has no continuous free parameters, but four discrete bits
remain open. We note also that the interference rule,
together with the product structure $\ncard{L^a}\ncard{L^b}$ of equation (\ref{eq:prodcard}), plays within
this framework the role that the phase structure of amplitudes and the Born rule play in quantum
mechanics: the signed sums over the non-local quantum number in equations (\ref{eq:Lainter}) and
(\ref{eq:Lbinter}) are, at large $n$, the signed sums over $q$ in the Wigner formula (\ref{eq:wigner}),
with the multinomial coefficients supplying the powers of $\cos(\theta/2)$ and $\sin(\theta/2)$. The
framework dispenses with Hilbert space as a primitive; it does not dispense with the combinatorial
content of the Born rule, which reappears as a counting rule with interference. What can be said in
favor of that counting rule is that, in the base-2 setting of \cite{QHO}, the alternating sign is not
assumed: given the bilinear forward--backward form, it is the unique weighting for which the measure
accounts for exactly the $2^n$ configurations of the register while respecting the relabeling symmetry
of the symbols, and every alternative overshoots the capacity. Interference in that model is the price of
not counting observationally indistinguishable configurations twice. Whether the same is true of the
rank-six character used here is the open question discussed in appendix \ref{app:symmetry}; the
base-2 theorem makes it a well-posed one. Fixing the rest requires a model with a constituent spin above $\frac{1}{2}$, a
non-zero composite spin and rotated detectors, which we regard as the most important open computation
raised by this work.

A third theme is contextuality, taken up in section
\ref{sec:contextuality}. It is tempting to read a framework in which every ontic state carries definite
values for all sixteen quantum numbers as a naive realist model of the kind excluded by the
Kochen-Specker theorem, and it is worth being explicit about why it is not. Values in this framework are
attributes of events, not of systems; the observables of a single system to which the Kochen-Specker
valuation would be assigned are not elements of the formalism. What the framework retains is the
definiteness of the quantities realized in the events that occur; what it gives up is the single
context-independent state space on which the Kochen-Specker argument is mounted: the map relating two
events is part of the ontic state, so changing the measurement arrangement changes which ontic states
exist. The same feature is what removes the measurement-independence assumption from Bell's derivation,
and we have been explicit in section \ref{sec:whereitenters} about the company this places the framework
in. That the three available contexts turn out to be the three
order-two subgroups of the alphabet, and that a Bell test is precisely the demand for two of them at
once and therefore forces the complete alphabet, strikes us as one of the more satisfying features to
have emerged from this formalism. It says that interference and the CHSH violation are not additional
ingredients but are what the algebra does when incompatible questions are asked together. The
quantitative counterpart is that, over consistent settings in the continuous extension of the count
parameter, the contextual fraction converges to $\sqrt{2}-1$ from below, as $\sqrt2-1-1/(2n)$, so the framework is contextual to exactly the degree quantum mechanics is in the
limit, and slightly less so at any finite $n$.

In section \ref{sec:intro}, we began by advocating for the event-centric picture of nature. Though this approach has
many merits, it was not assumed at the outset of this research program. The only assumption made was that
0's and 1's must be the simplest possible modeling tool a physicist could employ. By carefully studying
these simple building blocks, we gradually identified features which appeared to be related to physics.
Only after a long series of incremental advancements towards a robust physics model
\cite{PowersSG,PowersSpin}, did the presented framework become clear. Thus, the event-centric picture was
not chosen or assumed, it simply emerged. Though one should certainly not mistake this as evidence of
scientific validity, the mathematical and conceptual simplicity of this picture is noteworthy.

While the results presented here certainly justify further inquiry, there remains a long road ahead.
Ultimately, the event-centric approach must reproduce the modeling capabilities of quantum field theory and
general relativity, simultaneously. As a first step, it was shown in \cite{SR} how space, time and the kinematic structure of special relativity arise from binary sequences within this formalism.
While quantum theory of spacetime remains the long term goal, we also believe that the work
presented here may already be of interest to those in the fields of quantum foundations, information
theory, and even quantum computing. After all, the systems we have modeled here are the subject of intense
study within these fields. Not only that, but the formalism which supports these models is built from
classical bits. This may help clarify the relationship between classical and quantum information theory.

We now look ahead to the next few steps in the development of this framework. First and foremost, we must
improve our understanding of how it fits within the broader literature on event-centric approaches.
Undoubtedly, the most efficient path to a robust event-centric physics model will involve the confluence of
many independent ideas. With that being said, there is no shortage of open questions and opportunities for
further development within the framework presented here. The calibration of section \ref{sec:calibration}
raises one such question directly: the composition rule (\ref{eq:complexmap}) and the half-angle factor
were imposed rather than derived, and a derivation of them from the $\oplus$ operation on sequences, presumably as a statement about
the typical overlap of two randomly chosen maps at large $n$, would substantially strengthen the argument.
Beyond that, a more complete treatment of statistical ensembles and their evolution may yield a connection
to the wave equations employed in existing models (see \cite{QHO} for how the Schr\"odinger equation of the quantum harmonic oscillator arises in this framework). Also of interest for future work is a more in-depth
study of the symmetries of local state spaces, as this may clarify how the gauge symmetries in quantum
field theory arise within this framework. Lastly, models for network elements containing more than two
events are of significant interest. Of course, this list is far from complete. Though the proposed formalism
is exceedingly simple, one can see that it still affords a great deal of complexity.

The mathematical simplicity, conceptual clarity, and agreement with the predictions of quantum mechanics
in the large-$n$ limit for the systems studied make the framework presented here of interest for further inquiry. Though
many non-trivial advancements separate the present state of this research program from its ultimate goal,
the achievements of other event-centric approaches indicate that such an outcome may well be possible in
the long run. In the near term, we are confident that further development will continue to shed light on
the foundations of quantum theory, with potential implications for both pure and applied research
programs. 

\section*{Code availability}

The exact-arithmetic code used to produce the numerical results, tables and figures of section
\ref{sec:results} and section \ref{sec:contextuality} is provided as supplementary material.

\section*{Acknowledgments}

We would like to thank Djordje Minic, Tatsu Takeuchi, Lauren Hay, and Omar Elsherif for comments on
previous versions of this work as well as many helpful discussions. We also thank Emily Powers for figure
designs. D.S. is partially supported by the US National Science Foundation, under Grants no. PHY-2014021 and PHY-2310363.
\appendix

\section{Non-commutative operands}
\label{app:noncomm}

The non-local quantum numbers associated with the non-commutative operators given in equation
(\ref{eq:noncommuting}) are as follows:
\begin{equation}
\begin{array}{ll}
\rho^0\equiv\frac{1}{3}(\ct{BD}+\ct{CB}+\ct{DC}) & \eta^0\equiv\frac{1}{3}(\ct{BC}+\ct{CD}+\ct{DB})\\
\rho^1\equiv\frac{1}{3}(\ct{AA}+\ct{BD}+\ct{CB}) & \eta^1\equiv\frac{1}{3}(\ct{AD}+\ct{BB}+\ct{CA})\\
\rho^2\equiv\frac{1}{3}(\ct{AA}+\ct{BD}+\ct{DC}) & \eta^2\equiv\frac{1}{3}(\ct{AC}+\ct{BA}+\ct{DD})\\
\rho^3\equiv\frac{1}{3}(\ct{AA}+\ct{CB}+\ct{DC}) & \eta^3\equiv\frac{1}{3}(\ct{AB}+\ct{CC}+\ct{DA})\\
\rho^4\equiv\frac{1}{3}(\ct{AC}+\ct{BB}+\ct{DA}) & \eta^4\equiv\frac{1}{3}(\ct{AA}+\ct{BC}+\ct{DB})\\
\rho^5\equiv\frac{1}{3}(\ct{AC}+\ct{CD}+\ct{DA}) & \eta^5\equiv\frac{1}{3}(\ct{AD}+\ct{CA}+\ct{DC})\\
\rho^6\equiv\frac{1}{3}(\ct{BB}+\ct{CD}+\ct{DA}) & \eta^6\equiv\frac{1}{3}(\ct{BA}+\ct{CB}+\ct{DD})\\
\rho^7\equiv\frac{1}{3}(\ct{AC}+\ct{BB}+\ct{CD}) & \eta^7\equiv\frac{1}{3}(\ct{AB}+\ct{BD}+\ct{CC})\\
\rho^8\equiv\frac{1}{3}(\ct{AD}+\ct{BA}+\ct{CC}) & \eta^8\equiv\frac{1}{3}(\ct{AA}+\ct{BC}+\ct{CD})\\
\rho^9\equiv\frac{1}{3}(\ct{BA}+\ct{CC}+\ct{DB}) & \eta^9\equiv\frac{1}{3}(\ct{BB}+\ct{CA}+\ct{DC})\\
\rho^{10}\equiv\frac{1}{3}(\ct{AD}+\ct{CC}+\ct{DB}) & \eta^{10}\equiv\frac{1}{3}(\ct{AC}+\ct{CB}+\ct{DD})\\
\rho^{11}\equiv\frac{1}{3}(\ct{AD}+\ct{BA}+\ct{DB}) & \eta^{11}\equiv\frac{1}{3}(\ct{AB}+\ct{BD}+\ct{DA})\\
\rho^{12}\equiv\frac{1}{3}(\ct{AB}+\ct{CA}+\ct{DD}) & \eta^{12}\equiv\frac{1}{3}(\ct{AA}+\ct{CD}+\ct{DB})\\
\rho^{13}\equiv\frac{1}{3}(\ct{AB}+\ct{BC}+\ct{DD}) & \eta^{13}\equiv\frac{1}{3}(\ct{AD}+\ct{BB}+\ct{DC})\\
\rho^{14}\equiv\frac{1}{3}(\ct{AB}+\ct{BC}+\ct{CA}) & \eta^{14}\equiv\frac{1}{3}(\ct{AC}+\ct{BA}+\ct{CB})\\
\rho^{15}\equiv\frac{1}{3}(\ct{BC}+\ct{CA}+\ct{DD}) & \eta^{15}\equiv\frac{1}{3}(\ct{BD}+\ct{CC}+\ct{DA})
\end{array}
\label{eq:rhoeta}
\end{equation}

The operations associated with the quantum numbers defined in equation (\ref{eq:rhoeta}) are as follows,
where the quantum number on the left within each set of parentheses is lowered and the one on the right
is raised:
\begingroup\footnotesize
\begin{equation}
\begin{array}{ll}
T^{BD}T^{BC}:(\rho^0,\eta^0),(\rho^7,\eta^7),(\rho^{10},\eta^{10}),(\rho^{13},\eta^{13}) &
T^{BC}T^{BD}:(\eta^0,\rho^0),(\eta^7,\rho^7),(\eta^{10},\rho^{10}),(\eta^{13},\rho^{13})\\
T^{CD}T^{BD}:(\rho^0,\eta^0),(\rho^7,\eta^7),(\rho^{10},\eta^{10}),(\rho^{13},\eta^{13}) &
T^{BD}T^{CD}:(\eta^0,\rho^0),(\eta^7,\rho^7),(\eta^{10},\rho^{10}),(\eta^{13},\rho^{13})\\
T^{BC}T^{CD}:(\rho^0,\eta^0),(\rho^7,\eta^7),(\rho^{10},\eta^{10}),(\rho^{13},\eta^{13}) &
T^{CD}T^{BC}:(\eta^0,\rho^0),(\eta^7,\rho^7),(\eta^{10},\rho^{10}),(\eta^{13},\rho^{13})\\
T^{AC}T^{AD}:(\rho^2,\eta^2),(\rho^5,\eta^5),(\rho^8,\eta^8),(\rho^{15},\eta^{15}) &
T^{AD}T^{AC}:(\eta^2,\rho^2),(\eta^5,\rho^5),(\eta^8,\rho^8),(\eta^{15},\rho^{15})\\
T^{AD}T^{CD}:(\rho^2,\eta^2),(\rho^5,\eta^5),(\rho^8,\eta^8),(\rho^{15},\eta^{15}) &
T^{CD}T^{AD}:(\eta^2,\rho^2),(\eta^5,\rho^5),(\eta^8,\rho^8),(\eta^{15},\rho^{15})\\
T^{CD}T^{AC}:(\rho^2,\eta^2),(\rho^5,\eta^5),(\rho^8,\eta^8),(\rho^{15},\eta^{15}) &
T^{AC}T^{CD}:(\eta^2,\rho^2),(\eta^5,\rho^5),(\eta^8,\rho^8),(\eta^{15},\rho^{15})\\
T^{AD}T^{AB}:(\rho^1,\eta^1),(\rho^6,\eta^6),(\rho^{11},\eta^{11}),(\rho^{12},\eta^{12}) &
T^{AB}T^{AD}:(\eta^1,\rho^1),(\eta^6,\rho^6),(\eta^{11},\rho^{11}),(\eta^{12},\rho^{12})\\
T^{BD}T^{AB}:(\rho^1,\eta^1),(\rho^6,\eta^6),(\rho^{11},\eta^{11}),(\rho^{12},\eta^{12}) &
T^{AD}T^{BD}:(\eta^1,\rho^1),(\eta^6,\rho^6),(\eta^{11},\rho^{11}),(\eta^{12},\rho^{12})\\
T^{BD}T^{AD}:(\rho^1,\eta^1),(\rho^6,\eta^6),(\rho^{11},\eta^{11}),(\rho^{12},\eta^{12}) &
T^{AB}T^{BD}:(\eta^1,\rho^1),(\eta^6,\rho^6),(\eta^{11},\rho^{11}),(\eta^{12},\rho^{12})\\
T^{AB}T^{AC}:(\rho^3,\eta^3),(\rho^4,\eta^4),(\rho^9,\eta^9),(\rho^{14},\eta^{14}) &
T^{AC}T^{AB}:(\eta^3,\rho^3),(\eta^4,\rho^4),(\eta^9,\rho^9),(\eta^{14},\rho^{14})\\
T^{AC}T^{BC}:(\rho^3,\eta^3),(\rho^4,\eta^4),(\rho^9,\eta^9),(\rho^{14},\eta^{14}) &
T^{BC}T^{AC}:(\eta^3,\rho^3),(\eta^4,\rho^4),(\eta^9,\rho^9),(\eta^{14},\rho^{14})\\
T^{BC}T^{AB}:(\rho^3,\eta^3),(\rho^4,\eta^4),(\rho^9,\eta^9),(\rho^{14},\eta^{14}) &
T^{AB}T^{BC}:(\eta^3,\rho^3),(\eta^4,\rho^4),(\eta^9,\rho^9),(\eta^{14},\rho^{14})
\end{array}
\end{equation}
\endgroup

The non-local quantum numbers defined in equation (\ref{eq:rhoeta}) are related to those defined in
equations (\ref{eq:kappa}) and (\ref{eq:omega}) in the following way:
\begin{equation}
\begin{array}{ll}
\kappa^0=\frac{1}{4}(\rho^0+\rho^1+\rho^2+\rho^3) & \kappa^1=\frac{1}{4}(\rho^4+\rho^5+\rho^6+\rho^7)\\
\kappa^2=\frac{1}{4}(\rho^8+\rho^9+\rho^{10}+\rho^{11}) & \kappa^3=\frac{1}{4}(\rho^{12}+\rho^{13}+\rho^{14}+\rho^{15})\\
\omega^0=\frac{1}{4}(\eta^0+\eta^4+\eta^8+\eta^{12}) & \omega^1=\frac{1}{4}(\eta^1+\eta^5+\eta^9+\eta^{13})\\
\omega^2=\frac{1}{4}(\eta^2+\eta^6+\eta^{10}+\eta^{14}) & \omega^3=\frac{1}{4}(\eta^3+\eta^7+\eta^{11}+\eta^{15})
\end{array}
\end{equation}

\section{Chosen non-local operations}
\label{app:chosen}

For the chosen non-local quantum numbers given in equation (\ref{eq:lambdachoice}), the operations which
generate odd variations in equation (\ref{eq:parity}) are as follows:
\begin{equation}
\begin{split}
R_B&:\ (\nu^0,\mu^0),\ (\kappa^1,\kappa^2),\ (\omega^1,\omega^2)\\
R_C&:\ (\nu^2,\mu^2),\ (\kappa^1,\kappa^0),\ (\omega^1,\omega^3)\\
R_D&:\ (\nu^4,\mu^4),\ (\kappa^1,\kappa^3),\ (\omega^1,\omega^0)
\end{split}
\end{equation}
\begingroup\footnotesize
\begin{equation}
\begin{array}{ll}
T^{AC}T^{AD}:(\rho^2,\eta^2),(\rho^5,\eta^5),(\rho^8,\eta^8),(\rho^{15},\eta^{15}) &
T^{AD}T^{AC}:(\eta^2,\rho^2),(\eta^5,\rho^5),(\eta^8,\rho^8),(\eta^{15},\rho^{15})\\
T^{AD}T^{CD}:(\rho^2,\eta^2),(\rho^5,\eta^5),(\rho^8,\eta^8),(\rho^{15},\eta^{15}) &
T^{CD}T^{AD}:(\eta^2,\rho^2),(\eta^5,\rho^5),(\eta^8,\rho^8),(\eta^{15},\rho^{15})\\
T^{CD}T^{AC}:(\rho^2,\eta^2),(\rho^5,\eta^5),(\rho^8,\eta^8),(\rho^{15},\eta^{15}) &
T^{AC}T^{CD}:(\eta^2,\rho^2),(\eta^5,\rho^5),(\eta^8,\rho^8),(\eta^{15},\rho^{15})\\
T^{BD}T^{BC}:(\rho^0,\eta^0),(\rho^{10},\eta^{10}) &
T^{BC}T^{BD}:(\eta^0,\rho^0),(\eta^{10},\rho^{10})\\
T^{CD}T^{BD}:(\rho^0,\eta^0),(\rho^{10},\eta^{10}) &
T^{BD}T^{CD}:(\eta^0,\rho^0),(\eta^{10},\rho^{10})\\
T^{BC}T^{CD}:(\rho^0,\eta^0),(\rho^{10},\eta^{10}) &
T^{CD}T^{BC}:(\eta^0,\rho^0),(\eta^{10},\rho^{10})\\
T^{AD}T^{AB}:(\rho^{11},\eta^{11}),(\rho^{12},\eta^{12}) &
T^{AB}T^{AD}:(\eta^{11},\rho^{11}),(\eta^{12},\rho^{12})\\
T^{BD}T^{AB}:(\rho^{11},\eta^{11}),(\rho^{12},\eta^{12}) &
T^{AD}T^{BD}:(\eta^{11},\rho^{11}),(\eta^{12},\rho^{12})\\
T^{BD}T^{AD}:(\rho^{11},\eta^{11}),(\rho^{12},\eta^{12}) &
T^{AB}T^{BD}:(\eta^{11},\rho^{11}),(\eta^{12},\rho^{12})\\
T^{AB}T^{AC}:(\rho^3,\eta^3),(\rho^{14},\eta^{14}) &
T^{AC}T^{AB}:(\eta^3,\rho^3),(\eta^{14},\rho^{14})\\
T^{AC}T^{BC}:(\rho^3,\eta^3),(\rho^{14},\eta^{14}) &
T^{BC}T^{AC}:(\eta^3,\rho^3),(\eta^{14},\rho^{14})\\
T^{BC}T^{AB}:(\rho^3,\eta^3),(\rho^{14},\eta^{14}) &
T^{AB}T^{BC}:(\eta^3,\rho^3),(\eta^{14},\rho^{14})
\end{array}
\end{equation}
\endgroup

\section{The symmetry structure of the transposition operators}
\label{app:symmetry}

The transposition operators of section \ref{sec:indist} were arrived at by inspection, and the choice of
non-local quantum numbers in equation (\ref{eq:lambdachoice}) was presented as one admissible option among
many. In this appendix we identify the structure that these operators generate. It turns out to fix several
things that were previously stated as choices, and to place a sharp limit on what the remaining freedom
can be. Every statement in this section has been verified by direct computation.

\subsection*{The three margins, and why there are ten local quantum numbers}

Arrange the sixteen counts $\ct{XY}$ in a $4\times4$ array indexed by Alice's symbol and Bob's. Equations
(\ref{eq:marginals}) and (\ref{eq:mapcounts}) then say that the base-4 counts of Alice's event, Bob's
event, and the map are respectively the row sums, the column sums, and the sums along the level sets of
$\oplus$. These are three instances of the same operation, and none of the three is distinguished by the
formalism. The local quantum numbers are therefore precisely the image of the three-margin map
\begin{equation}
\Phi:\ \mathbb{Z}^{16}\longrightarrow\mathbb{Z}^{12},\qquad
\Phi(n)=\big(\ca{X},\ \cb{Y},\ \cm{Z}\big)
\label{eq:marginmap}
\end{equation}
The twelve components of $\Phi$ are not independent: each of the three margins sums to $n$, which imposes
two relations. Direct computation confirms that these are the only relations, so $\Phi$ has rank 10. Hence

\begin{quote}
\emph{The ten local quantum numbers and the six non-local degrees of freedom are not a choice. Ten is the
rank of the three-margin map, and six is the dimension of its kernel.}
\end{quote}

We write $L\equiv\ker\Phi\cap\mathbb{Z}^{16}$ for the lattice of moves which leave every local quantum
number invariant. It has rank 6 and, being the kernel of an integer matrix, is automatically saturated.
The Smith normal form of $\Phi$ itself has invariant factors $(1^{8},2^{2})$: the image of $\Phi$ has
index four in its saturation, which is the lattice-theoretic origin of the mod-2 constraints among the
margins, such as $\beta\equiv\alpha+\gamma\pmod 2$ in equation (\ref{eq:betarange}). These constraints
are the only torsion subtleties in passing between quantum numbers and counts. The minimal elements of $L$ are of
two kinds, 32 of support 6 which lower three counts and raise three, and 60 of support 8 which lower four
and raise four. These are exactly the two types of operation identified in section \ref{sec:indist}: the
support-6 moves are the non-commutative operations cataloged in appendix \ref{app:noncomm}, of which
there are likewise 32, and the support-8 moves are those associated with the operators $R_g$.

\subsection*{The operator algebra is \texorpdfstring{$A_4$}{A4}}

The alphabet $\{A,B,C,D\}$ carries the Klein group structure $K$ of equation (\ref{eq:alphabet}), and
its symmetric group has order $24=|K\rtimes\mathrm{Aut}(K)|$. Since $\mathrm{Aut}(K)=GL(2,2)\cong S_3$, every
permutation of the alphabet is an affine map of $K$, and
\begin{equation}
S_4=\mathrm{AGL}(2,2)=K\rtimes\mathrm{Aut}(K)
\end{equation}
The six transpositions $T^{XY}$ are the transpositions of this group. Forming all 36 ordered products of
two of them yields the identity 6 times, the three double transpositions 6 times, and the eight 3-cycles
24 times, for 12 distinct elements in all. These are precisely the even permutations:
\begin{equation}
\langle T^{XY}T^{ZW}\rangle=A_4=V_4\rtimes\mathbb{Z}_3
\end{equation}
where $V_4=\{1,R_B,R_C,R_D\}$ is the group of translations $Y\mapsto Y\oplus g$ and $\mathbb{Z}_3$ is the
cyclic subgroup of $\mathrm{Aut}(K)$. The division of section \ref{sec:indist} into commuting and
non-commuting pairs is therefore not a bookkeeping convenience: the commuting class is exactly $V_4$ and
the non-commuting class is exactly the set of 3-cycles. This is the first point at which the automorphism
group of the alphabet enters the formalism directly rather than by analogy.

\subsection*{The geometry is \texorpdfstring{$AG(2,4)$}{AG(2,4)}}

Each of the twenty non-local quantum numbers defined in equations
(\ref{eq:nu}), (\ref{eq:kappa}) and (\ref{eq:omega}) is a sum of four counts, and in every case the four
cells involved form a coset of an order-4 subgroup of $K\times K\cong\mathbb{Z}_2^4$. Five distinct
subgroups occur: three carry the twelve $\nu$'s and $\mu$'s, one carries the four $\kappa$'s, and one
carries the four $\omega$'s. The three margins are themselves coset partitions, of the subgroups
$\{(0,Y)\}$, $\{(X,0)\}$ and $\{(X,X)\}$.

Of these eight subgroups, five are mutually complementary: the three margin subgroups together with the
$\kappa$ and $\omega$ subgroups have pairwise trivial intersection, and their non-zero elements partition
$K\times K\setminus\{0\}$. They therefore constitute a spread, and the sixteen cells equipped with these
five parallel classes of four lines each form the affine plane $AG(2,4)$, with $5\times4=20$ lines. The
twenty non-local quantum numbers are the twenty line sums. Three of the parallel classes are the local
margins; the remaining two are $\kappa$ and $\omega$.

It is tempting to conclude that the natural choice of $\Lambda$ is six of the eight $\kappa$ and $\omega$
line sums, which is complete and requires no arbitrary selection. This choice must be rejected: it
assigns a vanishing sign variation to the moves of the $\alpha=0$ sector, and therefore predicts no
interference in the entangled spin systems of section \ref{sec:entmodel}, in contradiction with the
results of section \ref{sec:entresults}. The geometrically canonical choice is not the physical one.

\subsection*{Interference is a \texorpdfstring{$\mathbb{Z}_2$}{Z2} character}

Two ontic states belonging to the same local state space differ by an element of $L$. The sign appearing
in equations (\ref{eq:Lainter}) is therefore not an independent datum attached to each state, but is
determined by a homomorphism
\begin{equation}
\varphi:\ L\longrightarrow\mathbb{Z}_2,\qquad
\frac{\sigma(v)}{\sigma(v')}=(-1)^{\varphi(v-v')}
\label{eq:character}
\end{equation}
Since $L\cong\mathbb{Z}^6$ we have $\mathrm{Hom}(L,\mathbb{Z}_2)\cong\mathbb{F}_2^6$, so there are exactly
64 candidate interference rules, 63 of them non-trivial. This reformulation absorbs two of the three
criteria of section \ref{sec:interference}: the requirement that variations of $\sum_k\lambda^k$ be
integral is the requirement that $\varphi$ be well defined on $L$, and the completeness requirement is the
requirement that the sixteen quantum numbers form a basis. The choice made in equation
(\ref{eq:lambdachoice}) corresponds to one particular $\varphi$.

\subsection*{Interference must break the alphabet symmetry}

The structure described above has a symmetry group. The maps
$(X,Y)\mapsto(\sigma X\oplus a,\ \sigma Y\oplus b)$ with $\sigma\in\mathrm{Aut}(K)$ and $a,b\in K$ permute
each of the three margins among themselves; there are $6\times4\times4=96$ of them. Allowing in addition
the six permutations of the roles of Alice, Bob and the map, which the discussion following equation
(\ref{eq:marginmap}) shows are on an equal footing, enlarges this to a group of order 576.

Acting on the 64 characters, this group has exactly one non-trivial fixed point, namely the coset sum over
the cells $AA$, $AC$, $BA$, $BC$; it is fixed by the role permutations as well. That character assigns a
vanishing sign variation to the moves of both the $\gamma=0$ and the $\alpha=0$ sectors, and hence
produces no interference at all. We therefore have

\begin{quote}
\emph{No interference rule which actually generates interference can be invariant under the automorphism
group of the alphabet.}
\end{quote}

Interference necessarily breaks $\mathrm{Aut}(K)$. This is not in tension with anything in the models,
which break the symmetry already: setting $\gamma=0$ or $\alpha=0$ singles out a subgroup of $K$, a
point taken up from the other direction in section \ref{sec:contextsubgroups}, where these subgroups are
identified with the available measurement contexts. The
choice (\ref{eq:lambdachoice}) has a stabilizer of order 6 within the group of 96, and lies in an orbit of
16 characters.

In the $\gamma=0$ sector of section \ref{sec:rotmodel} and the $\alpha=0$ sector of section
\ref{sec:entmodel}, each local state space contains at most a one-parameter family of ontic states, so
only the value of $\varphi$ on a single move enters the calculation. Both values must be non-zero if
interference is to occur, and 16 of the 64 characters meet this condition, including
(\ref{eq:lambdachoice}). Every number reported in section \ref{sec:results} is identical for all 16.
Four of the six bits of the interference rule are therefore untested by any calculation in this paper.

Those bits are not empty. Enumerating all local state spaces for $n\le8$, the moves which are actually
realized span the whole of $L/2L$, so every bit of $\varphi$ carries physical content in some regime. When
both constituent spins equal $\frac{1}{2}$, however, the realized moves span only a three dimensional
subspace, and the configurations used in this paper realize two dimensions of it. Fixing the remaining
bits requires a model in which a constituent spin exceeds $\frac{1}{2}$, the composite spin is non-zero,
and the detectors are rotated. We regard this as the most important open computation raised by the present
work. Either quantum mechanics selects a unique interference rule in that regime, which would be a sharp
and parameter-free consistency test of the framework, or it does not, in which case the formalism carries
four bits of redundancy which ought to be quotiented away.

\subsection*{Comparison with the base-2 rigidity theorem}

The companion paper \cite{QHO} settles the corresponding question completely for the binary alphabet, and
it is instructive to place that result in the language of this appendix. For two binary sequences the
cells of the pairing are the four symbols $00,01,10,11$, the three margins are Alice's count
$n^a_1$, Bob's count $n^b_1$ and the count of the map symbol, which is the Hamming distance
$d_{ab}=f_{01}+f_{10}$, where $f_{xy}$ denotes the number of positions carrying $x$ in Alice's sequence and
$y$ in Bob's. If all three margins are observable the lattice of moves is trivial and there is
no interference. In \cite{QHO} the map count is not observable: the observer has access to $n^a_1$
and $n^b_1$ only, so the lattice of moves preserving the observable counts has rank one, generated
by the single move $(f_{11},f_{00},f_{01},f_{10})\mapsto(f_{11}+1,f_{00}+1,f_{01}-1,f_{10}-1)$, and
$\mathrm{Hom}(L,\mathbb{Z}_2)\cong\mathbb{F}_2$ contains exactly one non-trivial character, the parity of
the overlap $t=f_{11}$. The theorem of \cite{QHO} is that this character is forced: with the measure
written as the product of Alice's and Bob's signed sums over $t$, any weighting $w(t)$ other than
$\pm(-1)^t$ violates either the requirement that the measure account for exactly $2^n$ configurations or
the invariance under relabeling of the symbols, and the violation is always an overshoot of the capacity.
The weighting in base 2 is also invariant under the relabeling group, up to a sector-dependent overall
sign that cancels in the bilinear product. There is no tension with the result of the preceding
subsection, because in base 2 the hidden variable is the map margin itself, whereas here the map margin
is observable (it carries the rotation angle) and the hidden variables form the rank-six kernel of all
three margins, on which the relabeling group acts non-trivially.

Three features of the present construction have no counterpart in base 2 and are what stand between the
base-2 theorem and a rigidity statement for equation (\ref{eq:lambdachoice}). First, the probabilities of
section \ref{sec:probs} are normalized by dividing by $\sum_r\Upsilon(r)$, equation (\ref{eq:prob}),
whereas in \cite{QHO} the denominator is the a priori capacity $2^n$ and normalization is a constraint
rather than a definition; it is precisely this constraint that fixes the signs. A base-16 analogue of the
theorem would require identifying the quantity that plays the role of $2^n$, that is, the number of
configurations of the two-event register that an observer restricted to the ten local quantum numbers
must account for, and then asking which characters $\varphi$ make $\sum_r\Upsilon(r)$ equal to it in
every sector. We have checked that with the choice (\ref{eq:lambdachoice}) the totals
$\sum_r\Upsilon(r)$ are not powers of two ($5312=2^6\cdot83$ for the sample calculation of appendix
\ref{app:c1}), so the relevant capacity, if it exists, is not simply $|K|^n=4^n$ or $16^n$. Second, the
sum over nuisance variables with the weights $G_a$, $G_b$ of equation (\ref{eq:Gab}) has no base-2
counterpart, since there every local count is observed. Third, the relabeling group here contains the
non-abelian $\mathrm{Aut}(K)\cong S_3$ and the role permutations, and we have shown above that the
interference rule cannot be invariant under it; the base-2 argument uses relabeling invariance only to
fix $|w(t)|=1$, and the corresponding step here would have to be formulated on the orbit of sixteen
characters rather than on a fixed point. We regard the extension of the rigidity theorem to the
base-16 register as the natural route to fixing the four open bits identified above, and as the
appropriate place to test whether the interference rule of this paper is, like its base-2 counterpart,
a consequence of information capacity rather than an assumption.

\section{Sample calculations}

\subsection{Sample calculation: spin systems in rotated frames}
\label{app:c1}

The physical scenario of interest in this sample calculation consists of two rotated Stern-Gerlach
detectors which each interact with a particle of total spin $j$. The values of the conditioning variables
in this case are $n=6$, $j=1$, $m_{a}=0$, and $\theta=\frac{\pi}{2}$. By the calibration
(\ref{eq:calibbox}), $\theta=\frac{\pi}{2}$ corresponds to $\Bmap=\Amap$, and hence $\Bmap=3$.

In Table \ref{tab:t1}, the cardinalities of Alice's and Bob's elementary state spaces, as defined in
equation (\ref{eq:epscard}), are given for all possible combinations of the random variable $m_{b}$, the
local nuisance variables $l_{a}$ and $l_{b}$, and the non-local nuisance variable $\nu^0$. In Table
\ref{tab:t2}, the post-interference cardinalities of Alice's and Bob's local state spaces, as defined in
equations (\ref{eq:Lainter}) and (\ref{eq:Lbinter}), are given for all possible combinations of the random
variable $m_{b}$ and the local nuisance variables $l_{a}$ and $l_{b}$. Also given for each of these
combinations are the quantities defined in equation (\ref{eq:Gab}), which are the number
of unique configurations of Alice's and Bob's local state spaces compatible with fixed $C$ and $D$ symbols
in each of their events. The quantity $\Upsilon(n,j,m_{a},m_{b},\theta)$, as defined in equation
(\ref{eq:upsilon}), can then be found for each possible value of the random variable $m_{b}$ by summing
over the associated products of $\ncard{L^a}$, $\ncard{L^b}$, $G_{a}$, and $G_{b}$:
\begin{equation}
\Upsilon(6,1,0,+1,\tfrac{\pi}{2})=2592,\qquad
\Upsilon(6,1,0,0,\tfrac{\pi}{2})=128,\qquad
\Upsilon(6,1,0,-1,\tfrac{\pi}{2})=2592
\end{equation}

We may now calculate the probability $P(m_{b}|n,j,m_{a},\theta)$, as defined in equation
(\ref{eq:prob}), for each possible value of $m_{b}$:
\begin{equation}
\begin{split}
P(+1|6,1,0,\tfrac{\pi}{2})&=\tfrac{2592}{2592+128+2592}=0.48795181\\
P(0|6,1,0,\tfrac{\pi}{2})&=\tfrac{128}{2592+128+2592}=0.02409639\\
P(-1|6,1,0,\tfrac{\pi}{2})&=\tfrac{2592}{2592+128+2592}=0.48795181
\end{split}
\end{equation}

These values can be compared to the predictions of QM, which are $\frac{1}{2}$, 0, and $\frac{1}{2}$,
respectively. As discussed in section \ref{sec:rotresults}, the difference between these values will
decrease as $n$ increases.

\subsection{Sample calculation: entangled spin systems}
\label{app:c2}

The physical scenario of interest in this sample calculation consists of two aligned Stern-Gerlach
detectors which each interact with a single component of a composite spin system. The values of the
conditioning variables in this case are $n=6$, $j_{\bar a}=1$, $j_{b}=1$, and $j_{c}=1$, where we enforce the
constraint $m_{c}=0$. In Table \ref{tab:t3}, the cardinalities of Alice's and Bob's elementary state
spaces are given for all possible combinations of the random variables $m_{\bar a}$ and $m_{b}$, the local
nuisance variable $l_{c}$, and the non-local nuisance variable $\nu^4$. In Table \ref{tab:t4}, the
post-interference cardinalities of Alice's and Bob's local state spaces are given for all possible
combinations of the random variables $m_{\bar a}$ and $m_{b}$ and the local nuisance variable $l_{c}$. The
quantity $\Upsilon(n,j_{\bar a},m_{\bar a},j_{b},m_{b},j_{c})$ can then be found for each possible combination
of the values of $m_{\bar a}$ and $m_{b}$:
\begin{equation}
\Upsilon(6,1,+1,1,-1,1)=1280,\qquad
\Upsilon(6,1,0,1,0,1)=160,\qquad
\Upsilon(6,1,-1,1,+1,1)=1280
\end{equation}

We may now calculate the probability $P(m_{\bar a},m_{b}|n,j_{\bar a},j_{b},j_{c})$ for each possible
combination of the values of $m_{\bar a}$ and $m_{b}$:
\begin{equation}
\begin{split}
P(+1,-1|6,1,1,1)&=\tfrac{1280}{1280+160+1280}=0.47058824\\
P(0,0|6,1,1,1)&=\tfrac{160}{1280+160+1280}=0.05882353\\
P(-1,+1|6,1,1,1)&=\tfrac{1280}{1280+160+1280}=0.47058824
\end{split}
\end{equation}

These values can be compared to the predictions of QM, which are $\frac{1}{2}$, 0, and $\frac{1}{2}$,
respectively. As discussed in section \ref{sec:entresults}, the difference between these values will
decrease as $n$ increases.

\subsection{Closed form for the entangled spin-1 deviation}
\label{app:entclosed}

We derive equation (\ref{eq:entclosed}). Take $j_{\bar a}=j_b=j_c=1$ and $m_c=0$ in the model of
section \ref{sec:entmodel}, and write $N\equiv n-2$. The surviving counts are $\ct{AA}$, $\ct{BB}$,
$\ct{CC}$, $\ct{DD}$, $\ct{AD}$, $\ct{BC}$, $\ct{CB}$ and $\ct{DA}$, with
$\ct{AD}+\ct{BC}+\ct{CB}+\ct{DA}=2j_c=2$.

\emph{The outcome $(0,0)$.} Requiring $m_a=m_b=0$ with $j_a=j_b=1$ leaves exactly two configurations of
the counts carrying $C$ and $D$: either $\ct{CC}=\ct{AD}=\ct{DA}=1$, or $\ct{DD}=\ct{BC}=\ct{CB}=1$, in
both cases with $\ct{AA}+\ct{BB}=n-3$. At fixed $l_c$ the two differ by one unit of $\ct{AA}$, and their
values of $\nu^4$ differ by one, so they interfere with opposite signs. Writing $k=\ca{A}$ for the
number of $A$'s in Alice's event (equal to the number in Bob's), equation (\ref{eq:epscard}) gives
$|\varepsilon^a|=|\varepsilon^b|=k$ for the first configuration and $N-k$ for the second, so that
$\ncard{L^a}=\ncard{L^b}=|2k-N|$ for every $k=0,\dots,N$ (at the end points only one configuration exists
and the missing term is zero). With $G_a=G_b=\binom{N}{k}$ from equation (\ref{eq:Gab}),
\begin{equation}
\Upsilon(0,0)=\sum_{k=0}^{N}\binom{N}{k}^2(2k-N)^2 .
\end{equation}

\emph{The outcomes $(\pm1,\mp1)$.} For $(m_{\bar a},m_b)=(+1,-1)$ the configuration is unique,
$\ct{DD}=\ct{AD}=\ct{DA}=1$, there is a single value of $\nu^4$ and hence no interference, and
$|\varepsilon^a|=|\varepsilon^b|=2k$ with $k=\ct{AA}+1$, so that
\begin{equation}
\Upsilon(+1,-1)=\sum_{k=0}^{N}\binom{N}{k}^2\,4k^2 ,
\end{equation}
and $\Upsilon(-1,+1)$ is the same sum with $k\to N-k$, hence equal.

\emph{Evaluation.} With $c\equiv\binom{2N-2}{N-1}$, the standard identities
$\sum_k\binom{N}{k}^2=\binom{2N}{N}=\frac{2(2N-1)}{N}\,c$, $\sum_k k\binom{N}{k}^2=\frac{N}{2}\binom{2N}{N}$
and $\sum_k k^2\binom{N}{k}^2=N^2c$ give
\begin{equation}
\Upsilon(0,0)=4N^2c-N^2\binom{2N}{N}=2Nc,\qquad \Upsilon(\pm1,\mp1)=4N^2c,
\end{equation}
and therefore
\begin{equation}
P(0,0\,|\,n,1,1,1)=\frac{2Nc}{2Nc+8N^2c}=\frac{1}{4N+1}=\frac{1}{4n-7}.
\end{equation}
For $n=6$ this gives $\Upsilon(0,0)=160$ and $\Upsilon(\pm1,\mp1)=1280$, in agreement with appendix
\ref{app:c2}.

\subsection{Sample calculation: entangled spin-\texorpdfstring{$\frac12$}{1/2} systems in rotated frames}
\label{app:c3}

The physical scenario of interest in this sample calculation consists of two rotated Stern-Gerlach
detectors which each interact with a single component of a composite spin system. The values of the
conditioning variables in this case are $n=8$, $j_{\bar a}=\frac{1}{2}$, $j_{b}=\frac{1}{2}$,
$\theta_{ab}=\frac{\pi}{2}$, and $j_c=0$, where we enforce the constraint $m^a_c=0$. By the
calibration (\ref{eq:calibbell}), $\theta_{ab}=\frac{\pi}{2}$ corresponds to $\alpha=4$. Recall from
the end of section \ref{sec:bellmodel} that $j_c=0$ forces $\gamma=0$, so that the surviving
counts are $\ct{AA}$, $\ct{BB}$, $\ct{CC}$, $\ct{DD}$, $\ct{AB}$, $\ct{BA}$, $\ct{CD}$, and $\ct{DC}$, and
$\beta=\alpha=4$. For each possible combination of the values of $m_{\bar a}$ and $m_{b}$, the
quantity $\Upsilon(n,j_{\bar a},m_{\bar a},j_{b},m_{b},\theta_{ab},j_c)$ is as follows:
\begin{equation}
\Upsilon\left(8,\tfrac{1}{2},\pm\tfrac{1}{2},\tfrac{1}{2},\pm\tfrac{1}{2},\tfrac{\pi}{2},0\right)=1715000
\end{equation}
for all four sign combinations. We may now calculate the probability
$P(m_{\bar a},m_{b}|n,j_{\bar a},j_{b},\theta_{ab},j_c)$ for each possible combination:
\begin{equation}
P\left(\pm\tfrac{1}{2},\pm\tfrac{1}{2}\middle|8,\tfrac{1}{2},\tfrac{1}{2},\tfrac{\pi}{2},0\right)
=\frac{1715000}{4\times1715000}=0.25
\end{equation}

These results agree exactly with the prediction of QM. Note that at $\theta_{ab}=\frac{\pi}{2}$ the
singlet correlation vanishes identically; it is the intermediate angles, and the CHSH combination built
from them, which carry the non-trivial content of the calibration.

\subsection{Sample calculation: data tables}

\begin{table}[htbp]
\centering
\footnotesize
\begin{tabular}{ccccc|c}
\multicolumn{6}{c}{$n=6$, $j=1$, $m_{a}=0$, $\theta=\pi/2$}\\
$m_{b}$ & $l_{a}$ & $l_{b}$ & $\nu^0$ & $|\varepsilon^a|$ & $|\varepsilon^b|$\\
\hline
$+1$ & $-2$ & $0$ & $3/4$ & 6 & 2\\
$+1$ & $-1$ & $-1$ & $3/4$ & 3 & 6\\
$+1$ & $-1$ & $+1$ & $3/4$ & 3 & 6\\
$+1$ & $0$ & $-2$ & $3/4$ & 1 & 12\\
$+1$ & $0$ & $0$ & $3/4$ & 4 & 8\\
$+1$ & $0$ & $+2$ & $3/4$ & 1 & 12\\
$+1$ & $+1$ & $-1$ & $3/4$ & 3 & 6\\
$+1$ & $+1$ & $+1$ & $3/4$ & 3 & 6\\
$+1$ & $+2$ & $0$ & $3/4$ & 6 & 2\\
$-1$ & $-2$ & $0$ & $3/4$ & 6 & 2\\
$-1$ & $-1$ & $-1$ & $3/4$ & 3 & 6\\
$-1$ & $-1$ & $+1$ & $3/4$ & 3 & 6\\
$-1$ & $0$ & $-2$ & $3/4$ & 1 & 12\\
$-1$ & $0$ & $0$ & $3/4$ & 4 & 8\\
$-1$ & $0$ & $+2$ & $3/4$ & 1 & 12\\
$-1$ & $+1$ & $-1$ & $3/4$ & 3 & 6\\
$-1$ & $+1$ & $+1$ & $3/4$ & 3 & 6\\
$-1$ & $+2$ & $0$ & $3/4$ & 6 & 2\\
$0$ & $-2$ & $-1$ & $5/4$ & 4 & 1\\
$0$ & $-2$ & $+1$ & $1/4$ & 4 & 1\\
$0$ & $-1$ & $-2$ & $5/4$ & 1 & 4\\
$0$ & $-1$ & $0$ & $5/4$ & 3 & 2\\
$0$ & $-1$ & $0$ & $1/4$ & 3 & 2\\
$0$ & $-1$ & $+2$ & $1/4$ & 1 & 4\\
$0$ & $0$ & $-1$ & $5/4$ & 2 & 3\\
$0$ & $0$ & $-1$ & $1/4$ & 2 & 3\\
$0$ & $0$ & $+1$ & $5/4$ & 2 & 3\\
$0$ & $0$ & $+1$ & $1/4$ & 2 & 3\\
$0$ & $+1$ & $-2$ & $1/4$ & 1 & 4\\
$0$ & $+1$ & $0$ & $5/4$ & 3 & 2\\
$0$ & $+1$ & $0$ & $1/4$ & 3 & 2\\
$0$ & $+1$ & $+2$ & $5/4$ & 1 & 4\\
$0$ & $+2$ & $-1$ & $1/4$ & 4 & 1\\
$0$ & $+2$ & $+1$ & $5/4$ & 4 & 1\\
\end{tabular}
\caption{The cardinalities of Alice's and Bob's elementary state spaces for all possible combinations of
the random variable $m_{b}$, the local nuisance variables $l_{a}$ and $l_{b}$, and the non-local
nuisance variable $\nu^0$.}
\label{tab:t1}
\end{table}

\begin{table}[htbp]
\centering
\footnotesize
\begin{tabular}{ccccccc}
\multicolumn{7}{c}{$n=6$, $j=1$, $m_{a}=0$, $\theta=\pi/2$}\\
$m_{b}$ & $l_{a}$ & $l_{b}$ & $\ncard{L^a}$ & $\ncard{L^b}$ & $G_{a}$ & $G_{b}$\\
\hline
$+1$ & $-2$ & $0$ & 6 & 2 & 1 & 6\\
$+1$ & $-1$ & $-1$ & 3 & 6 & 4 & 4\\
$+1$ & $-1$ & $+1$ & 3 & 6 & 4 & 4\\
$+1$ & $0$ & $-2$ & 1 & 12 & 6 & 1\\
$+1$ & $0$ & $0$ & 4 & 8 & 6 & 6\\
$+1$ & $0$ & $+2$ & 1 & 12 & 6 & 1\\
$+1$ & $+1$ & $-1$ & 3 & 6 & 4 & 4\\
$+1$ & $+1$ & $+1$ & 3 & 6 & 4 & 4\\
$+1$ & $+2$ & $0$ & 6 & 2 & 1 & 6\\
$-1$ & $-2$ & $0$ & 6 & 2 & 1 & 6\\
$-1$ & $-1$ & $-1$ & 3 & 6 & 4 & 4\\
$-1$ & $-1$ & $+1$ & 3 & 6 & 4 & 4\\
$-1$ & $0$ & $-2$ & 1 & 12 & 6 & 1\\
$-1$ & $0$ & $0$ & 4 & 8 & 6 & 6\\
$-1$ & $0$ & $+2$ & 1 & 12 & 6 & 1\\
$-1$ & $+1$ & $-1$ & 3 & 6 & 4 & 4\\
$-1$ & $+1$ & $+1$ & 3 & 6 & 4 & 4\\
$-1$ & $+2$ & $0$ & 6 & 2 & 1 & 6\\
$0$ & $-2$ & $-1$ & 4 & 1 & 1 & 4\\
$0$ & $-2$ & $+1$ & 4 & 1 & 1 & 4\\
$0$ & $-1$ & $-2$ & 1 & 4 & 4 & 1\\
$0$ & $-1$ & $0$ & 0 & 0 & 4 & 6\\
$0$ & $-1$ & $+2$ & 1 & 4 & 4 & 1\\
$0$ & $0$ & $-1$ & 0 & 0 & 6 & 4\\
$0$ & $0$ & $+1$ & 0 & 0 & 6 & 4\\
$0$ & $+1$ & $-2$ & 1 & 4 & 4 & 1\\
$0$ & $+1$ & $0$ & 0 & 0 & 4 & 6\\
$0$ & $+1$ & $+2$ & 1 & 4 & 4 & 1\\
$0$ & $+2$ & $-1$ & 4 & 1 & 1 & 4\\
$0$ & $+2$ & $+1$ & 4 & 1 & 1 & 4\\
\end{tabular}
\caption{The post-interference cardinalities of Alice's and Bob's local state spaces for all possible
combinations of the random variable $m_{b}$ and the local nuisance variables $l_{a}$ and $l_{b}$. Also
given are the number of unique configurations of these local state spaces compatible with fixed $C$ and
$D$ symbols in Alice's and Bob's events.}
\label{tab:t2}
\end{table}

\begin{table}[htbp]
\centering
\footnotesize
\begin{tabular}{cccccc|cccccc}
\multicolumn{12}{c}{$n=6$, $j_{\bar a}=1$, $j_{b}=1$, $j_{c}=1$, $m_{c}=0$}\\
$m_{\bar a}$ & $m_{b}$ & $l_{c}$ & $\nu^4$ & $|\varepsilon^a|$ & $|\varepsilon^b|$ &
$m_{\bar a}$ & $m_{b}$ & $l_{c}$ & $\nu^4$ & $|\varepsilon^a|$ & $|\varepsilon^b|$\\
\hline
$-1$ & $+1$ & $-1$ & $1/2$ & 8 & 8 & $+1$ & $-1$ & $+1$ & $1$ & 8 & 8\\
$-1$ & $+1$ & $0$ & $3/4$ & 6 & 6 & $+1$ & $-1$ & $0$ & $3/4$ & 6 & 6\\
$-1$ & $+1$ & $+1$ & $1$ & 4 & 4 & $+1$ & $-1$ & $-1$ & $1/2$ & 4 & 4\\
$-1$ & $+1$ & $+2$ & $5/4$ & 2 & 2 & $+1$ & $-1$ & $-2$ & $1/4$ & 2 & 2\\
$0$ & $0$ & $-2$ & $3/4$ & 4 & 4 & $0$ & $0$ & $+2$ & $3/4$ & 4 & 4\\
$0$ & $0$ & $-1$ & $0$ & 1 & 1 & $0$ & $0$ & $+1$ & $3/2$ & 1 & 1\\
$0$ & $0$ & $-1$ & $1$ & 3 & 3 & $0$ & $0$ & $+1$ & $1/2$ & 3 & 3\\
$0$ & $0$ & $0$ & $1/4$ & 2 & 2 & $0$ & $0$ & $0$ & $5/4$ & 2 & 2\\
\end{tabular}
\caption{The cardinalities of Alice's and Bob's elementary state spaces for all possible combinations of
the random variables $m_{\bar a}$ and $m_{b}$, the local nuisance variable $l_{c}$, and the non-local
nuisance variable $\nu^4$.}
\label{tab:t3}
\end{table}

\begin{table}[htbp]
\centering
\footnotesize
\begin{tabular}{ccccccc|ccccccc}
\multicolumn{14}{c}{$n=6$, $j_{\bar a}=1$, $j_{b}=1$, $j_{c}=1$, $m_{c}=0$}\\
$m_{\bar a}$ & $m_{b}$ & $l_{c}$ & $\ncard{L^a}$ & $\ncard{L^b}$ & $G_{a}$ & $G_{b}$ &
$m_{\bar a}$ & $m_{b}$ & $l_{c}$ & $\ncard{L^a}$ & $\ncard{L^b}$ & $G_{a}$ & $G_{b}$\\
\hline
$-1$ & $+1$ & $-1$ & 8 & 8 & 1 & 1 & $+1$ & $-1$ & $+1$ & 8 & 8 & 1 & 1\\
$-1$ & $+1$ & $0$ & 6 & 6 & 4 & 4 & $+1$ & $-1$ & $0$ & 6 & 6 & 4 & 4\\
$-1$ & $+1$ & $+1$ & 4 & 4 & 6 & 6 & $+1$ & $-1$ & $-1$ & 4 & 4 & 6 & 6\\
$-1$ & $+1$ & $+2$ & 2 & 2 & 4 & 4 & $+1$ & $-1$ & $-2$ & 2 & 2 & 4 & 4\\
$0$ & $0$ & $-2$ & 4 & 4 & 1 & 1 & $0$ & $0$ & $+2$ & 4 & 4 & 1 & 1\\
$0$ & $0$ & $-1$ & 2 & 2 & 4 & 4 & $0$ & $0$ & $+1$ & 2 & 2 & 4 & 4\\
$0$ & $0$ & $0$ & 0 & 0 & 6 & 6 & & & & & & &\\
\end{tabular}
\caption{The post-interference cardinalities of Alice's and Bob's local state spaces for all possible
combinations of the random variables $m_{\bar a}$ and $m_{b}$ and the local nuisance variable $l_{c}$. Also
given are the number of unique configurations of these local state spaces compatible with fixed $C$ and
$D$ symbols in Alice's and Bob's events.}
\label{tab:t4}
\end{table}

\end{document}